\documentclass[11pt]{article}

\usepackage{a4}
\usepackage{amstext}
\usepackage{amsfonts}
\usepackage{amssymb,amsmath}
\usepackage{color}
\usepackage{cite}
\usepackage{epsfig}
\usepackage{mathtools}
\usepackage[normalem]{ulem}
\usepackage{caption}         
\usepackage{subcaption}   
\usepackage{framed}
\usepackage{multirow}
\usepackage{booktabs}
\usepackage{braket,slashed,bbm,hyperref}
\usepackage[section]{placeins}

\usepackage{placeins}
\usepackage[english]{babel}
\usepackage[babel]{csquotes}
\usepackage{xspace}
\newcommand*\xbar[1]{%
	\hbox{%
		\vbox{%
			\hrule height 0.5pt % The actual bar
			\kern0.5ex%         % Distance between bar and symbol
			\hbox{%
				\kern-0.1em%      % Shortening on the left side
				\ensuremath{#1}%
				\kern-0.1em%      % Shortening on the right side
			}%
		}%
	}%
} 
\DeclareMathAlphabet{\boldmathe}{T1}{cmr}{bx}{it}
\newcommand{\vsig}{\vert\hskip.4mm\xbar{\sigma}\hskip.2mm\vert}
\newcommand{\sig}{\hskip.4mm\xbar{\sigma}\hskip.2mm}
\renewcommand{\eqref}[1]{Eq.~(\ref{#1})}

\newcommand{\mbf}[1]{\boldmathe{#1}}

\newcommand{\Ns}{\ensuremath{N_s}}
\newcommand{\Nt}{\ensuremath{N_t}}
\newcommand{\Nf}{\ensuremath{N_{\mathrm{f}}}}

\newcommand{\e}{\ensuremath{\mathrm{e}}}
\newcommand{\D}{\ensuremath{\mathrm{d}}}
\newcommand{\ii}{\ensuremath{\mathrm{i}}}

\newcommand{\Z}{\ensuremath{\mathbb{Z}}}

\newcommand{\fdi}{\slashed{\partial}}
\newcommand{\cO}{\mathcal{O}}
\newcommand{\cC}{\mathcal{C}}
\newcommand{\vx}{\mbf{x}}
\newcommand{\vy}{\mbf{y}}
\newcommand{\ve}{\mbf{e}}
\newcommand{\vk}{\mbf{k}}
\newcommand{\vm}{\mbf{m}}

\def\ring{\mathaccent"7017}

\begin{document}

% ********************
% ********************
% ********************
% ********************
% ********************

\begin{center}

{\huge \bf Inhomogeneous phases in the}

{\huge \bf $\phantom{g}$ Gross-Neveu model in $1+1$ dimensions $\phantom{g}$}

{\huge \bf at finite number of flavors}

\vspace{0.5cm}

\textbf{Julian Lenz$^1$, Laurin Pannullo$^2$, Marc Wagner$^2$, Bj\"orn Wellegehausen$^1$, \\ Andreas Wipf$^1$}

$^1$ Friedrich Schiller-Universit\"at Jena, Theoretisch Physikalisches Institut, Fr\"obelstieg~1, D-07743 Jena, Germany

$^2$ Goethe-Universit\"at Frankfurt am Main, Institut f\"ur Theoretische Physik, Max-von-Laue-Stra{\ss}e 1, D-60438 Frankfurt am Main, Germany

\vspace{0.5cm}

April 1, 2020

\end{center}

\begin{tabular*}{16cm}{l@{\extracolsep{\fill}}r} \hline \end{tabular*}

\vspace{-0.4cm}
\begin{center} \textbf{Abstract}

We explore the thermodynamics of the $1+1$-dimensional Gross-Neveu (GN) model at finite number of fermion flavors $\Nf$, finite temperature and finite chemical potential using lattice field theory. In the limit $\Nf \rightarrow \infty$ the model has been solved analytically in the continuum. In this limit three phases exist: a massive phase, in which a homogeneous chiral condensate breaks chiral symmetry spontaneously, a massless symmetric phase with vanishing condensate and most interestingly an inhomogeneous phase with a condensate, which oscillates in the spatial direction. In the present work we use chiral lattice fermions (naive fermions and SLAC fermions) to simulate the GN model with $2$, $8$ and $16$ flavors. The results obtained with both discretizations are in agreement. Similarly as for $\Nf \rightarrow \infty$ we find three distinct regimes in the phase diagram, characterized by a qualitatively different behavior of the two-point function of the condensate field. For $\Nf = 8$ we map out the phase diagram in detail and obtain an inhomogeneous region smaller as in the limit $\Nf \rightarrow \infty$, where quantum fluctuations are suppressed. We also comment on the existence or absence of Goldstone bosons related to the breaking of translation invariance in $1+1$ dimensions.

\end{center}
\vspace{-0.4cm}

\begin{tabular*}{16cm}{l@{\extracolsep{\fill}}r} \hline \end{tabular*}

\thispagestyle{empty}

% ********************
% ********************
% ********************
% ********************
% ********************

\newpage

\setcounter{page}{1}

\section{Introduction}\label{sec:intro}
The GN model describes Dirac fermions with $\Nf$ flavors
interacting via quartic interactions in $1+1$ 
dimensions. It was originally introduced as a toy model that shares
several fundamental features with QCD \cite{Gross:1974jv}: it is
renormalizable, asymptotically free, exhibits dynamical symmetry
breaking of the $\Z_2$ chiral symmetry, and has a
large $\Nf$ limit that behaves like the 't~Hooft
large $N_\mathrm{c}$ limit of QCD. The particle
spectrum and thermodynamics
of the theory  in the $\Nf\to\infty$ limit is known analytically.
Similarly, the $1$-flavor model is equivalent to the
$1$-flavor Thirring model which can be solved analytically
in the massless limit \cite{Sachs:1995dm} (it has a vanishing
$\beta$-function).
But for intermediate numbers of flavors $1<\Nf <\infty$ there is
-- despite many analytical and numerical studies -- 
no complete understanding of the thermodynamics and 
particle spectrum.

The GN model and related four-Fermi theories  
in $1+1$ dimensions have been used in 
particle physics, condensed matter physics and
quantum information theory. For example, in condensed
matter physics the GN model describes the charge-soliton-conducting
to metallic phase transition in polyacetylene (CH)$_x$ as a function
of a doping parameter \cite{Chodos:1993mf}. It is equivalent
to the Takayama-Lin-Liu-Maki model \cite{Takayama:1980zz}
which describes the electron-phonon interactions in CH 
in an effective low-energy continuum description,
see Ref.\ \cite{Caldas:2011fm}.
The multi-flavor chiral GN model is related to the
interacting Su-Schrieffer-Heeger model  \cite{Kuno:2018pcp}
which is used to investigate cold atoms in an optical lattice.
Four-Fermi models are intensively studied to better understand
and classify symmetry-protected topological phases of strongly 
interacting systems. For more details we refer to
the nice summary in Ref.\ \cite{Bermudez:2018eyh}. 
%A discretized
%version of the one-flavor GN-model can be realized 
%\todo{XXXXX Andreas: more. XXXXX}

Recently we have seen a renewed interest
in the physics of the GN model at low temperature and high baryon density,
because it is the region of the QCD phase diagram which
is particularly challenging for 
first-principles QCD approaches. Corresponding results 
are only available at asymptotically high densities, where the QCD 
coupling constant is small so that perturbation theory can be applied, at 
vanishing density, where lattice
QCD does not suffer from the sign problem, or for unphysically large quark masses,
where effective theories exist that mitigate the sign problem. At moderate densities
and realistic values of the quark masses, i.e.\ the regime, which is probed by heavy-ion experiments
and which is relevant for supernovae and compact stars, neither approach can be applied. In this regime our current
picture of the QCD phase diagram is, thus, mostly based on QCD-inspired models, e.g.\ the GN model, the Nambu-Jona-Lasinio (NJL) model or the quark-meson model. In early calculations within these models it was assumed that the chiral condensate is homogeneous, i.e.\ constant with respect to the spatial coordinate(s). However, allowing for spatially varying condenates it turned out that there exist regions in
the phase diagram, where inhomogeneous chiral phases are favored \cite{Thies:2003kk,Schnetz:2004vr}. The majority of the existing calculations have been performed in the limit $\Nf \rightarrow \infty$ or, equivalently, the mean-field approximation (see Ref.\ \cite{Buballa:2014tba} for a review and Refs.\ \cite{Basar:2009fg,Nickel:2009wj,Carignano:2010ac,Carignano:2012sx,Heinz:2013hza,Braun:2015fva,Buballa:2018hux,Carignano:2019ivp} for examples of recent work). Using lattice field theory and related numerical methods and considering $\Nf \rightarrow \infty$, the GN model has been explored in $1+1$ and $2+1$ dimensions, the chiral GN model in 1+1 dimensions and the NJL model in $1+1$ and $3+1$ dimension \cite{deForcrand:2006zz,Wagner:2007he,Heinz:2015lua,Winstel:2019zfn,Narayanan:2020uqt}. However, a full lattice simulation and investigation of the phase diagram of any of these models at finite number of fermion flavors, where quantum fluctuations are taken into account, is still missing. The main goal of the present work is to make a step in this direction and to explore, whether such 
inhomogeneous phases also exist in the $1+1$-dimensional GN model at finite $\Nf$.

In this work we shall use naive fermions
and SLAC fermions to study the multi-flavor GN model.
These fermion discretizations are all chiral and no fine tuning
is required to end up with a chirally symmetric continuum limit.
But the theorem of Nielsen and Ninomyia \cite{Nielsen:1981hk} tells us
that we have to pay a price for using (strictly) chiral fermions.
And indeed, with naive fermions we can only simulate $4,8,\dots$
flavors. With SLAC fermions we can simulate $1,2,\dots$ flavors,
but the associated Dirac operator is non-local. It has been
argued elsewhere that there is no problem with SLAC fermions
for lattice systems without local symmetries, see for 
example Ref.\ \cite{Wellegehausen:2017goy}. We shall consider
GN models with $2,8$ and $16$ flavors which have no
sign-problem. The results for $8$ and $16$ flavors can be
compared with the results obtained with naive fermions. We find full
agreement of the results obtained with both fermion species.
Note that with Wilson fermions the full chiral 
symmetry cannot be restored
in the continuum limit with just one bare coupling. 
One needs to introduce a bare mass plus two bare 
couplings and fine tune these three parameters to arrive at a chirally 
symmetric continuum limit \cite{Aoki:1985jj}. An alternative would
be to use fermions which obey the Ginsparg-Wilson relation.
We did not use such fermions, because we sample the 
full $\mu$-$T$ parameter space and carefully check for 
discretization and finite size effects. With Ginsparg-Wilson fermions 
this would be would be too time-consuming.

This paper is organized as follows. In section~\ref{sec:theory}
we summarize some known features of the GN model that are relevant 
for its thermodynamical properties.
These include properties of the fermion determinant
in the continuum and on the lattice, homogeneous and
inhomogeneous phases in the $\Nf \rightarrow \infty$ limit and some
comments concerning the spontaneous symmetry breaking (SSB)
of translation invariance. In section~\ref{sec:lattice}
we discuss different lattice discretizations, the scale setting and some 
details of the simulations. Our numerical results 
are presented in section~\ref{sec:numerics}.
The main focus concerns the behavior of the two-point
function of the order parameter for chiral symmetry
breaking and the resulting consequences for the phase
diagram in the plane spanned by the chemical potential $\mu$ and the temperature $T$. We shall see that the GN model with $\Nf=2$, $8$ and $16$ flavors behaves qualitatively similar
to the model in the $\Nf \rightarrow$ limit. In particular we
localize three regions in $\mu$-$T$ parameter space, where the
two-point function shows a qualitatively different dependence
on the spatial separation. The model with $\Nf=2$ is also simulated on rather large lattices with spatial extent $N_s$ up
to $725$ lattice points to carefully investigate the 
long-range behavior of the correlator. In appendix~\ref{APP567}
we discuss, why the lattice GN model with naive fermions
may have an incorrect continuum limit, and how to modify
the interaction term to end up with an (almost) naive fermion discretization with correct continuum limit.

\clearpage

\section{\label{sec:theory}Theoretical basics}
% ********************

\subsection{The Gross-Neveu model}

The Gross-Neveu model (GN model) is a relativistic quantum field theory describing 
$\Nf$ flavors of Dirac fermions with a four fermion interaction. 
In this work we investigate this asymptotically free model in
2 spacetime dimensions. The fermions are described by a field $\psi=(\psi_1,\dots,\psi_{\Nf})$, 
the components of which are two-component Dirac spinors. Originally it has
been studied in the $1/\Nf$ expansion.
The action and partition function are
\begin{equation}
\label{gn1} S_\psi = \int \mathrm{d}^2x \, \bigg(\bar{\psi} \ii \fdi \psi + \frac{g^2}{2 \Nf} (\bar{\psi} \psi)^2\bigg) \, , \qquad
Z = \int \mathcal{D}\bar{\psi} \mathcal{D}\psi \, \e^{-S_\psi} \, ,
\end{equation}
where the fermion bilinears contain sums over flavor indices, e.g.\ $\bar{\psi} \psi = \sum_i \bar{\psi}_i \psi_i$.

To be able to perform the fermion integration one 
follows Hubbard and Stratonovich by introducing
a fluctuating auxiliary scalar field $\sigma$ to linearize the
operator $\bar{\psi} \psi$ in the interaction term,
\begin{equation}
\label{lsigma} S_\sigma = \int \mathrm{d}^2x \, \bigg(\bar{\psi} \ii D \psi + \frac{\Nf}{2g^2} \sigma^2\bigg) \, , \qquad
Z = \int \mathcal{D}\bar{\psi} \mathcal{D}\psi \, \mathcal{D}\sigma \, e^{-S_\sigma} \, ,
\end{equation}
where
\begin{equation}
\label{Dirac_op} D = \fdi + \sigma + \mu \gamma^0
\end{equation}
is the Dirac operator. The four-Fermi term in (\ref{gn1}) is recovered after
eliminating $\sigma$ by its equation of motion or 
equivalently by integrating over $\sigma$ in the
functional integral. In eqs.\ (\ref{lsigma}) and (\ref{Dirac_op}) we also introduced a chemical potential $\mu$ to study
the system at finite fermion density. Expectation values
of operators $\cO(\psi,\bar{\psi},\sigma)$ in the grand canonical 
ensemble are given by
\begin{equation}
\langle \cO \rangle = \frac{1}{Z} \int \mathcal{D}\bar{\psi} \mathcal{D}\psi \, \mathcal{D}\sigma \, \e^{-S_\sigma} \cO(\psi,\bar{\psi},\sigma) \, .
\end{equation}
Note that the integration is over fermion fields, which are anti-periodic in the Euclidean time direction,
with period $\beta = 1 / T$, while the auxiliary scalar field is periodic.

Integrating over the fermion fields leads to
\begin{equation}
\label{eff-action} S_\textrm{eff} = \frac{1}{2 g^2} \int d^2x \, \sigma^2 - \log\det D \, , \qquad Z =
\int \mathcal{D}\sigma \, e^{-\Nf S_\textrm{eff}}
\end{equation}
with expectation values of operators $\cO(\sigma)$ given by
\begin{equation}
\langle \cO \rangle = \frac{1}{Z} \int \mathcal{D}\sigma \, \e^{-\Nf S_\textrm{eff}} \cO(\sigma) \, .
\end{equation}

Of particular interest in the present work is the
chiral condensate, which distinguishes the different
phases of the GN model.  Translation invariance of the 
integral over $\D\sigma_x$ in the (well-defined) functional
integral on the lattice implies 
a Ward-identity which states, that  the condensate is proportional 
to the average auxiliary field,
\begin{equation}
\langle \bar{\psi}(\vx)\psi(\vx)\rangle
=\frac{\ii\Nf}{g^2} \langle \sigma(\vx)\rangle \, , \qquad
\vx=(x^\mu)=\left(t\atop x\right) \, .
\end{equation}
Also of interest is the Ward-identity relating the 
two-point function of the condensate to the two-point
function of the auxiliary field,
\begin{equation}
\langle (\bar{\psi}\psi)(\vx)(\bar{\psi}\psi)(\vy)\rangle
=\frac{\Nf}{g^2}\,\delta^2(\vx-\vy)
-\bigg(\frac{\Nf}{g^2}\bigg)^2\, \langle\sigma(\vx)\sigma(\vy)\rangle\,.
\end{equation}
In our analysis of the phase diagram at finite temperature and 
density,
the two-point function of the auxiliary field on the right hand side
will play a crucial role (see section~\ref{SEC330}).

% ********************

\subsection{\label{SEC22}The fermion determinant}

In this section we study some relevant spectral properties
of the Euclidean Dirac operator $D$ with auxiliary field and chemical
potential as defined in eq.\ (\ref{Dirac_op}).

Clearly, in the continuum the
free massless Dirac operator $\fdi=\gamma^\mu\partial_\mu$ 
and the partial derivatives $\partial_\mu$
have the following properties:
\begin{enumerate}\itemsep=-1mm
	\item[a)] the differential operator $\fdi$ is anti-hermitean and
	anti-commutes with $\gamma_*=\ii\gamma^0\gamma^1$,
	\item[b)] the partial derivatives $\partial_\mu$ are real,
	$\partial_\mu^*=\partial_\mu$.
\end{enumerate}
Later on we shall discretize Euclidean spacetime on a
lattice such that $\partial_\mu$ turns into a difference operator.
For most discretizations one does not retain the above
properties without introducing doublers -- 
this is what the celebrated Nielsen-Ninomiya theorem tells us \cite{Nielsen:1981hk}.
In the present work, however, we shall use chiral lattice fermions with 
the above properties, naive fermions (having doublers) and SLAC fermions.
Now we investigate the spectral properties of the 
neither hermitian nor anti-hermitian operator $D$ with eigenvalue equation
\begin{equation}
D\psi=
(\fdi+\sigma+\mu\gamma^0)\psi=\lambda\psi\label{2dgn3}
\end{equation}
in the continuum or on the lattice under the assumption that
a) and b) hold true.

\textit{Charge conjugation:}
In $2$ Euclidean spacetime dimensions there exists a symmetric charge
conjugation matrix $\cC$ with $\cC^{-1}\gamma^\mu\cC
=\gamma^{\mu T}$. 
%For example, in a Majorana representation
%with $\gamma^0=\sigma_1$ and $\gamma^1=\sigma_3$
%we may choose $\cC$ to be the identity matrix.
Since the (Euclidean) $\gamma$-matrices are
hermitian we have $\gamma^{\mu *}=\gamma^{\mu T}$, and 
property b) implies
\begin{equation}
D^*=
\gamma^{\mu *}\partial_\mu+\sigma+\mu\gamma^{0*}=
\cC^{-1} D\,\cC\,.
\label{2dgn5}
\end{equation}
%Hence, if $\psi$ is eigenvector of $D$
%with eigenvalue $\lambda$  then $\cC\psi^*$ is 
%eigenvector with complex conjugated eigenvalue $\lambda^*$.
It follows that all non-real eigenvalues come in complex
conjugated pairs $(\lambda,\lambda^*)$ such that 
$\det D$ is real. 
%The determinant can only change sign 
%if an odd number of real eigenvalues change sign. 
Hence there is no sign problem for an 
even number of flavors, since then  $(\det D)^{\Nf}$
is non-negative.

\textit{Chiral symmetry:} The four-Fermi term breaks the
$\textrm{U}_A(\Nf)$ chiral symmetry of the kinetic term. But a discrete
discrete $\Z_2$ chiral symmetry still remains under which $\bar{\psi}\psi$
and $\sigma$ change their signs. Under this discrete chiral
symmetry the Dirac operator is conjugated with $\gamma_*=
\ii\gamma^0\gamma^1$:
\begin{equation}
\gamma_*D\gamma_*=-\fdi+\sigma-\mu\gamma^0.\label{2dgn7}
%\Longrightarrow
%\det D_{\mu,\sigma}=\det D_{\mu,-\sigma}\,.\label{2dgn7}
\end{equation}
%Finally we make a parity transform $x^1\to -x^1$ (the lattice 
%should be symmetric with respect to $x_1=0$).
%This means that the Dirac operator changes sign
%when we conjugate with $\gamma_*$ and at the same
%time flip the sign of the auxiliary field.
Since (on a finite lattice) the number of eigenvalues of $D$ is
even, we conclude that the determinant is an even
function of the auxiliary field,
\begin{equation}
(\det D)[\sigma,\mu]=(\det D)[-\sigma,\mu]\,.\label{2dgn9}
\end{equation}

\textit{Hermitean conjugation:}
The Dirac operator in eq.\ (\ref{2dgn3}) is the sum of
one anti-hermitean and two hermitean terms and
\begin{equation}
D^\dagger[\sigma,\mu]=-D[-\sigma,-\mu]\,.\label{2dgn11}
\end{equation}
Since the determinant is real (and the number
of eigenvalues is even) it follows
that $\det D$ is invariant under a simultaneous
sign change of $\sigma$ and $\mu$,
\begin{equation}
(\det D)[\sigma,\mu]=(\det D)[-\sigma,-\mu]\,.
\end{equation}
Together with (\ref{2dgn9})  this leads to 
a determinant which is an even function of the chemical 
potential,
\begin{equation} 
(\det D)[\sigma,\mu]=
(\det D)[\sigma,-\mu]\,.\label{2dgn13}
\end{equation}

Note that for most lattice fermions not all of the above properties
hold true, for example for Wilson fermions the $\Z_2$ chiral symmetry
is explicitly broken.

% ********************

\subsection{Summary of existing results in large\,-$\Nf$ limit}
\label{s:anal} 
Before 2003 most field-theoreticians and particle
physicists took for granted that in thermal equilibrium
translation invariance is realized such that the chiral condensate $\langle\bar{\psi}\psi\rangle$ is constant.
Assuming translation invariance one can analytically determine
the phase diagram of the GN model at finite temperature
and fermion density in the large\,-$\Nf$
limit \cite{Wolff:1985av}. But in the condensed matter
community it has been known for a while that the Peierls instability
may trigger a breaking of translation invariance. This explains,
for example, the inhomogenous Fulde-Ferrell-Larkin-Ovchinnikov equilibrium
state for ultracold fermions \cite{Fulde:1964zz,larkin:1964zz}
(see \cite{Kinnunen:2017kox} for a recent review).
Subsequently it was shown that for $\Nf\to\infty$ the relativistic 
GN model exhibits an inhomogeneous
condensate at low temperature and high density.
In a series of interesting papers \cite{Thies:2003kk,Schnetz:2004vr}
an explicit expression for the condensate in terms of Jacobi elliptic functions  
has been derived.

\subsubsection{Homogeneous phases in large $\Nf$-limit}

For $\Nf\to\infty$ the saddle point approximation to the 
functional integral with integrand \\ $\exp(-\Nf S_\mathrm{eff})$
in eq.\ (\ref{eff-action}) becomes exact. This means that the
condensate $\langle\sigma\rangle$ is identical to the 
field $\sigma$ which minimizes $S_\mathrm{eff}$. In particular,
if we assume translation invariance then we may minimize
$S_\mathrm{eff}[\sigma]$ on the set of constant fields. 
But for a constant $\sigma$ the regularized action is proportional
to the Euclidean spacetime volume,
\begin{equation}
S_\mathrm{eff}=(\beta L)\, U_\mathrm{eff},\quad
U_\mathrm{eff}=\frac{1}{2g_\Lambda^2}\sigma^2
-\frac{1}{\beta L}\log {\det}_\Lambda D\,.
\end{equation}
We renormalize the theory such that the
minimum of the effective potential 
at zero temperature and zero chemical potential
is at some $\sigma_0>0$. This determines the bare
coupling $g_\Lambda$ as function of the dimensional
parameter $\sigma_0$ and the momentum-cutoff $\Lambda$,
\begin{equation}
\frac{1}{g_\Lambda^2}=\frac{1}{2\pi}
\log\bigg(\frac{2\Lambda}{\sigma_0}\bigg)^2\,.\label{gn2d13}
\end{equation}
The renormalized potential has the simple form 
\begin{equation}
U_\mathrm{eff}=
\frac{\sigma^2}{4\pi}
\Big(\log \frac{\sigma^2}{\sigma_0}-1\Big)
-\frac{1}{\pi}\int_0^\infty\D k \;\frac{k^2}{\varepsilon_k}
\,\bigg(\frac{1}{1+\e^{
		\beta(\varepsilon_k+\mu)}}
+\frac{1}{1+\e^{
		\beta(\varepsilon_k-\mu)}}\bigg)
\end{equation}
with one-particle energies $\epsilon_k=\sqrt{k^2+\sigma^2}$.
In accordance with our previous discussion it is an even function of the 
auxiliary field and of the chemical potential.

%For zero temperature only states up to the
%Fermi-momentum are filled and
%\begin{equation}
%U_\mathrm{eff}\big\vert_{T=0}=\begin{cases}
%\frac{\sigma^2}{4\pi}
%\Big(\log \frac{\sigma^2}{\sigma_0^2}-1\Big)&\; \sigma^2>\mu^2\\
%\frac{\sigma^2}{4\pi}
%\Big(\log \frac{\mu^2}{\sigma_0^2}-1\Big)
%-\frac{\mu^2}{2\pi}\sqrt{1-\frac{\sigma^2}{\mu^2}}
%+\frac{\sigma^2}{2\pi}
%\log\Big(1+\sqrt{1-\frac{\sigma^2}{\mu^2}}\,\Big)
%&\;\sigma^2<\mu^2.
%\end{cases}
%\end{equation}

\begin{figure}
	\centering
	\includegraphics[width=0.467\linewidth]{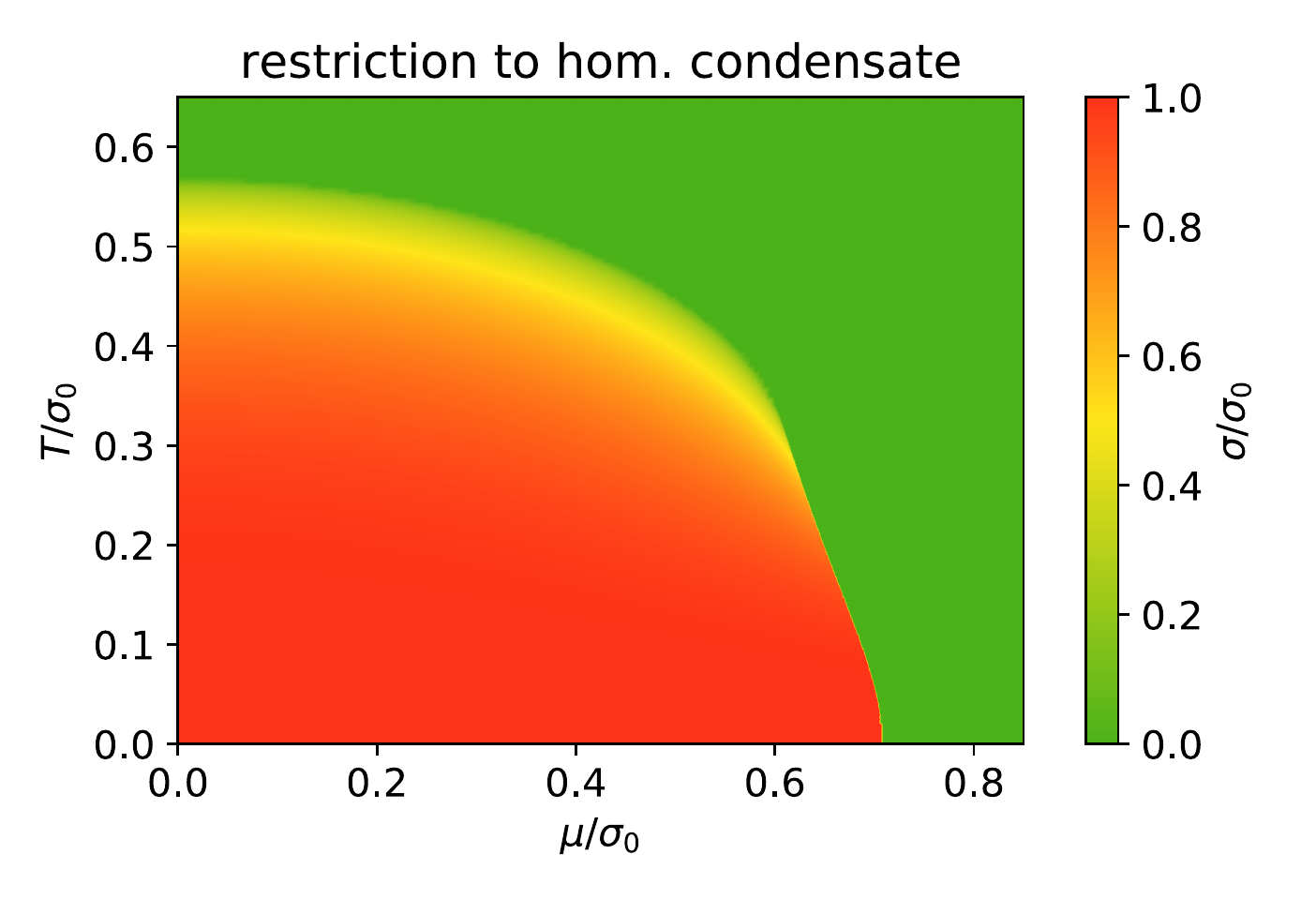}
	\includegraphics[width=0.513\linewidth]{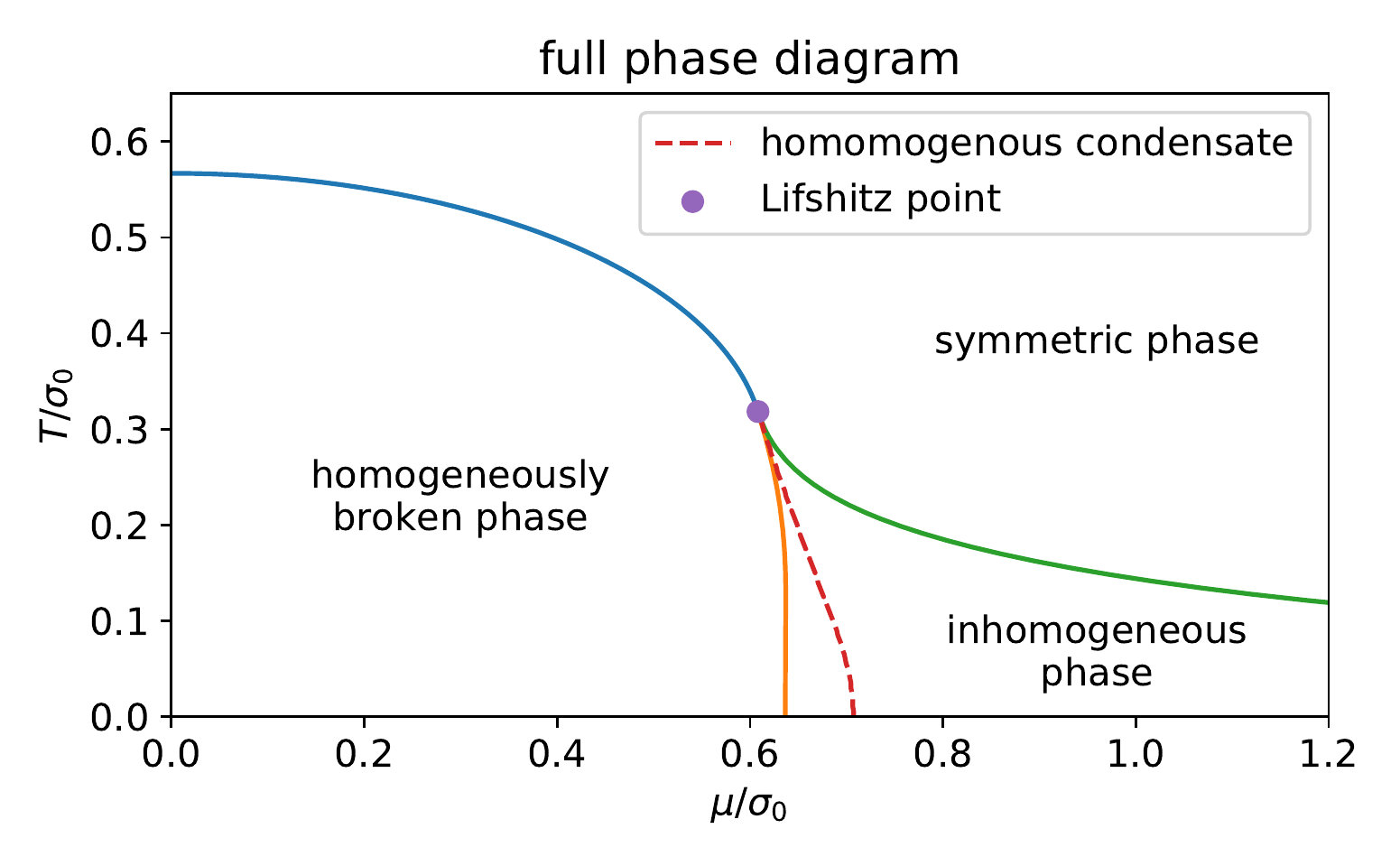}
	\caption{\label{FIG945}(left)~The symmetric phase and the broken phase of the large-$\Nf$ GN model assuming a homogeneous condensate $\sigma$. The value of the condensate in units of $\sigma_0$ is color-coded. For large temperature or large chemical potential (the green region) the condensate vanishes. (right)~The corrected phase diagram of the large-$\Nf$ GN model with homogeneous and inhomogeneous condensates. The first order line from the Lifshitz point to $T=0$ (red dashed line) obtained, when assuming a homogeneous condensate (see plot on the left side), turns into two second order lines (orange and green). In the region with large $\mu$ and low $T$ the condensate is inhomogeneous.}
\end{figure}

The minimizing field as function of the temperature
and chemical potential is depicted in Figure \ref{FIG945}, left. 
Throughout the present work we use $\sigma_0$ to set the scale 
and thus measure the chemical potential, temperature and condensate field
in units of $\sigma_0$. The phase diagram 
shows a symmetric phase with vanishing condensate 
and a broken phase with homogeneous condensate.

The system undergoes a phase transition from the
symmetric phase at high temperature or large
chemical potential to the 
broken phase at low $T$ and small $\mu$
\cite{Wolff:1985av,Barducci:1994cb}. At vanishing
chemical potential the transition happens at
$T_c=e^\gamma/\pi\approx 0.567$.
The second order line extends up to the Lifshitz
point at $(\mu_0,T)\approx (0.318,0.608)$,
where it turns into a first order line. The latter
intersects the zero-temperature axis at 
$\mu_c=1/\sqrt{2}\approx 0.707$.
At $\mu_c$ and $T = 0$ the condensate jumps from $1$
to $0$.

\subsubsection{Inhomogeneous phase in large $\Nf$-limit}

At low temperature and large chemical potential
the minimum of $S_\mathrm{eff}$ does not correspond to a homogeneous but to a spatially inhomogeneous
condensate $\langle\sigma(x)\rangle$. For a time-independent but
spatially varying auxiliary field $\sigma$ the eigenfunctions 
of $D$ have the form $\psi_{nm}(\vx)=e^{\ii\omega_n t}\psi_m(x)$
with Matsubara frequency $\omega_n$. Summing over
these frequencies in $\log\det D$ one arrives
at the renormalized effective action
%\begin{align}
%S_\mathrm{eff}[\sigma]&=
%\beta\lim_{\Lambda\to\infty} \bigg(\frac{1}{4\pi}\log\Big(\frac{2\Lambda}{\sigma_0}
%\Big)^2\int d x\,\sigma^2+\frac{L}{2\pi}\Lambda^2
%-\mathrm{tr}_{>} B\bigg)
%\nonumber\\
%&-\mathrm{tr}_{>}\left( \log\big(1+e^{-\beta (B+\mu)}\big)
%+\log\big(1+e^{-\beta (B-\mu)}\big)\right)\,,
%\label{ar21}
%\end{align}
%xxx
\begin{align}
S_\mathrm{eff}[\sigma]&=
\frac{\beta L}{4\pi}\,{\xbar{\sigma}}^{\,2}
\Big(\log\frac{{\xbar{\sigma}}^{\,2}}{\sigma_0^2}-1\Big)
+\beta\Big(\sum_{n:\varepsilon_m<0}\varepsilon_m
-\sum_{m:\bar\varepsilon_m<0}\bar\varepsilon_m\Big)
\nonumber\\
%-\vert h_{\bar{\sigma}}\vert\big)\nonumber\\
&-\sum_{m:\varepsilon_m>0}\left( \log\big(1+e^{-\beta (\varepsilon_m+\mu)}\big)
+\log\big(1+e^{-\beta (\varepsilon_m-\mu)}\big)\right)\,,
\label{ar21}
\end{align}
where the same renormalization prescription as in the
homogeneous case has been adopted.
The $\varepsilon_m$ are the real eigenvalues
of the hermitean Dirac-Hamiltonian 
\begin{equation}
h_\sigma\psi_m=\varepsilon_m\psi_m,\qquad
h_\sigma
=\gamma^0\gamma^1\partial_x+\gamma^0\sigma\label{diraceq1}
\end{equation}
which appears in the decomposition
\begin{equation}
\gamma^0 D=\partial_0+\mu+h_\sigma
\,.\label{ar1}
\end{equation}
%It is the well-known Hamiltonian in 
%the Dirac theory. 
The $\bar{\varepsilon}_m$
are the eigenvalues of the Dirac Hamiltonian 
with constant auxiliary field $\bar{\sigma}$ given
by
\begin{equation}
{\bar{\sigma}}^2=\frac{1}{L}\int \D x\,\sigma^2(x)\,.
\label{comp_field}
\end{equation}
%The non-zero eigenvalues come in pairs 
%$(\varepsilon_n,-\varepsilon_n)$ and 
%this symmetry has been used in deriving the result (\ref{ar21}).
The two sums over the negative one-particle energies in 
the first line  of (\ref{ar21})
are easily identified as difference of two divergent 
vacuum energies: one for the prescribed auxiliary
field $\sigma(x)$ and the other for the constant
reference field $\xbar{\sigma}$ defined in (\ref{comp_field}).
A heat kernel regularization reveals that the first line
in (\ref{ar21}) is UV-finite if and only if the reference field
is chosen as in (\ref{comp_field}).
The $T$- and $\mu$-\,dependent traces in the second line in (\ref{ar21}) are manifestly UV-finite and represent the 
finite temperature and density corrections. 
%Since the eigenvalues of
%the first-quantized Hamilontians are symmetrical
%around the origin we could replace $\vert h_\sigma\vert$ in %(\ref{ar21}) by $h_\sigma$  and only sum twice 
%over the positive eigenvalues of $h_\sigma$. 
%The $\varepsilon_n$ are the
%square roots of the eigenvalues of the supersymmetric
%Hamilton operator
%\begin{equation}
%h_\sigma^2=-\frac{\D^2}{\D x^2}+\sigma^2(x)-\gamma^1\sigma'(x)
%=\begin{pmatrix}
%AA^\dagger&0\\ 0&A^\dagger A
%\end{pmatrix},\quad
%A=-\frac{\D}{\D x}+\sigma(x)\,,
%\end{equation}
%where the last expression is correct in the Majorana representation with 
%$\gamma_0=\sigma_1$ and $\gamma_1=\sigma_3$.

In the large-$\Nf$ limit only auxiliary fields
which minimize $S_\mathrm{eff}[\sigma]$
contribute to the functional integral in (\ref{lsigma}).
With the Hellman-Feynman formula for the expectation
values $\varepsilon_m=\langle\psi_m\vert h_\sigma\vert\psi_m\rangle$
%\begin{equation}
%\delta\varepsilon_n =(\psi_n,\delta h_\sigma\psi_n)
%=(\psi_n,\gamma^0\delta\sigma\psi_n)\,,
%\end{equation}
%in orthonormal eigenmodes $\psi_m$
%of $h_\sigma$ (and similarly $\bar{\psi}_n$
%of $h_{\bar{\sigma}}$),
the variational derivative of 
$S_\mathrm{eff}$ with respect to $\sigma$ 
can be calculated and one ends up with the Gap equation
\begin{align}
\frac{1}{2\pi}\sigma(x)\log\frac{\bar{\sigma}^2}{\sigma_0^2}&+
\sum_{m:\varepsilon_m<0}\psi_m^\dagger(x)\gamma^0\psi_m(x)
-\sum_{m:\bar\varepsilon_m<0}\bar{\psi}^\dagger_m(x)\gamma^0
\bar{\psi}_m(x)\nonumber\\
&+ \sum_{m:\varepsilon_m>0}
\bigg(\frac{1}{1+e^{\beta(\varepsilon_m+\mu)}}
+\frac{1}{1+e^{\beta(\varepsilon_m-\mu)}}\bigg)\psi_m^\dagger(x)\gamma^0\psi_m(x)=0\,.\label{gap-equation}
\end{align}
%At zero temperature this self-consistency 
%equation simplifies considerably,
%\begin{equation}
%\frac{1}{2\pi}\sigma(x)
%\log\frac{\bar{\sigma}^2}{\sigma_0^2}+
%\sum_{n:\varepsilon_n<-\mu}\psi_n^\dagger(x)
%\gamma^0\psi_n(x)
%-\sum_{n:\bar\varepsilon_n<0}
%\bar{\psi}_n^\dagger(x)\gamma^0\bar{\psi}_n(x)
%=0\,.\label{gap-equation0}
%\end{equation}
%%For a homogeneous condensate $\sigma(x)
%%=\bar{\sigma}$ and the orthonormal 
%%solution of the Dirac Hamiltonian with negative energies
%%on a finite interval of extend $L$ read 
%%\begin{equation}
%%\psi_{n:\varepsilon_n<0}=
%%\frac{1}{\sqrt{2\vert\varepsilon_n\vert L}} 
%%\begin{pmatrix}
%%\sqrt{\sigma-\ii p_n}\\ -\sqrt{\sigma+\ii p_n}
%%\end{pmatrix}e^{\ii p_nx},\quad p_n=\frac{2\pi n}{L}\,.
%%\end{equation}
%%In the thermodynamic limit $L\to\infty$ the
%%gap equation (\ref{gap-equation0}) has $3$ (self-consistent) solutions:
%%\begin{align}
%%&\hbox{minimum at}\,\quad\sigma=0\\
%%&\hbox{minimum at}\,\quad\sigma=\sigma_0,\quad\hbox{ if }
%%\sigma_0^2\geq \mu^2\\
%%&\hbox{maximum at}\quad\sigma=\big(\mu^2-(\mu-\sigma_0)^2
%%\big)^{1/2}\,.
%%\end{align}
This renormalized self-consistency equation is a complicated
functional equation, whose solutions have been investigated at various
times in the literature. Most derivations given previously
derived the regularized gap equation from
the regularized trace of the Green-function with bare
coupling constant and cutoff parameter
\cite{Dashen:1975xh,Pausch:1991ee,Feinberg:2003qz,Basar:2008im}.
Here the point of departure is the renormalized effective
action (\ref{ar21}) with physical scale parameter $\sigma_0$
and only finite quantities enter the derivation of the gap equation.

To summarize, to calculate the chiral condensate at finite temperature
and finite density in the large-$\Nf$ limit one must solve the spectral problem for 
the $\sigma$-dependent Dirac Hamiltonian (\ref{diraceq1})
%\begin{equation}
%h_\sigma\psi_n(x)=\varepsilon_n\psi_n(x)\,\label{mode-equ}
%\end{equation}
and find a self-consistent solution $\sigma(x)$ of the
gap equation (\ref{gap-equation}).
At zero temperature and fermion density Dashen et al.\
indeed could solve the coupled system for the modes $\psi_n(x)$
and the scalar field $\sigma(x)$ by
using powerful inverse scattering methods \cite{Dashen:1975xh}.
They observed that a scalar field
could only solve the gap equation if the solutions
of the Dirac equation $\psi_n$ are not reflected.
Their space-dependent solutions describe
$n$-particle bound states with filled Dirac sea and masses
\begin{equation}
m_B=\frac{2\sigma_0}{\pi}\left(\frac{\sin\theta}{\theta}\right),\qquad
\theta=\frac{n\pi}{2\Nf},\quad
\quad n=2,\dots,\Nf-1\,.
\end{equation}
Self-consistent solutions at finite temperature and fermion density
have been constructed by Thies et al.\ \cite{Thies:2003kk,Schnetz:2004vr}
by some (nonlinear) superposition of kink-antikink solutions.
They succeeded to construct periodic solutions 
$\sigma(x)$ with associated Bloch waves $\psi_n(x)$
of the coupled system (\ref{diraceq1}) and (\ref{gap-equation})
in a certain region of the $(T,\mu)$ phase diagram. 
The Bloch waves (they are solutions of the Lam\'e
equation) and the scalar field $\sigma(x)$ are 
given in terms of Jacobi's elliptic functions, see Ref.\ \cite{Schnetz:2004vr}.
The associated Dirac-Hamiltonian has one gap in the 
spectrum and the periodic and anti-periodic states at the
band-edges are given by particular simple Jacobi functions. Thus, the
property that $h_\sigma$ shows no reflection for baryon
excitations above the vacuum is replaced by the property of having 
exactly one band gap in the spectrum if the system has high density.

In the large-$\Nf$ limit where the saddle point approximation
to the functional integral (\ref{eff-action}) becomes exact 
the inhomogeneous condensate  $\langle \sigma(x)\rangle$ 
minimizes the effective action (\ref{ar21}) and thus
is given by the solution of the gap equation, i.e.\ by a
Jacobi elliptic function. For points in the phase diagram
where the inhomogeneous solution has a lower effective action
as any homogeneous solution the system is in a inhomogeneous
phase. The correct phase diagram in the large-$\Nf$ limit
is depicted in Figure \ref{FIG945}, right.
Note that the metastable phases and first order transition 
line (to guide the eye this
line is kept as dashed line) disappear and are replaced by 
two second order transition lines. At low temperature and
small chemical potential there is a homogeneous phase
with broken chiral symmetry, at sufficiently high temperature
we are in the homogeneous symmetric phase and at low temperature
and large chemical potential we are in the inhomogeneous
phase with an oscillating chiral condensate.

The wave length and amplitude of the condensate in the inhomogeneous 
phase are determined by the chemical potential or equivalently by 
the Fermi-momentum and by the temperature. If one moves within the
inhomogeneous phase towards the symmetric phase, the amplitude
of the condensate vanishes. If one moves towards the homogeneously
broken phase, then the wave length of the condensate increases.
In this work we mainly address the question whether there exists an
inhomogeneous phase for a finite number of flavors $\Nf$
or whether such a phase is an artifact of the large-$\Nf$ limit.

\subsection{\label{SEC499}Spontaneous breaking of a continuous symmetry in 1+1 dimensions}

A well-known theorem by N.\ D.\ Mermin and H.\ Wagner in statistical mechanics
states that a continuous symmetry cannot be
spontaneously broken at finite temperature in 1- and 2-dimensional
statistical systems with short range interaction \cite{Mermin:1966fe}.
A similar theorem has been proven by S.\ Coleman for relativistic 
quantum field theory in $d\leq 2$ dimensions \cite{Coleman:1973ci}. 
Indeed, if spontaneous symmetry breaking of a continuous
symmetry would occur, then as a consequence of the Goldstone theorem
\cite{Goldstone:1961eq,Nambu:1960xd} one would expect 
to find massless Nambu-Goldstone bosons (NGBs) in the particle 
spectrum. But massless scalars with a relativistic dispersion
relation have an infrared divergent 
correlation function in lower dimensions and thus should not exist.

When proving the Goldstone theorem one makes basic
assumptions: the theory should be Lorentz invariant, the Hilbert
space should be positive and a global symmetry group $G$ should be broken
to a subgroup $H$. Then there exists one massless scalar
particle for each broken symmetry (or broken generator)
such that there are
$n_\text{BG}=\text{dim}(G/H)$ massless Goldstone bosons.
In non-relativistic systems and for spacetime
symmetries the situation is more intricate, since sometimes 
NGBs have unusual dispersion relations or 
they are even redundant. 
\begin{itemize} \itemsep=0mm
	\item For example, in a ferromagnet
	and antiferromagnet we may have the same spontaneous symmetry breaking O(3)$\to$O(2),
	but in the first case we have only one NGB
	(the magnon) whereas in the second case there are two NGB. Since quantum field theories at high densities 
	may be described by quasi-excitations with 
	non-relativistic dispersion relations a similar 
	reduction of NGB may happen in the high-density
	GN model.
	\item 
	In addition, for a breaking of spacetime symmetries
	the simple counting rule does not apply.
	For example, crystals have phonons for spontaneously
	broken translations but no gapless excitations for
	equally spontaneously broken rotations. Again a
	reduction of the number of NGB may happen
	if we are dealing with spacetime symmetries instead of 
	inner symmetries. 
	\item Finally, it may happen that the 
	NGB completely decouple from the rest
	of the system. Then one may evade the conclusion
	of Coleman's theorem about the non-existence of
	NGB in $2$ spacetime dimensions. This seems to happen
	in the large $\Nf$-limit of the GN model \cite{Shei:1976mn},
	where translation invariance is definitely broken
	for high fermion density.
\end{itemize}

In 1976,  Nielsen and Chadha \cite{Nielsen:1975hm} presented a
general counting rule of NGBs valid either with 
or without relativistic invariance.
They divided the modes into two classes, based on the
behavior of their dispersion relations for small $\vert k\vert$:
\begin{equation}
\varepsilon_k\propto \begin{cases}\;
\vert k\vert^{2n+1}&n\geq 0,\quad \text{type I}\\
\;\vert k\vert^{2n}&n>0,\quad \text{type II}\,.\label{cg1}
\end{cases}
\end{equation}
Relativistic modes are of type I and non-relativistic
modes are of type II.
By examining analytic properties of correlation
functions they showed that
\begin{equation} 
n_\text{NGB}\leq n_\text{BG}\leq n_\mathrm{I}+2n_\mathrm{II}\,,\label{cg3}
\end{equation}
where $n_\text{NGB}$ is the total number of 
NGBs and $n_\mathrm{I}$ and $n_\mathrm{II}$ are the 
number of type I and type II NGBs.
The number of broken generators $n_\text{BG}=\text{dim}(G/H)$ 
agrees with the number of “flat directions” of fluctuations of 
the order parameter.

In passing we note that there exists a related, but
in general slightly different division of NGBs
into type A and B \cite{Watanabe:2011ec,Watanabe:2012hr}.
It is an algebraic classification based on the Lie algebra
of symmetry generators. To each pair of non-commuting
symmetry generators $\hat Q_i,\hat Q_j$ (the conserved
charges belonging to the symmetry group $G$)
a NGB of type B is associated. The NGBs of type A tend to be 
linearly dispersive for small $\vert k\vert$. There is a 
simple counting for these NGBs:
\begin{equation}
n_A=n_\text{BG}-\text{rank}\,\rho,\quad 
n_B=\frac{1}{2}\text{rank}\,\rho\quad\text{such that}
\quad n_\text{NGB}=n_\text{BG}-\frac{1}{2}\text{rank}\,\rho\,,\label{cg5}
\end{equation}
where $\rho$ is the Watanabe-Brauner matrix build from
the conserved charges related to the symmetry group
$G$, $\rho_{ij}\propto \langle [\hat Q_i,\hat Q_j]\rangle$.
For more details we refer to Refs.\ \cite{Watanabe:2019xul,Ammon:2019wci}. 

Strictly speaking the above results hold for internal 
symmetries only. But it is believed that the NGBs originating
from a spontaneously broken translation symmetry can be treated
in essentially the same way as those associated with
internal symmetries \cite{Watanabe:2019xul}.
This leaves us with the following scenarios for the 
finite-temperature GN model  at high density:
\begin{itemize}
	\item
	Only the (abelian) spatial translation symmetry is broken such that
	the Watanabe-Brauner matrix $\rho$ vanishes. Then the
	results (\ref{cg5}) would imply that there is just one
	type A NGB. If this would be -- as expected -- a NGB of type
	I with relativistic dispersion relation then we would be
	confronted with infrared divergences. The way out could
	be that it is not of type I but of type II or that it
	fully decouples from the system.
	\item
	Alternatively, if the above results do not apply to 
	the breaking of translation invariance at high density
	systems, then we may as well find no NGB or a NGB of type II with
	non-relativistic dispersion relation $\varepsilon_k\sim 
	\vert k\vert^2$. Its correlation function is not infrared 
	divergent and the problem with the spontaneous breaking
	would go away.
\end{itemize}
Since the inhomogeneous condensate appears
(at least in the large-$\Nf$ limit) at high density, a
non-relativistic dispersion relation seems to be more
likely than a relativistic one. Unfortunately, with
the available ensembles on lattices of spatial extent
up to $\Ns = 725$ lattice sites we cannot reliably
measure the dispersion relation of the NGB (if it exists)
in the model with $\Nf=2$ flavors
in the infrared and thus cannot decide whether the NGB has
a non-relativistic or a relativistic dispersion relation.
It may even be that for finite $\Nf$ there is no SSB 
of translation invariance
in the strict sense and that the model behaves like a simple 
atomic liquid, for example as liquid argon. Indeed, the correlator 
of the condensate on large lattices as presented
in section \ref{sec:numerics} resembles the radial pair correlation 
function in an atomic fluid, see the reviews \cite{Barker:1976zza,chandler}.
In a forthcoming accompanying publication we will
further substantiate, by studying
the baryon number as function of the chemical potential, that 
the GN model at high density is either a crystal or an
extremely viscous fluid.

\clearpage

\section{\label{sec:sim}Lattice field theory techniques}
\label{sec:lattice}

% ********************

\subsection{\label{SEC468}Notation}

The number of lattice sites in temporal and spatial direction are denoted by $\Nt$ and $\Ns$, respectively. Consequently, the temperature is given by $T = 1 / \Nt a$ and the extent of the periodic spatial direction by $L = \Ns a$, where $a$ is the lattice spacing.

In the following we consider spacetime averages of observables $O[\sigma]$ for given field configurations $\sigma(\vx)$ with $\vx=(t,x)$,
\begin{equation}
\xbar{O} = \frac{1}{\Nt \Ns} \sum_{\vx} O[\sigma] \, .
\end{equation}
We also compute ensemble averages,
\begin{equation}
\label{EQN080} \big\langle O \big\rangle = \frac{1}{N_\textrm{conf}} 
\sum_{\sigma} O[\sigma] \approx \frac{1}{Z}\int D\sigma \, e^{-S_\textrm{eff}[\sigma]} O[\sigma]\,,
\end{equation}
where the sum over $\sigma$ extends over $N_\textrm{conf}$ field configurations
$\sigma(\vx)$ generated by Monte Carlo sampling according to
$e^{-S_\textrm{eff}[\sigma]}$. The ``$\approx$'' sign in eq.\ (\ref{EQN080}) 
becomes the identity ``$=$'' in the limit $N_\textrm{conf} \rightarrow \infty$.

Moreover, we use the discrete Fourier transform either with respect to spacetime
\begin{equation}
\label{EQN690} \tilde{F}(\vk) = \mathcal{F}(F)(\vk) = \frac{1}{\sqrt{\Nt \Ns}} \sum_{t,x} e^{-\ii \vk \cdot \vx} F(\vx) \, , \qquad \vk = \left(\omega \atop k\right)
\end{equation}
or space
\begin{equation}
\label{EQN690b} \tilde{F}(k) = \mathcal{F}_x(F)(k) = \frac{1}{\sqrt{\Ns}} \sum_{x} e^{-\ii k x} F(x) \, ,
\end{equation}
with time and space restricted to the lattice sites, i.e.\
\begin{equation}
\vx=a \vm,\quad m_0=0,1,\dots,\Nt-1,\quad m_1=0,1,\dots,\Ns-1\,.
\end{equation}
The corresponding momenta are
\begin{equation}
k_0\equiv \omega\in\left\{\frac{2\pi}{a\Nt}n_0\right\},\quad 
k_1\equiv k\in \left\{\frac{2\pi }{a\Ns}n_1\right\}\,.
\end{equation}
For fermions we impose antiperiodic boundary conditions (BC) in 
$t$-direction
such that the integer-spaced $n_0$ are half-integer valued.
For bosons we impose periodic BC in $t$-direction such that
the integer-spaced $n_0$ are integer valued. Both bosons and
fermions are periodic in $x$-direction such that
the integer-spaced $n_1$ are integer valued. 

% ********************

\subsection{Lattice discretizations of fermions}

We use two different lattice discretizations of fermions, naive fermions and SLAC fermions. Both discretizations have certain advantages but come also with subtleties, which are discussed in the following.

In section~\ref{SEC588} we present numerical results for both discretizations and find agreement, which we consider an important cross check. Another comparison of the two discretizations can be found in section~\ref{SEC853}, where we show the $\Nf \rightarrow \infty$ phase diagram with the restriction to homogeneous condensate $\sigma$.

% **********

\subsubsection{Naive fermions}

The naive discretization is at first glance the most straightforward lattice discretization of fermions (see e.g.\ the textbooks \cite{Rothe,Smit:2002ug,Wipf:2013vp}). 
In contrast to other common fermion discretizations, e.g.\ Wilson fermions, the 
free massless naive fermion action is chirally symmetric, which is an essential 
and necessary property in the context of this work. Naive fermions, however, 
lead to fermion doubling according to the Nielsen-Nynomia theorem
\cite{Nielsen:1981hk}. Thus, for most applications, e.g.\ QCD 
with 2, 2+1 or 2+1+1 quark flavors, naive fermions are not appropriate. 
In our case of $1+1$ spacetime dimensions the number of fermion flavors 
is restricted to multiples of $4$. This is not a severe limitation, 
since we are not interested in a particular number of flavors $< 4$, 
but mostly in simulating finite numbers of flavors, e.g.\ $\Nf = 8$ 
or $\Nf = 16$.

Besides fermion doubling there are, however, further pitfalls, which might lead 
to a continuum limit different from the theory of interest. In the context of 
the GN model, this was first observed and discussed in Ref.\ \cite{Cohen:1983nr}. 
In appendix~\ref{APP567} we reproduce the arguments of Ref.\ \cite{Cohen:1983nr} 
and we derive a modification of the straightforward naive discretization of the 
GN model, which has the correct continuum limit.\footnote{At an 
	early stage of this work we used the straightforward naive
	discretization  \cite{Pannullo:2019bfn,Pannullo:2019prx}.} 
This modified naive lattice action is
\begin{align}
S_\textrm{GN}
= \sum_{i=1}^{\Nf/4} S_\textrm{free}[\chi_i,\bar{\chi}_i] 
+ \frac{\ii}{\sqrt{\Nt\Ns}}\sum_{i=1}^{\Nf/4} \sum_{\vx,\vy} \bar{\chi}_i(\vx) W(\vx-\vy)\sigma(\vy) \chi_i(\vx) %\\
+ \frac{\Nf}{2 g^2} \sum_{\vx} \sigma^2(\vx)\label{EQN892c}
\end{align}
with the well-known action for naive free fermions coupled
to a chemical potential
\begin{equation}
S_\textrm{free}[\chi,\bar{\chi}] = \sum_{\vx}
\sum_{\nu=0,1} \frac{\ii}{2a} \bar{\chi}(\vx)\, \gamma^\nu 
\left(e^{a\mu \,\delta_{\nu 0}} \chi(\vx+a\ve_\nu) - e^{-a\mu\,\delta_{\nu0}} \chi(\vx-a\ve_\nu)\right) \label{EQN744}
\end{equation}
and the auxiliary field summed over neighbors with separation $a$ and $\sqrt{2}a$
\begin{equation}
\frac{1}{\sqrt{\Nt\Ns}}\sum_{\vy}W(\vx-\vy)\sigma(\vy) =
\frac{1}{4}\sigma(\vx)
+\frac{1}{8}\sum_{\vy:\vert\vy-\vx\vert=a}\sigma(\vy)
+\frac{1}{16}\sum_{\vy:\vert\vy-\vx\vert=\sqrt{2}a}\sigma(\vy) \, , \label{EQN892b}
\end{equation}
where $\Nf$ is a multiple of $4$
and $\Nt,\Ns$ are even integers such that all doublers obey the same BC (for details see appendix~\ref{APP567}).
For all computations with naive fermions presented in the following 
sections we use the action (\ref{EQN892c}).

% **********

\subsubsection{\label{SEC432}SLAC fermions}

For SLAC-fermions the non-local derivatives in the Dirac operator are easily characterized in momentum space \cite{Drell:1976bq},
\begin{equation}
\mathcal{F}(\partial^\mathrm{slac}_\mu \psi)(\vk) = \ii k_\mu \mathcal{F}(\psi)(\vk)
\end{equation}
with the Fourier transform $\mathcal{F}$ as defined in eq.\ (\ref{EQN690}). 
We choose the discrete momenta $k_\mu$ symmetric to the origin to end up 
with a real and antisymmetric matrix $\partial_\mu$ \cite{Bergner:2007pu}. 
This means that in spatial direction (with periodic BC) 
the lattice has an odd number $\Ns$ of lattice sites, whereas in temporal 
direction (with antiperiodic BC) the lattice has an 
even number $\Nt$ of lattice sites \cite{Wipf:2013vp},
\begin{eqnarray}
\nonumber & & \hspace{-0.7cm} k_0 = \frac{2 \pi}{\Nt a}n_0 \quad , \quad n_0 = \frac{\Nt-1}{2} , \frac{\Nt-3}{2} , \ldots , \frac{1-\Nt}{2} \quad , \quad \Nt \textrm{ even} \\
& & \hspace{-0.7cm} k_1 = \frac{2 \pi}{\Ns a}n_1 \quad , \quad n_1 = \frac{\Ns-1}{2} , \frac{\Ns-3}{2} , \dots , 
\frac{1-\Ns}{2} \quad , \quad \Ns \textrm{ odd} .
\end{eqnarray}
Both the naive and the SLAC derivative define chiral fermions, for 
which $\ii \fdi$ is hermitean and anticommutes with $\gamma_*=\ii\gamma^0\gamma^1$.
In contrast to naive fermions, however, there are no doublers for SLAC-fermions.
Thus they can be used to study any positive integer number of fermion flavors.\footnote{Because of the sign problem our numerical
	simulations are restricted 
	to even $\Nf$.} We point out that the non-local SLAC-derivative must not 
be used in gauge theories, where the edge of the Brillouin zone 
(where $k_\mu$ jumps) is probed in the functional integral, which 
leads to a clash with Euclidean Lorentz invariance \cite{Karsten:1979wh}. 
But SLAC-fermions have been successfully
applied to non-gauge theories, for example scalar field theories
\cite{Wozar:2011gu}, non-linear sigma models \cite{Flore:2012xj}, 
supersymmetric Yukawa models \cite{Wozar:2011gu} and more recently 
to interacting fermion systems \cite{Wellegehausen:2017goy}.

For SLAC-fermions the chemical potential $\mu$ is introduced as in the continuum theory,
\begin{equation}
\bar{\psi}(x) \gamma_0 \Big(\partial_0 + \mu\Big) \psi(x) \ \ \rightarrow \ \ \bar{\psi}(x) \gamma_0 
\Big(\partial^\mathrm{slac}_0 + \mu\Big) \psi(x) .
\end{equation}
Note that the chemical potential $\mu$ enters linearly and not via $\exp(\pm a\mu)$
multiplying a hopping term as e.g.\ for naive fermions 
(see e.g.\ eq.\ (\ref{EQN744}))
%When one introduces a chemical potential as additive linear term
%then one needs to subtracting a finite $c\mu^{2}$ from the 
%free energy, similarly as for naive fermions 
%requires the same correction terrm 
%when $\beta/\Nt$ and $L/\Ns$ approach zero simultaneously.
%and subtracting $c\mu^{2}$ from the partition function.
%It has been demonstrated in \cite{Gavai:2014lia} that this
%method of divergence removal works in (quenched) QCD.
%\footnote{\todo{Maybe:
%		Comment on how this cures/not cures the problems with current conservation.}}
%This is different from earlier attempts to eliminate the divergencies
%by suitably modifying the
%lattice action
For some observables, for example the 
fermion density, this introduces an additional term 
$\propto\mu$ in the continuum limit, which can be easily subtracted, 
since (in $2$ spacetime dimensions) the term is finite and can be 
calculated analytically. We emphasize that this term is not a lattice
artifact -- it also exists in continuum theories when appropriate
care is taken in manipulating divergent integrals \cite{Gavai:2014lia}.
After the subtraction is performed, results obtained with SLAC-fermions 
converge much faster to the continuum limit than for other fermion 
discretizations.\footnote{These findings will be published elsewhere.} 
A similar observation has been made when using a linear $\mu$
for naive fermions in $4$ spacetime dimensions \cite{Gavai:2014lia}.
All observables considered in the present work need no such subtraction.
% **********

\subsubsection{\label{SEC853}Comparison of naive and SLAC fermions for $\Nf \rightarrow \infty$ and homogeneous condensate}

\begin{figure}
	\centering
	\includegraphics[width=0.5\linewidth]{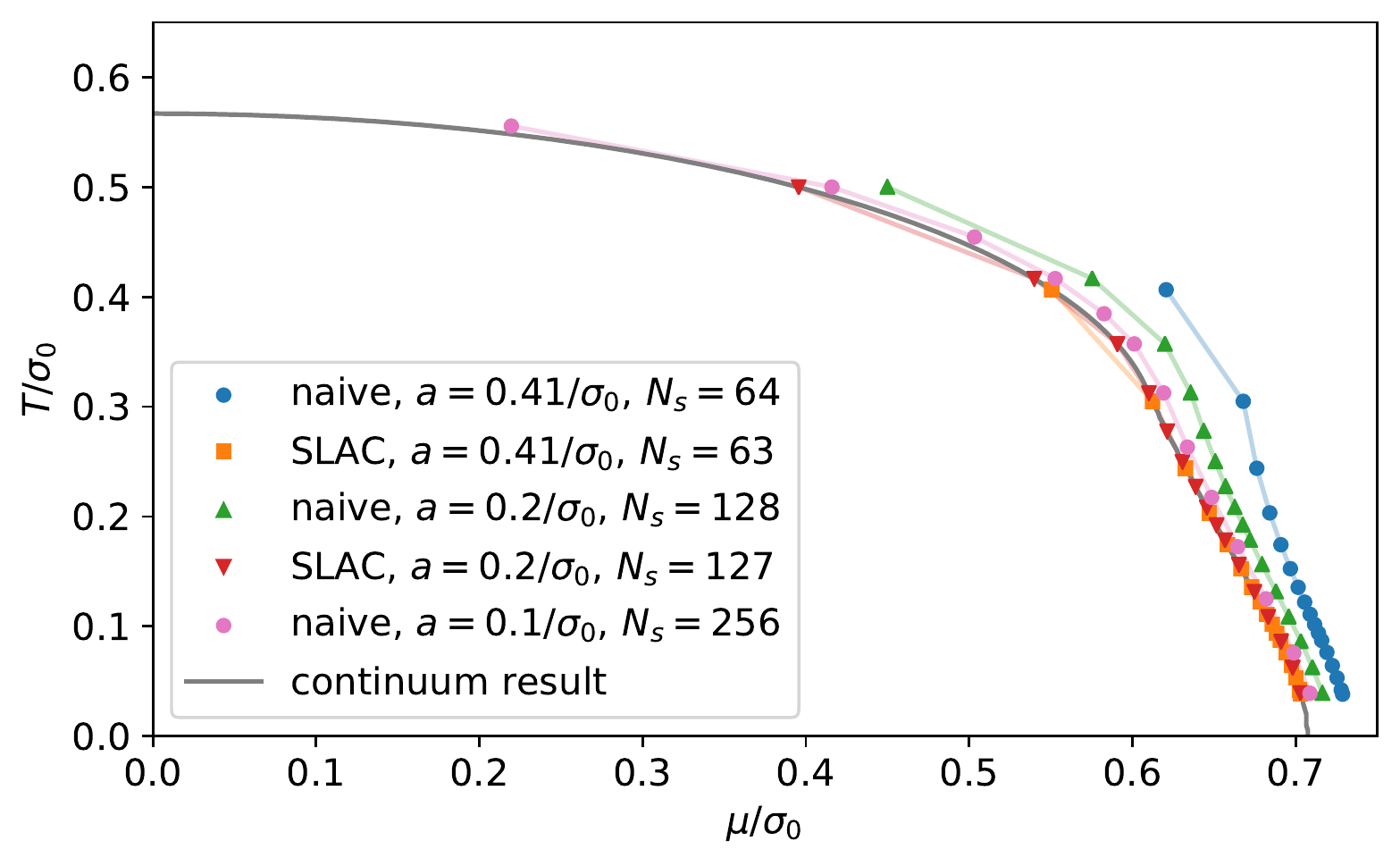}
	\caption{\label{FIG644}$N_f \rightarrow \infty$ phase diagram with the restriction
		to homogeneous condensate $\sigma$ for naive and for SLAC fermions (three
		different lattice spacings, 
		$a \approx 0.41 / \sigma_0 , 0.20 / \sigma_0 , 0.10 / \sigma_0$, similar spatial extent $L$). 
		The solid grey line represents the continuum result from 
		Ref.\ \cite{Wolff:1985av}.}
\end{figure}

In Figure~\ref{FIG644} we show the $N_f \rightarrow \infty$ phase diagram 
with the restriction to a homogeneous condensate $\sigma$ for naive and for 
SLAC fermions. The corresponding computations are straightforward and
computationally rather cheap, when using techniques similar to those 
discussed in Refs.\ \cite{deForcrand:2006zz,Wagner:2007he,Heinz:2015lua}. 
For both discretizations we performed computations for several significantly
different values of the lattice spacing, $a \approx 0.41 / \sigma_0$ 
and $a \approx 0.20 / \sigma_0$ (naive and SLAC) and 
$a \approx 0.10 / \sigma_0$ (only naive), but similar spatial 
extent $L$. When decreasing the lattice spacing, the results 
obtained with each of the two discretizations approach the continuum 
result from Ref.\ \cite{Wolff:1985av}. Note, however, that discretization 
errors for SLAC fermions are almost negligible, i.e.\ significantly smaller 
than discretization errors for naive fermions.

% ********************

\subsection{Simulation setup}

We use a standard RHMC (Rational Hybrid Monte Carlo) algorithm 
\cite{Clark:2003na} to perform numerical simulations. In detail 
we use the implementation described in Ref.\ \cite{bjorn_wellegehausen_phase_2012}, 
which was also used in Refs.\ \cite{Wellegehausen:2017goy,Lenz:2019qwu,August:2018esp}.

% **********

\subsubsection{Scale setting}

We assume that at chemical potential $\mu = 0$ and temperature $T = 0$ the system is in a homogeneously broken phase and use the (positive) expectation value
\begin{align}
\label{eq:sigma0} \sigma_0 = \lim_{L \rightarrow \infty} \langle \, \xbar{\sigma} \, \rangle = \lim_{L \rightarrow \infty} \langle \vsig \rangle\Big|_{\mu=0,T=0}
\end{align}
to set the scale. In other words, we express all dimensionful quantities in 
units of $\sigma_0$, e.g.\ we use for the chemical potential $\mu / \sigma_0$, 
for the temperature $T / \sigma_0$, etc. Setting the scale via $\sigma_0$ was 
also done in previous analytical and numerical studies of the phase diagram 
of the GN model in the $\Nf \rightarrow \infty$ limit (see e.g.\ \cite{Thies:2003kk,Schnetz:2004vr,deForcrand:2006zz,Wagner:2007he,Heinz:2015lua}),
i.e.\ expressing dimensionful quantities in units of $\sigma_0$ allows a
straightforward comparison of our results at finite $\Nf$ to existing $\Nf
\rightarrow \infty$ results.

\begin{figure}
	\centering
	\includegraphics[width=0.5\linewidth]{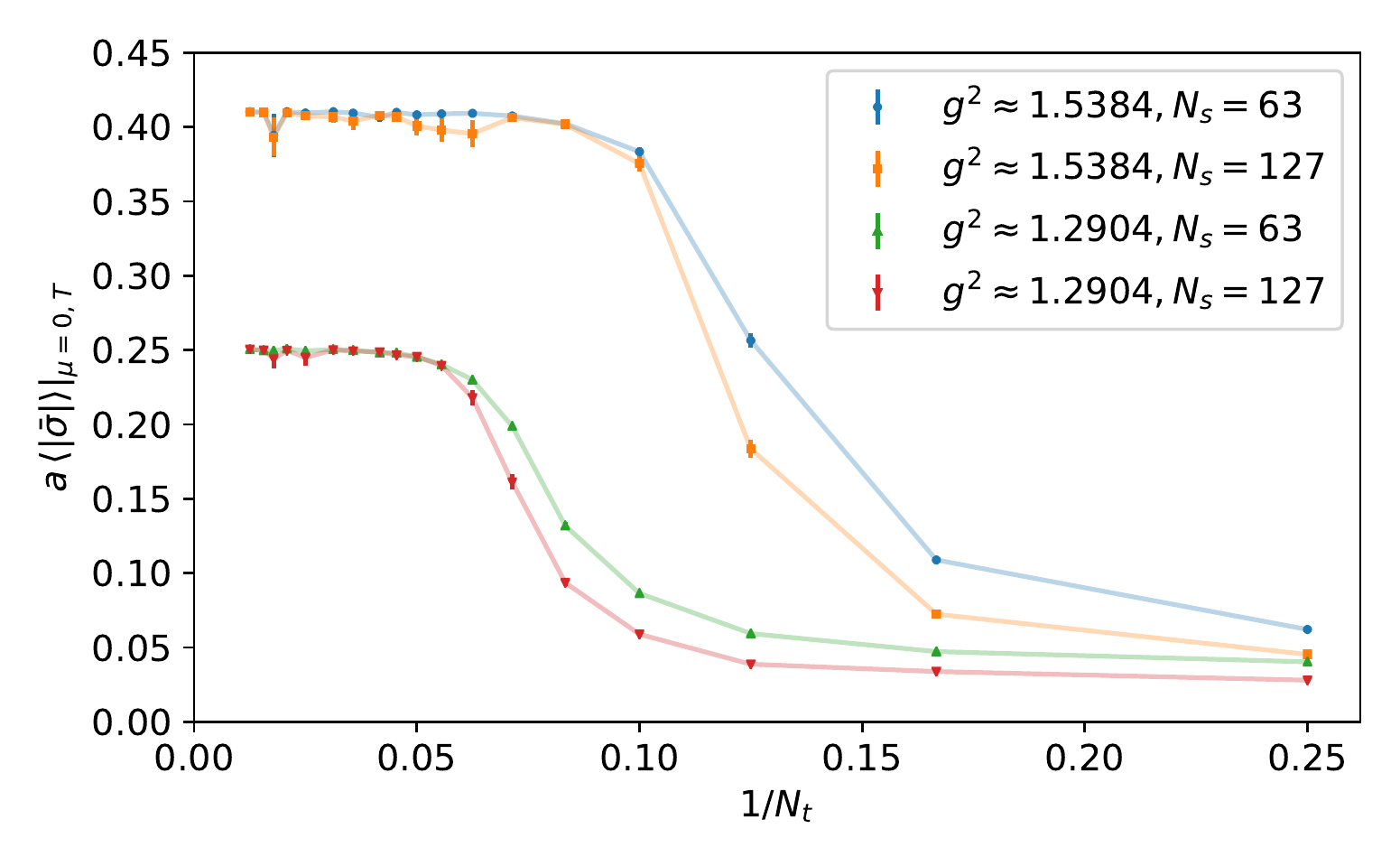}
	\caption{\label{f:scaleSet1}$a \langle \vsig \rangle|_{\mu=0,T}$ as a function
		of $1 / \Nt = T a$ for $\mu=0$, $\Nf = 8$, SLAC fermions, two different $g^2$ 
		and two different $\Ns = L / a$. The plateau values at small $1 / \Nt$ correspond
		to $a \sigma_0$.}
\end{figure}

The determination of $\sigma_0$ in lattice units is technically straightforward. When increasing the number of lattice sites in temporal direction $\Nt$ as well as in spatial direction $\Ns$ at fixed coupling $g^2$, the ensemble average $\langle \vsig \rangle|_{\mu=0,T}$ quickly approaches the constant $\sigma_0$. Thus, in practice, one just has to compute $\langle \vsig \rangle|_{\mu=0,T}$ on a lattice with sufficiently large $\Nt$ and $\Ns$, where $\langle |\xbar{\sigma}| \rangle |_{\mu=0,T} \approx \sigma_0$. This is illustrated in Figure~\ref{f:scaleSet1} for $\Nf = 8$, SLAC fermions and two different $g^2$.

As e.g.\ in lattice simulations of 4-dimensional Yang-Mills theory or QCD, 
the lattice spacing $a$ is a function of the dimensionless coupling $g^2$ 
and can be set by choosing appropriate values for $g^2$. This is reflected 
by the two plateau values at small $1 / \Nt$ in Figure~\ref{f:scaleSet1}
representing $a \sigma_0$ (the lattice spacing in units of $\sigma_0$),
which correspond to $g^2 = 0.192$ (larger lattice spacing) 
and $g^2 = 0.161$ (smaller lattice spacing).

% **********

\subsubsection{Ensembles of field configurations}

To explore the $\mu$-$T$ phase diagram of the $1+1$-dimensional GN model and its
dependence on the number of fermion flavors $\Nf$ and to exclude sizable lattice
discretization and finite volume corrections, we generated a large number of
ensembles of field configurations $\sigma(\vx)$. These ensembles are listed in
Table~\ref{t:ensembles}.

\begin{table}
	\centering
	
	\begin{tabular}{llllll}
		\hline \hline
		$\Nf$ & $\Ns = L / a$    & $\Nt = 1 / T a$ & $\mu/\sigma_0$ 
		&    $g^2$ & $a \sigma_0$     \\
		\hline
		\multicolumn{5}{l}{\rule[-1ex]{0pt}{5.5ex}\textbf{SLAC, thermodynamics}}\\
		\hline
		\rule[-1ex]{0pt}{6.5ex}$2$& $63$ & 
		\shortstack{$4,6,\dots,24,28,32,$\\$40,\dots,64,80$} & 
		$0.0,\ldots,1.4$& $1.9569$ & $0.4100(5)$ \\
		
		\hline
		\multirow{3}{*}{$8$}&\multirow{3}{*}{$31,47,63,127$}&\multirow{3}{*}{\shortstack{$4,6,\dots,24,28,32,$\\$40,\dots,64,80$}}&\multirow{3}{*}{$0.0,\ldots,1.4$}&$1.5384$&$0.4100(5)$\\ 
		
		&&&&$1.2904$&$0.2495(5)$\\
		&&&&$1.1680$&$0.195(5)$\\
		\hline
		\rule[-1ex]{0pt}{6.5ex}$16$ & $63$& 
		\shortstack{$4,6,\dots,24,28,32,$\\$40,\dots,64\phantom{,80}$}
		&$0$& $1.4953$ & $0.4100(5)$ \\
		\hline
		\multicolumn{5}{l}{\rule[-1ex]{0pt}{5.5ex}\textbf{Naive, thermodynamics}}\\
		\hline
		\multirow{3}{*}{$8$}&\multirow{2}{*}{$64$}&\multirow{2}{*}{\shortstack{$2,4,\dots,44,48,$\\$52,\dots,64$}}&$0.0,\ldots,1.4$  &$1.8132$&$0.4113(3)$\\
		&&&$0.0,\ldots,1.1$&$1.4172$&$0.2518(5)$\\\cline{2-6}
		&$128$&$100$&$0.7,0.9$&$1.0960$&$0.1253(3)$\\
		\hline
		\multicolumn{5}{l}{\rule[-1ex]{0pt}{5.5ex}\textbf{SLAC, long-range behavior of the correlation function}}\\
		\hline
		\rule[-1ex]{0pt}{6.5ex}$2$& 
		\shortstack{$65,125,185,255,$\\$375,525,725 $} & $80$ & $0.5$& $1.9569$ & $0.4100(5)$ \\
		\hline\hline
	\end{tabular}
	
	\caption{\label{t:ensembles}Ensembles of field configurations.}
\end{table}

For given coupling $g^2$, i.e.\ for fixed lattice spacing $a$, we vary the
temperature $T = 1 / \Nt a$ by changing $\Nt$, the number 
of lattice sites in temporal direction. Thus, at fixed $g^2$ the 
temperature $T$ can only be changed in discrete steps. The chemical 
potential $\mu$, on the other hand, is not restricted in such a way 
and can be set to any value.

The majority of simulations were carried out for $\Nf = 8$:
\begin{itemize}\itemsep=1mm
	\item We simulated at several different spatial extents with \mbox{$31 \leq \Ns \leq 128$} corresponding to $L = \Ns a$ to check for finite volume corrections.
	
	\item We simulated at several different values of the coupling $g^2$ corresponding to four different lattice spacings $a \approx 0.41 / \sigma_0 , 0.25 / \sigma_0 , 0.20 / \sigma_0 , 0.13 / \sigma_0$ (the lattice spacing is listed in units defined by $\sigma_0$ in the column ``$a \sigma_0$'' of Table~\ref{t:ensembles}).
	
	\item We simulated at many different values of the chemical potential, to explore the phase diagram.
	
	\item We carried out a sizable amount of these simulations using both fermion
	discretizations, i.e.\ SLAC fermions and naive fermions, to cross-check 
	our results (the corresponding coupling constants $g^2$ have been tuned 
	in such a way, that the simulated lattice spacings are almost identical).
\end{itemize}
Simulations at $\Nf = 2$ and $\Nf = 16$ were done with SLAC fermions, but not with naive fermions. For each ensemble between $300$ and $10 \, 000$ configurations were generated.

\clearpage
\section{\label{SEC588}Numerical results}\label{sec:numerics}

The majority of results shown in the following (section~\ref{SEC597} to section~\ref{SEC330}) correspond to $\Nf = 8$, the minimal number of flavors where computations are possible for both naive and SLAC fermions. Results for $\Nf = 2$ and $\Nf = 16$ are presented in section~\ref{SEC598} and section~\ref{SEC599}.

% **********

\subsection{\label{SEC690}Qualitative expectations}

\begin{figure}[t]
\centering
\includegraphics[width=0.49\linewidth]{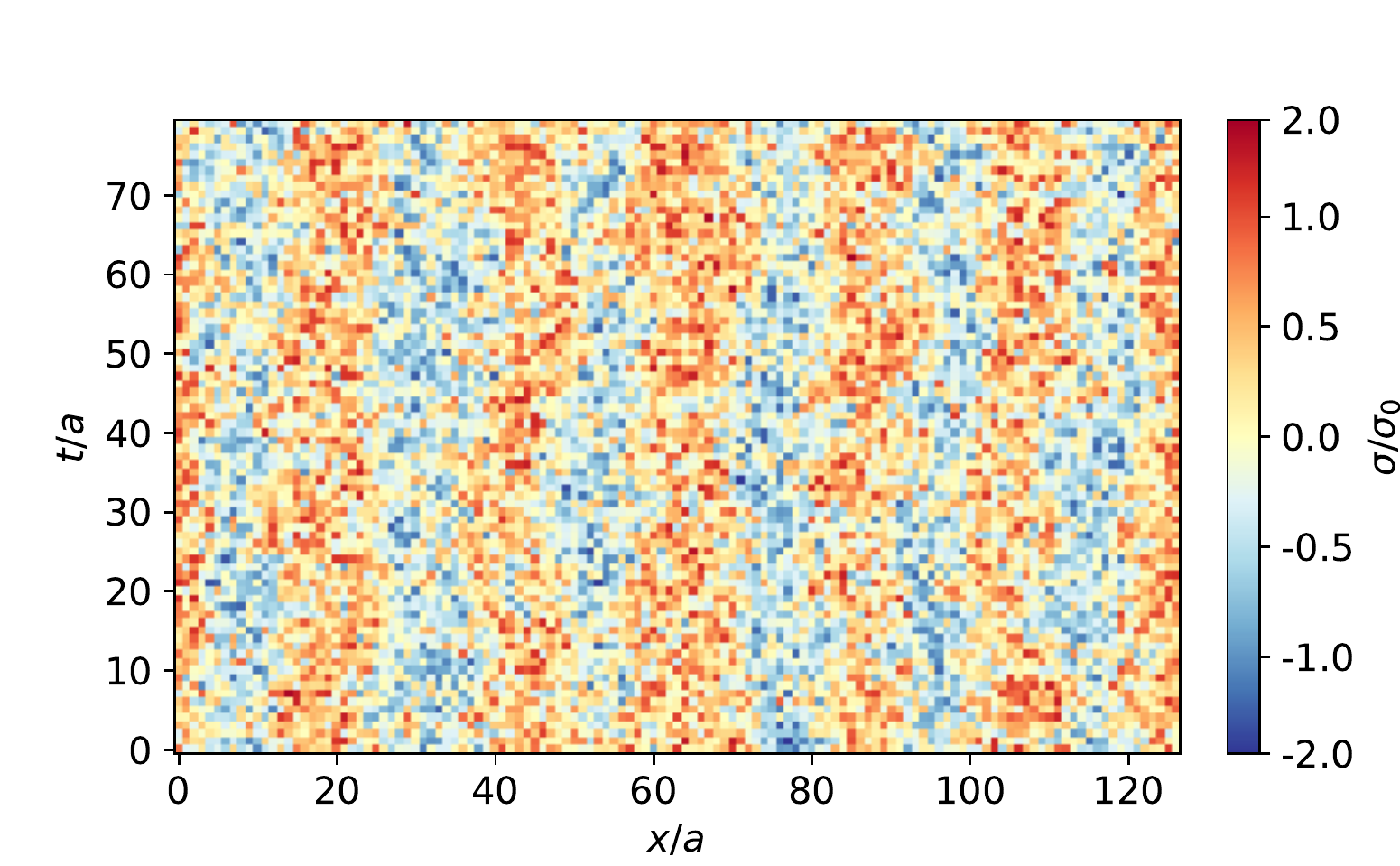}
\caption{\label{FIG_typical_sigma}A typical field configuration $\sigma(\vx) / \sigma_0$ generated by the HMC algorithm at large $\mu / \sigma_0 \approx 0.450$ and small $T/\sigma_0 \approx 0.030$, where an inhomogeneous phase is expected ($\Nf = 8$, SLAC fermions, $a \approx 0.410 / \sigma_0$, $\Ns = 127$). The clearly visible vertical stripes indicate six oscillations in spatial direction.}
\end{figure}

In the $1+1$-dimensional GN model in the limit $\Nf \rightarrow \infty$ there are three phases, a symmetric phase, a homogeneously broken phase and an inhomogeneous phase (see the discussion in section~\ref{s:anal}). The structure of the field configurations $\sigma(\vx)$ generated during our simulations by the HMC algorithm suggest that there is a similar phase structure at finite $\Nf$. At large $T$ the field $\sigma(\vx)$ mostly fluctuates around zero, while at small $T$ and small $\mu$ either $\sigma(\vx) \approx +\sigma_0$ or $\sigma(\vx) \approx -\sigma_0$. Most interestingly, however, at small $T$ and large $\mu$ the field $\sigma(\vx)$ exhibits spatial periodic oscillations similar to a cos-function, which might signal an inhomogeneous phase.\footnote{The existence of kink-antikink structures in simulations of the $1+1$-dimensional GN model with $\Nf = 12$ was observed already many years ago \cite{Karsch:1986hm}.} An example of a typical field configuration at $(\mu / \sigma_0 , T / \sigma_0) \approx (0.450 , 0.030)$ with such periodic oscillations is shown in Figure~\ref{FIG_typical_sigma}.

Thus, we expect that the field configurations $\sigma(\vx)$ generated by the Monte Carlo algorithm are crudely described by the following model:
\begin{itemize}
	\item Inside a symmetric phase
	\begin{equation}
	\label{EQN201} \sigma(\vx) = \varepsilon \eta(\vx) \, .
	\end{equation}
	
	\item Inside a homogeneously broken phase
	\begin{equation}
	\label{EQN202} \sigma(\vx) = \pm\,\zeta\sigma_0 + \varepsilon \eta(\vx) \, .
	\end{equation}
	
	\item Inside an inhomogeneous phase
	\begin{equation}
	\label{EQN203} \sigma(\vx) = A \cos\bigg(\frac{2 \pi (x + \delta x)}{\lambda}\bigg) + \varepsilon \eta(\vx) \, .
	\end{equation}
\end{itemize}
$\varepsilon \geq 0$, $\zeta > 0$ and $A \geq 0$ are real parameters, which depend on $\mu$ and $T$. $\eta(\vx)$ are independent continuous random variables with Gaussian probability distributions $p(\eta(\vx)) \propto \exp(-\eta(\vx)^2 / 2)$, which represent statistical fluctuations. $\lambda = L / (q + \delta q)$ is the wavelength of $\sigma$ in an inhomogeneous phase, where $q \geq 1$ is an integer parameter and $\delta q$ is an integer-valued discrete random variable with Gaussian probabilities $p(\delta q) \propto \exp(-\delta q^2 / 2 
\Delta q^2)$. $\Delta q \ll q$, the width of the Gaussian, is another real parameter. $\delta x \in [0 , L)$ is also a random variable, where it is a priori not clear, what kind of distribution to expect. The distribution could depend on the details of the HMC algorithm, and whether translation symmetry is spontaneously broken or not. Note, however, that the observables we are studying are constructed in such a way that they are independent of this distribution. To summarize, the model defined by eqs.\ (\ref{EQN201}) to (\ref{EQN203}) describes field configurations $\sigma(\vx)$, which fluctuate around $0$ in a symmetric phase, around $\pm \zeta\sigma_0$ in a homogeneously broken phase and around a cos-function with varying wavelength $\lambda$ in an inhomogeneous phase.

With this model in mind, which is based on existing results in the $\Nf \rightarrow \infty$ limit \cite{Thies:2003kk,Schnetz:2004vr}, we designed several observables, which are able to distinguish the three phases. Note that the sole purpose of this model is to provide some guidelines for the construction of observables and to develop expectations, in which way they characterize the three phases. The model is not used elsewhere in this work, in particular not for the analysis of our numerical results.

% **********

\subsection{\label{SEC597}Squared spacetime average of $\sigma(\vx)$}

A rather simple observable is
\begin{align}
\label{eq:sigma} \Sigma^2 = \frac{\langle \sig^2 \rangle}{\sigma_0^2} ,
\end{align}
the normalized ensemble average of the squared spacetime 
average of $\sigma(\vx)$. It is not suited to characterize 
an inhomogeneous phase, but still useful to distinguish a
homogeneously broken phase from a symmetric phase and 
an inhomogeneous phase. Note that our numerical 
	results do not allow to decide, whether these regions 
	in the $\mu$-$T$ plane are phases in a strict thermodynamical
	sense or rather regimes, which strongly resemble phases. 
	In any case, throughout this paper we denote these regions 
	as ``phases''.\label{remark1}

Within the model defined in section~\ref{SEC690} one finds
\begin{itemize}
	\item inside a symmetric phase
	\begin{align}
	\Sigma^2 = \varepsilon^2 / \Nt \Ns \sigma_0^2.
	\end{align}
	\item inside a homogeneously broken phase
	\begin{align}
	\Sigma^2 = 
	\zeta^2 + \varepsilon^2 / \Nt \Ns \sigma_0^2.
	\end{align}
	\item inside an inhomogeneous phase
	\begin{align}
	\Sigma^2 = \varepsilon^2 / \Nt \Ns \sigma_0^2.
	\end{align}
\end{itemize}
Thus, we expect $\Sigma^2 \approx 0$ both inside a chirally symmetric phase and inside an inhomogeneous phase, while it should be significantly larger, $\Sigma^2 \approx 1$, inside a homogeneously broken phase.

\begin{figure}
	\centering
	\includegraphics[width=0.49\linewidth]{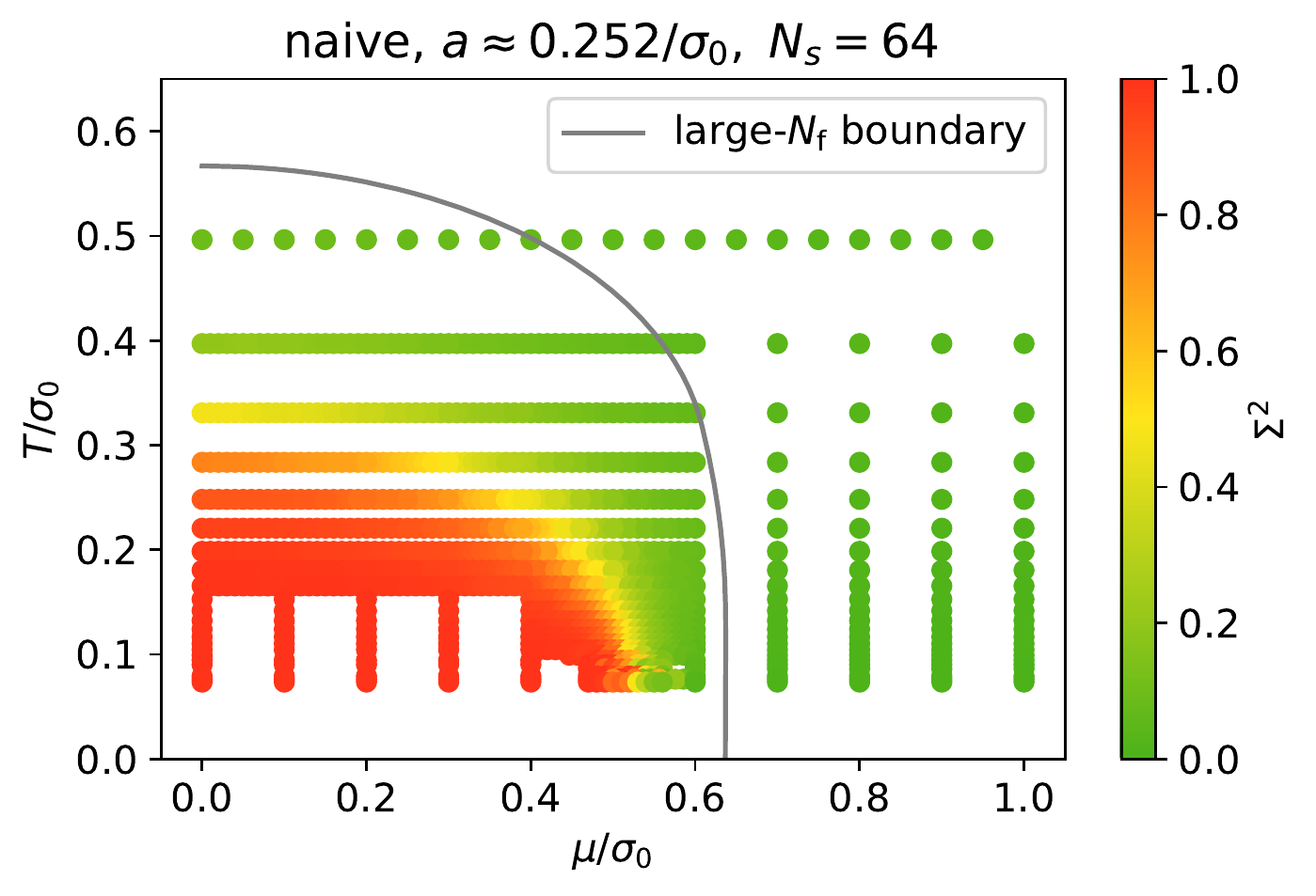}
	\includegraphics[width=0.49\linewidth]{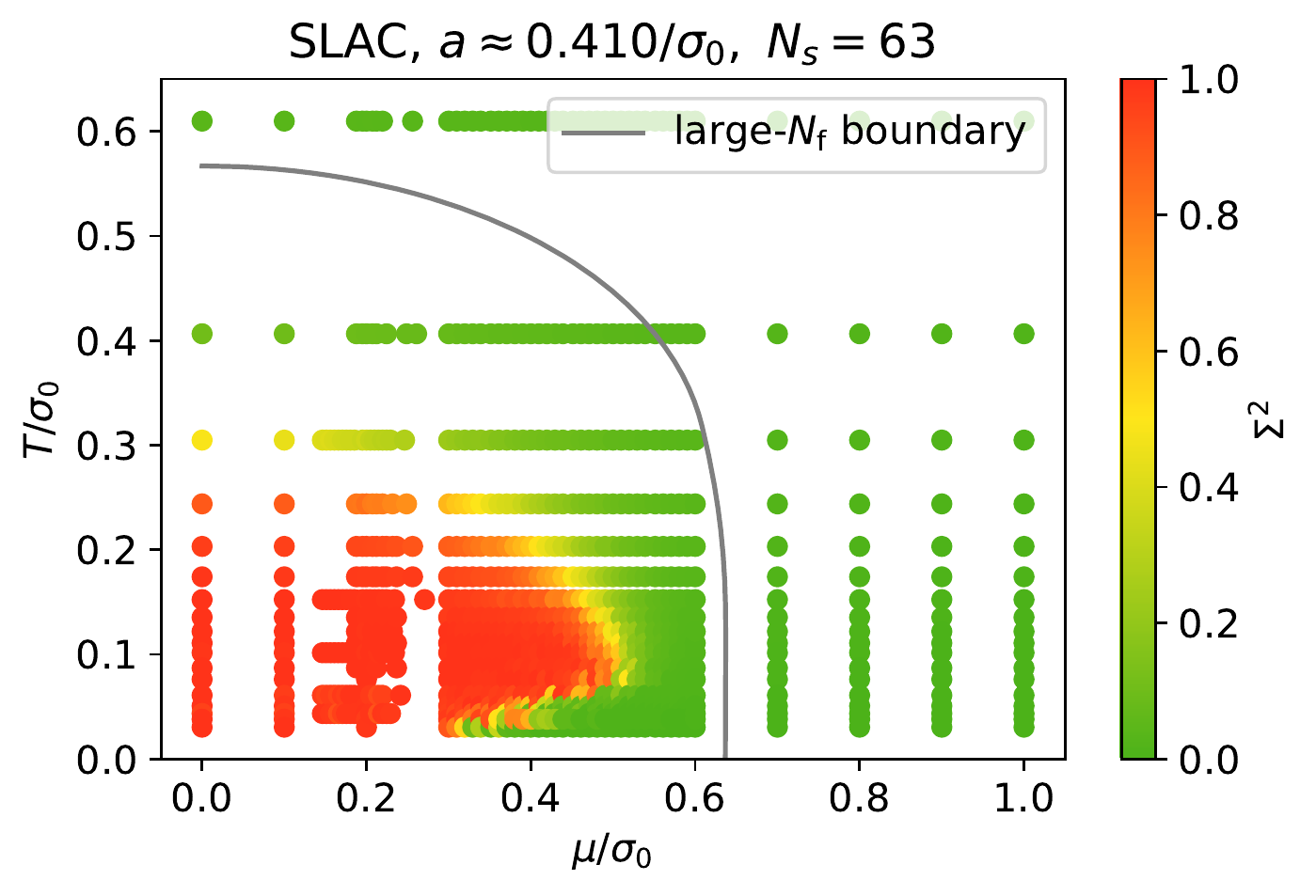}
	\caption{\label{FIG_hom_phase_diagram}$\Sigma^2$ in the $\mu$-$T$ plane for naive fermions (left plot) and SLAC fermions (right plot) ($\Nf = 8$). The red regions correspond to a homogeneously broken phase and the green regions to a symmetric and/or an inhomogenous phase. The gray lines represent the $\Nf \rightarrow \infty$ phase boundary of the homogeneously broken phase \cite{Thies:2003kk,Schnetz:2004vr}.}
\end{figure}

In Figure~\ref{FIG_hom_phase_diagram} we show $\Sigma^2$ in 
the $\mu$-$T$ plane for naive fermions and SLAC fermions. The 
red regions clearly indicate a homogeneously broken phase, while 
the green regions represents a symmetric and/or an inhomogenous
phase. The two plots are very similar. The main reason for the 
small discrepancies are lattice discretization errors, which are
expected to be significantly larger for naive fermions than 
for SLAC fermions (see section~\ref{SEC853}). To ease comparison
with $\Nf \rightarrow \infty$ results, we included the
corresponding phase boundary of the homogeneously broken 
phase from Refs.\ \cite{Thies:2003kk,Schnetz:2004vr}. 
It is obvious that the homogeneously broken phase at finite 
$\Nf = 8$ is of similar shape, but of smaller size than its 
analog at $\Nf \rightarrow \infty$. Such a reduction in size 
is expected, because at finite $\Nf$ there are fluctuations 
in $\sigma(\vx)$, which increase disorder and, thus, favor 
a symmetric phase. Note that at small $T$ the boundary 
between the red and the green region starts to deviate 
significantly from the $\Nf \rightarrow \infty$ boundary 
and turns towards $(\mu,T) = (0,0)$. Our numerical results 
indicate that this is caused by the finite lattice spacing 
(see e.g.\ Figure~\ref{Fig:Cmin_array}). A qualitatively 
similar behavior was observed in an $\Nf \rightarrow \infty$ 
lattice study of the GN model \cite{deForcrand:2006zz}.

% **********

\subsection{\label{SEC330}The spatial correlation function of $\sigma(\vx)$}

In the limit $\Nf \rightarrow \infty$ in the inhomogeneous phase, $\sigma(\vx)$ 
is a periodic function of the spatial coordinate $x$. It has a kink-antikink
structure with large wavelength close to the boundary to the homogeneously 
broken phase and is sin-like with smaller wavelength for larger $\mu$. We 
expect a similar behavior also at finite $\Nf$ (see also eq.\ (\ref{EQN203})).

Since the action $S_\textrm{eff}$ is invariant under spatial
translations, field configurations, which are spatially 
shifted relative to each other, i.e.\ $\sigma(t,x)$ and
$\sigma(t,x+\delta x)$, contribute with the same weight
$e^{-S_\textrm{eff}}$ to the partition function and, thus, 
should be generated with the same probability by the HMC 
algorithm (see also the discussion on the distribution 
of $\delta x$ in section~\ref{SEC690}). Consequently, 
simple observables like $\langle \sigma(\vx) \rangle$ 
are not suited to detect an inhomogeneous phase in a finite
system, because destructive interference should lead to 
$\langle \sigma(\vx) \rangle = 0$ in a finite system, 
even in cases, where 
all field configurations exhibit spatial oscillations with 
the same wavelength.\footnote{An alternative would be
	to break translation invariance explicitly, for example
	by imposing Dirichlet boundary conditions 
	on $\sigma(x)$.}
An observable, which does not suffer 
from destructive interference and is able to exhibit 
information about possibly present inhomogeneous structures, 
is the spatial correlation function of $\sigma(\vx)$, 
i.e.\@
\begin{align}
\label{eq:Cx} C(x) =\left\langle \sigma(t_0,x)\sigma(t_0,0)\right\rangle=
\frac{1}{\Nt \Ns} \sum_{t,y} \big\langle
\sigma(t,y+x) \sigma(t,y) \big\rangle \, .
\end{align}
The equality holds in thermal equilibrium and a finite 
box of length $L$ with periodic boundary conditions since
$\langle\sigma(t,x+y)\sigma(t,y)\rangle$ does neither depend
on $t$ nor on $y$.
% Actually, our HMC algorithm is so good as to produce $y$-independent expectation values $\langle \sigma(t,x+y)\sigma(t,y)\rangle$.
Actually, our HMC algorithm is able to sample all field configurations and, thus, to produce $y$-independent expectation values $\langle \sigma(t,x+y)\sigma(t,y)\rangle$.
We use the sum over $t$ and $y$ in \eqref{eq:Cx}
to decrease statistical errors in the Monte Carlo average. 

The correlator $C(x)$ is our main observable to detect and to distinguish the three expected 
phases, in particular an inhomogeneous phase.
In contrast to the typical exponential decay of correlation functions, $C(x)$ 
is expected to oscillate in an inhomogeneous phase, i.e.\ $C(x)$ should be 
positive, if $x / \lambda$ is close to an integer, and negative, if $x / \lambda$ 
is close to a half-integer, where $\lambda$ denotes the wavelength of the 
spatial periodic structure of $\sigma(\vx)$. Such oscillations are also 
found in the $\Nf \rightarrow \infty$ limit \cite{Thies_unpublished}. 
The expectation is also supported by analytical calculations within our 
model defined in section~\ref{SEC690}, where
\begin{itemize}
	\item inside a symmetric phase
	\begin{align}
	\label{EQN_C_SYM} C(x) = \varepsilon^2 \delta_{x,0}
	\end{align}
	\item inside a homogeneously broken phase
	\begin{align}
	\label{EQN_C_HOM} C(x) = (\zeta^2\sigma_0)^2 + \varepsilon^2 \delta_{x,0}
	\end{align}
	\item inside an inhomogeneous phase
	\begin{align}
	\label{EQN_C_INH} & \quad C(x) \approx \frac{A^2}{2} \frac{\vartheta(x/L , i/2 \pi \Delta q^2)}{\vartheta(0 , i/2 \pi \Delta q^2)} \cos\bigg(\frac{2 \pi q x}{L}\bigg) + \varepsilon^2 \delta_{x,0}
	\end{align}
\end{itemize}
with the Jacobi $\vartheta$ function
\begin{align}
\label{EQN410} \vartheta(z,\tau) = 1 + 2 \sum_{n = 1}^\infty e^{i \pi n^2 \tau} \cos(2 \pi n z) \, .
\end{align}
The cos-term in eq.\ (\ref{EQN_C_INH}) leads to oscillations with wave length $L/q$, while the factor including the $\vartheta$ function causes a damping of these oscillations for increasing separations $x$. This damping is due to the random fluctuations of the wave number $q + \delta q$ entering eq.\ (\ref{EQN203}) via $\lambda$, which cause destructive interference for larger $x$. The damping is strong for large fluctuations, i.e.\ large $\Delta q$, and not present in the limit $\Delta q \rightarrow 0$. Note that the result for $C(x)$ inside an inhomogeneous phase, i.e.\ the right hand side of eq.\ (\ref{EQN_C_INH}), is independent of the distribution of the random variable $\delta x$ introduced in section~\ref{SEC690}.

Of similar interest as $C(x)$ is its Fourier transform
\begin{eqnarray}
\tilde{C}(k) = \mathcal{F}_x(C)(k)
\end{eqnarray}
(see eq.\ (\ref{EQN690b})). The expected behavior is the following:
\begin{itemize}\itemsep=1mm
	\item inside a symmetric phase $\tilde{C}(k)$ is rather small and smooth without any pronounced peak.
	
	\item inside a homogeneously broken phase $\tilde{C}(k)$ has a pronounced peak at $k = 0$ and is rather small and smooth at $k \neq 0$.
	
	\item inside an inhomogeneous phase $\tilde{C}(k)$ has pronounced peaks at $k = \pm q$, where $q$ is related to the wavelength of the spatial oscillations of $C(x)$ via $\lambda = L / q$.
\end{itemize}
This expectation is in agreement with results obtained within our model from section~\ref{SEC690}, where
\begin{itemize}
	\item inside a symmetric phase
	\begin{align}
	\label{EQN_Ck_SYM} \tilde{C}(k) = \frac{1}{\sqrt{N_s}} \varepsilon^2,
	\end{align}
	\item inside a homogeneously broken phase
	\begin{align}
	\label{EQN_Ck_HOM} \tilde{C}(k) = \sqrt{N_s} (\zeta\sigma_0)^2 \delta_{k,0} + \frac{1}{\sqrt{N_s}} \varepsilon^2,
	\end{align}
	\item inside an inhomogeneous phase
	\begin{align}
	\label{EQN_Ck_INH} \tilde{C}(k) \approx \frac{\sqrt{N_s} A^2}{4 \vartheta(0 , i/2 \pi \Delta q^2)} \bigg(\exp\bigg(-\frac{(k-q)^2}{2 \Delta q^2}\bigg) + \exp\bigg(-\frac{(k+q)^2}{2 \Delta q^2}\bigg)\bigg) + \frac{1}{\sqrt{N_s}} \varepsilon^2.
	\end{align}
\end{itemize}

\begin{figure}[p]
	\vspace{-.5cm}
	\centering
	
	{\scriptsize symmetric phase}
	
	\includegraphics[width=0.49\linewidth]{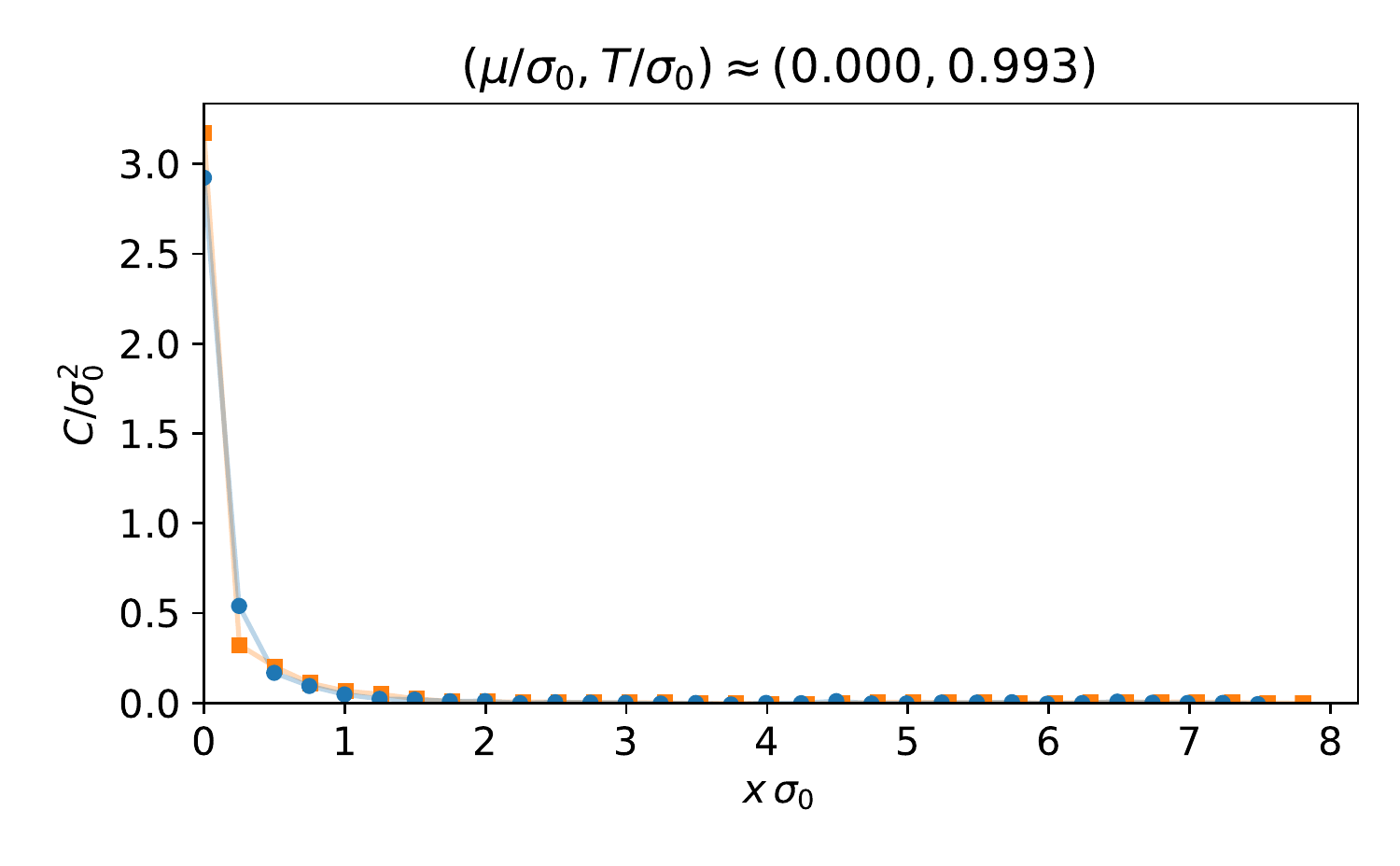}
	\includegraphics[width=0.49\linewidth]{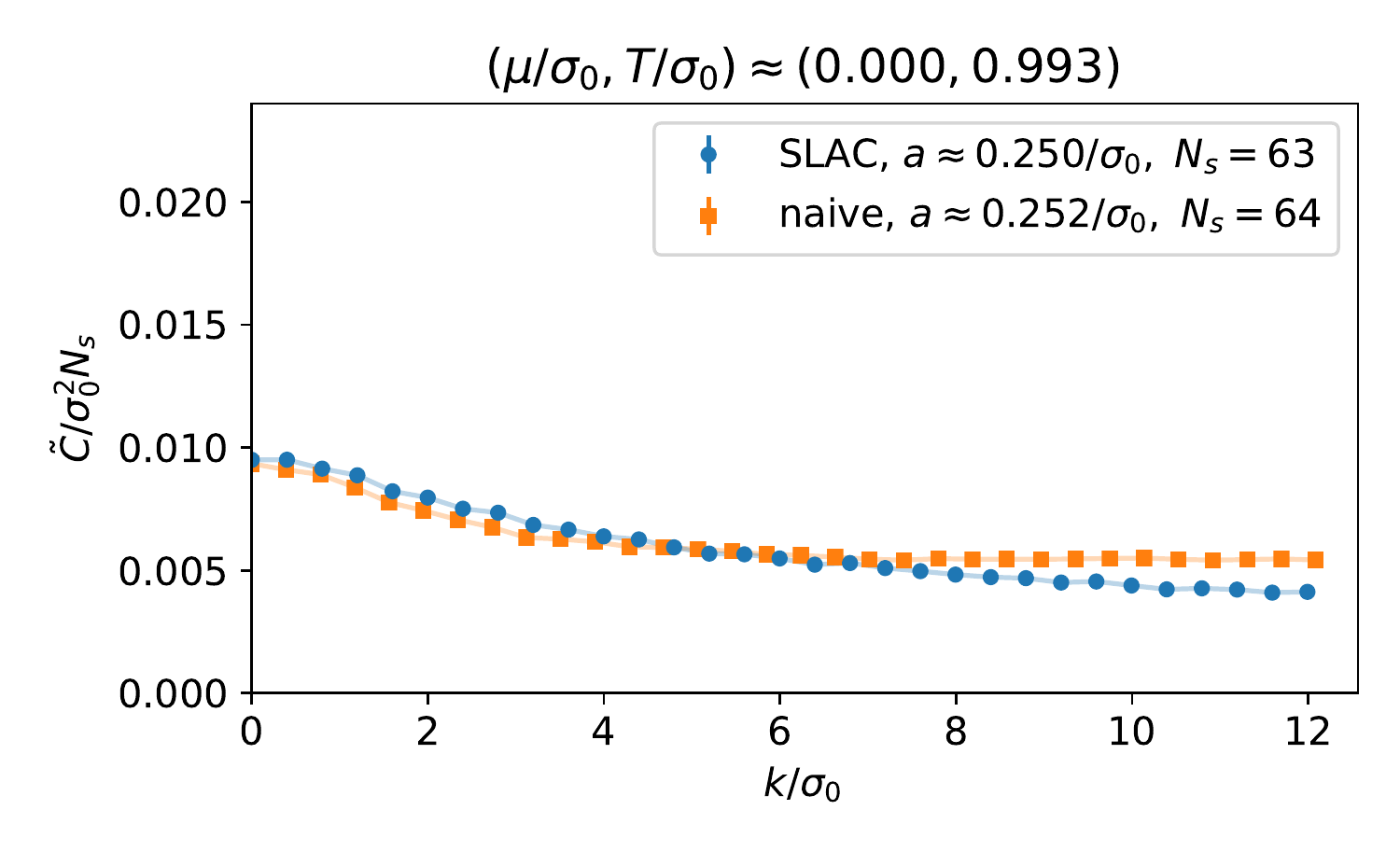}
	
	{\scriptsize homogeneously broken phase}
	
	\includegraphics[width=0.49\linewidth]{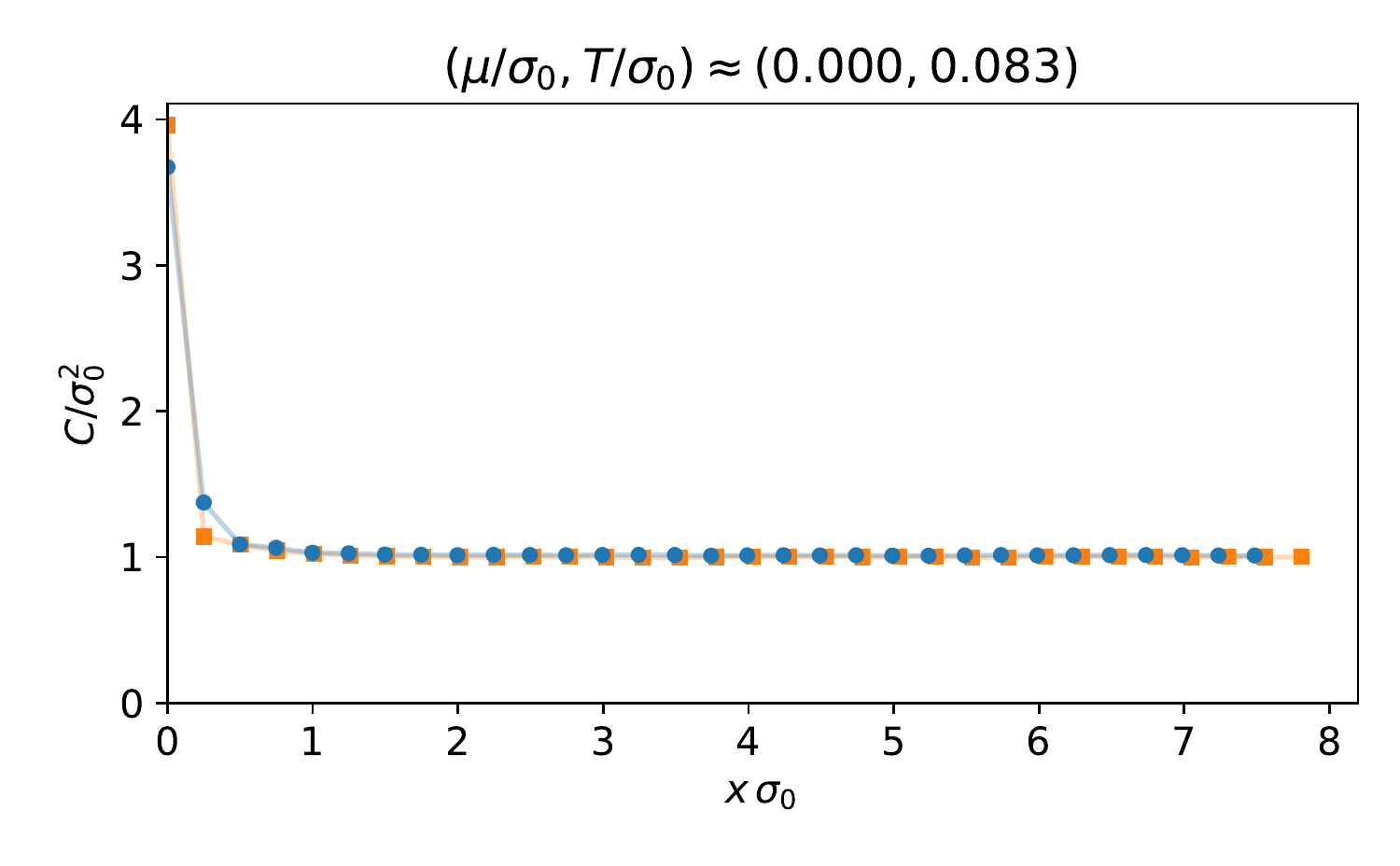}
	\includegraphics[width=0.49\linewidth]{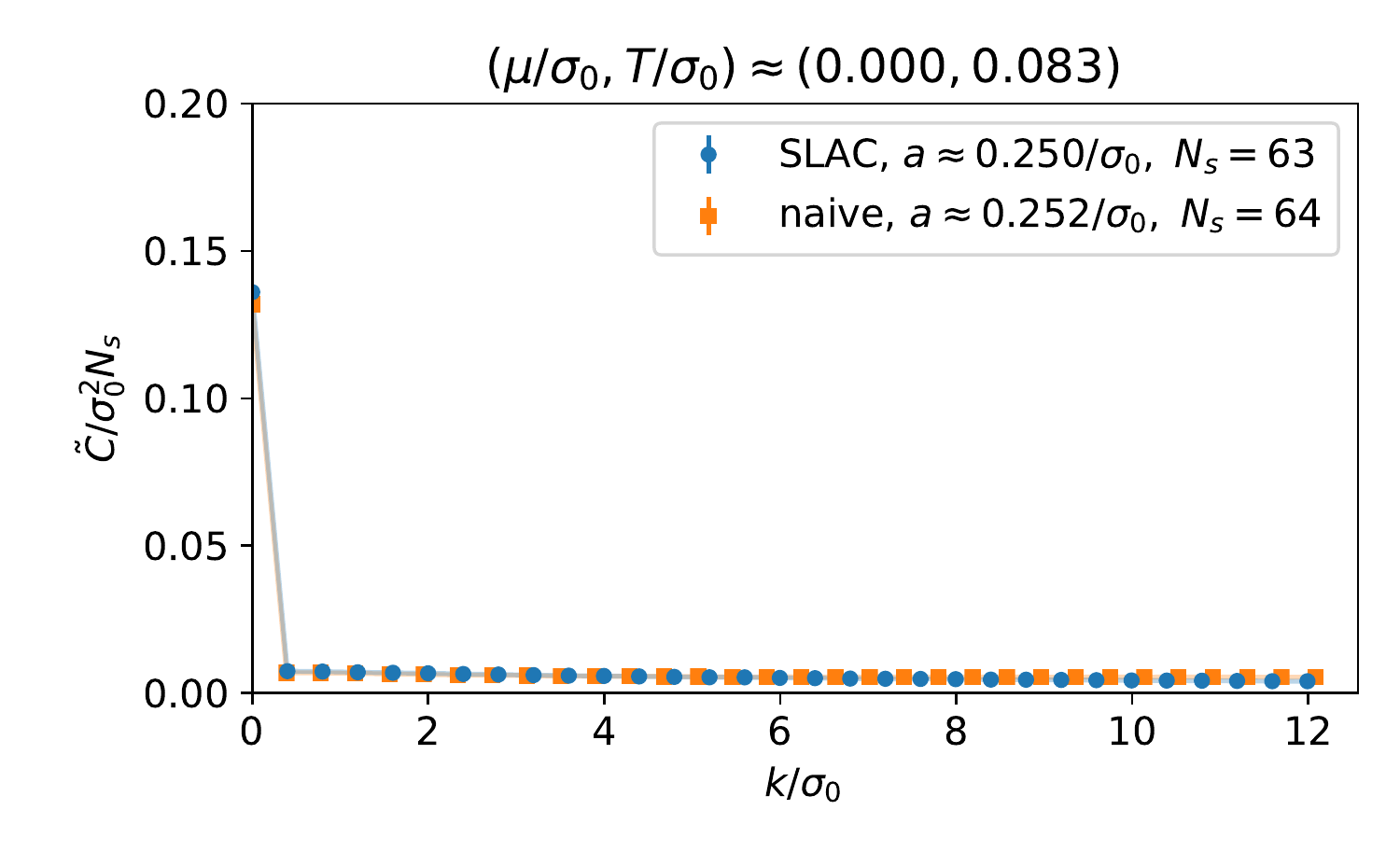}
	
	{\scriptsize inhomogeneous phase}
	
	\includegraphics[width=0.49\linewidth]{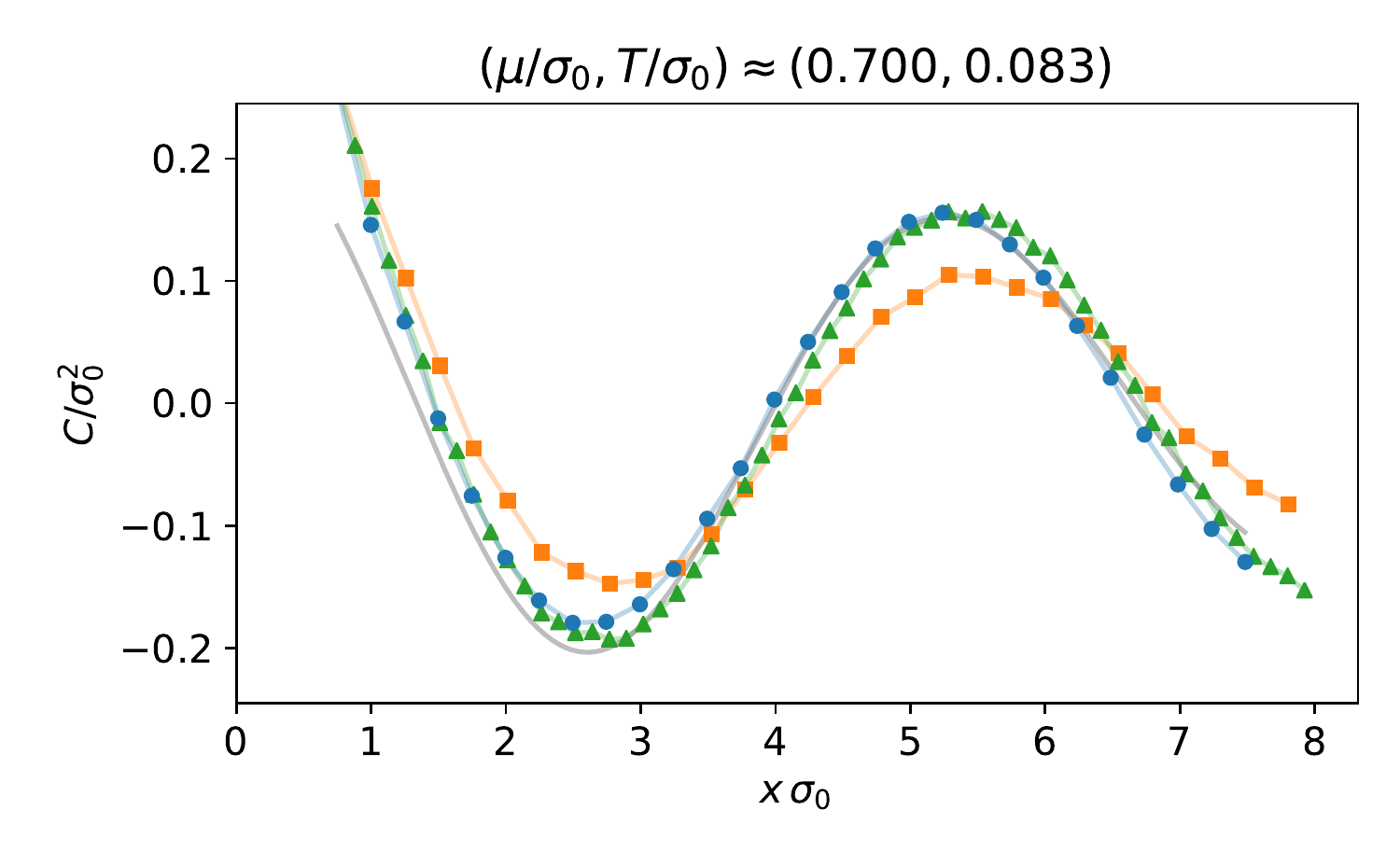}
	\includegraphics[width=0.49\linewidth]{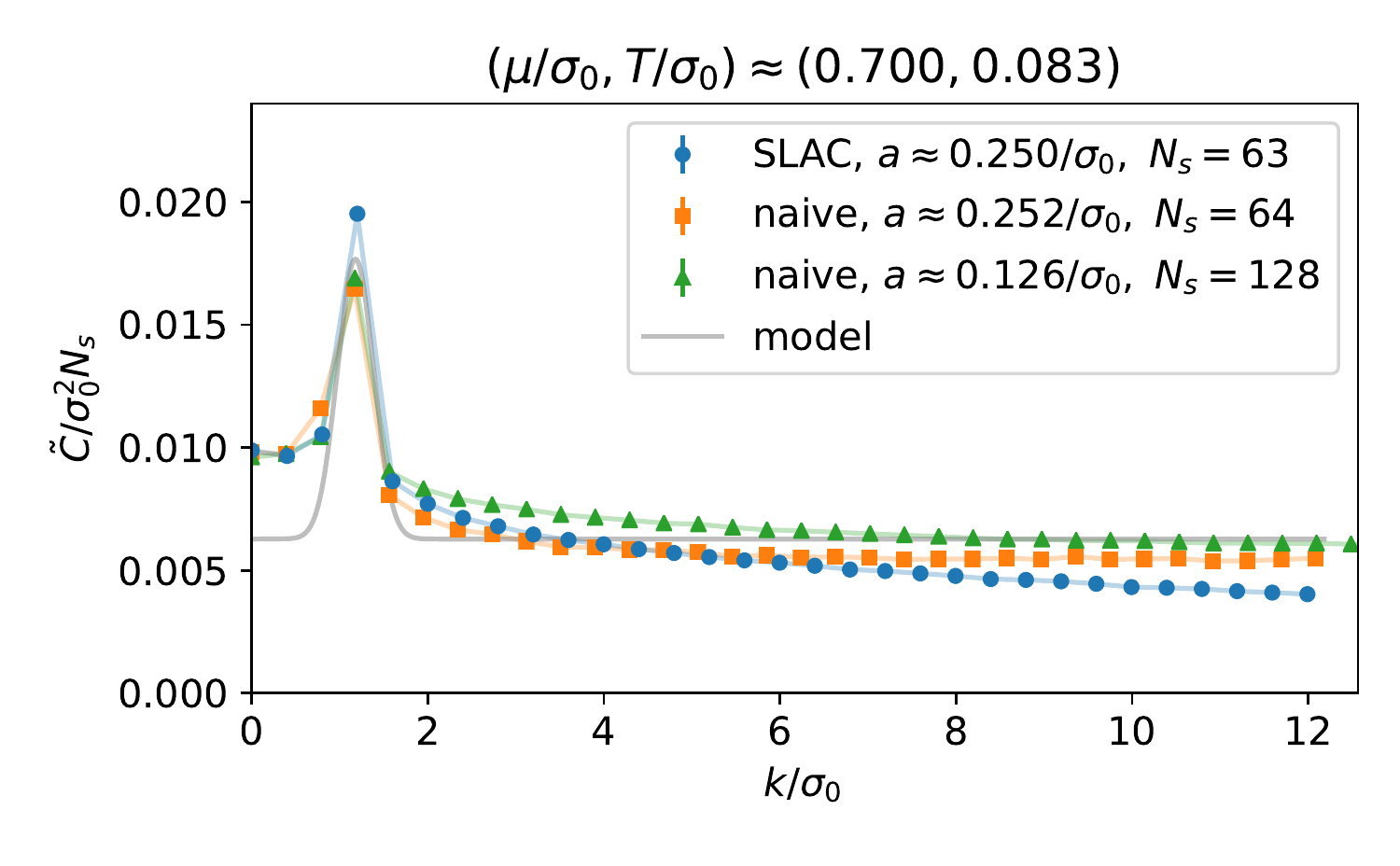}
	
	\includegraphics[width=0.49\linewidth]{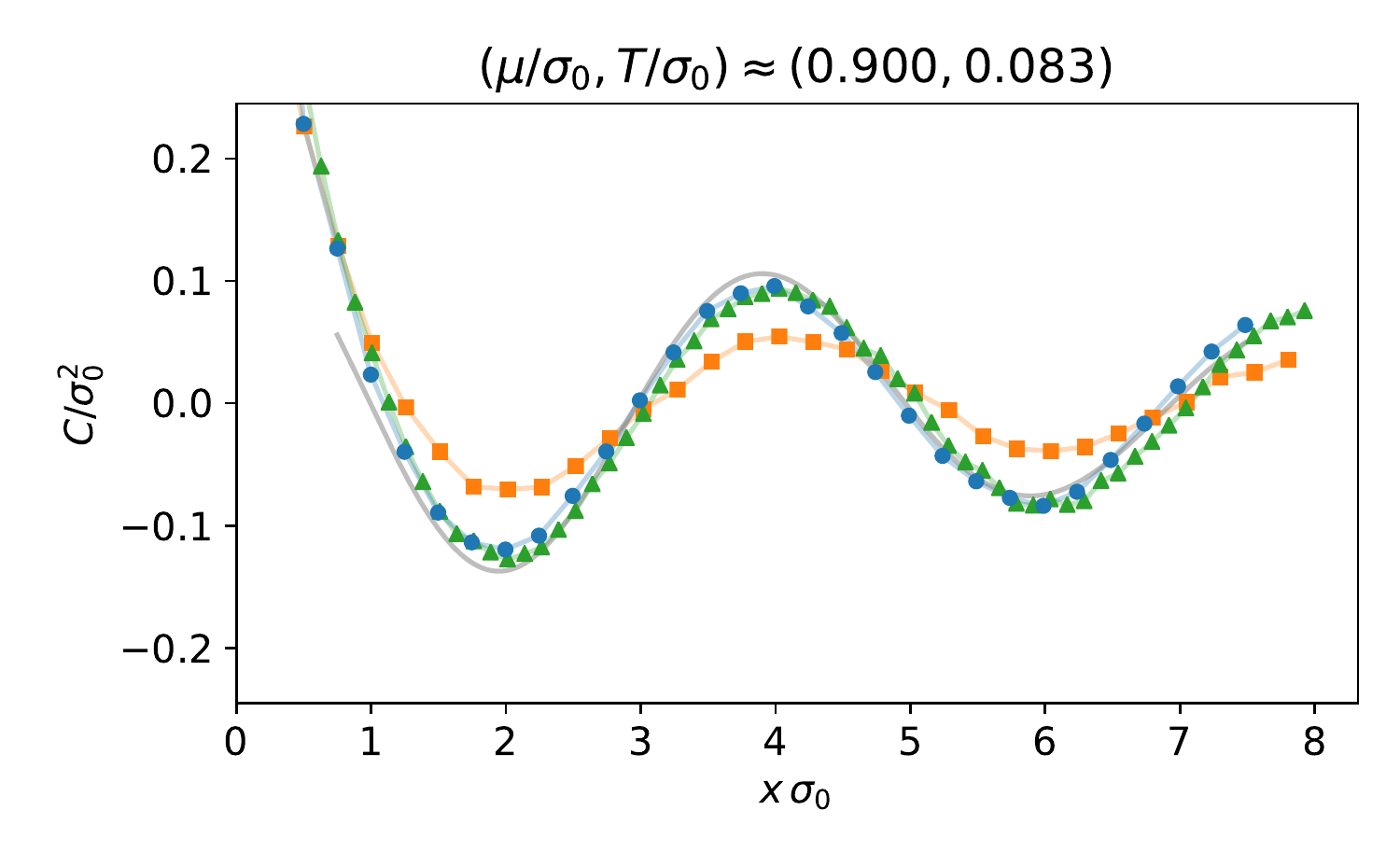}
	\includegraphics[width=0.49\linewidth]{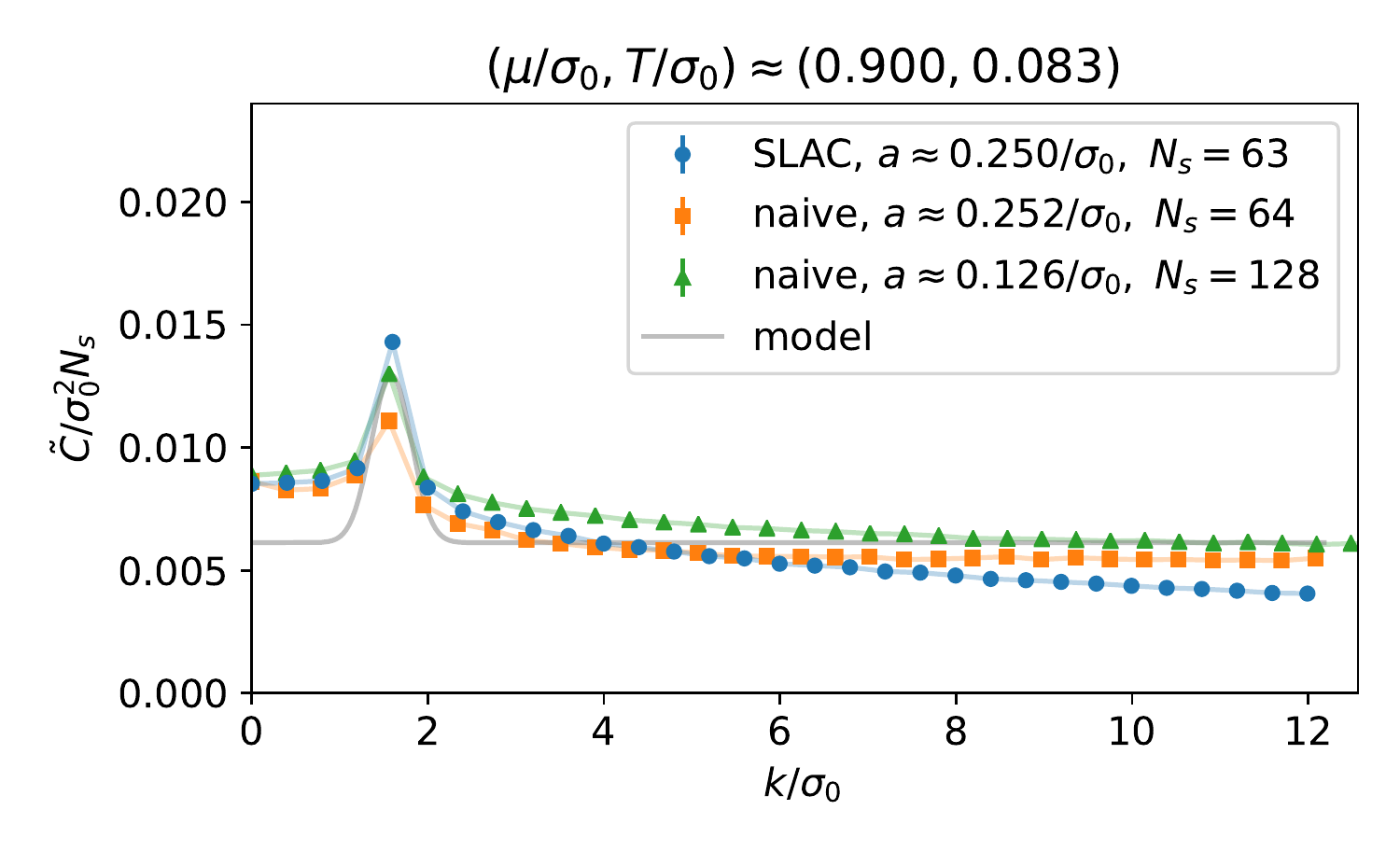}
	
	\caption{\label{FIG_C}The spatial correlation function $C(x) / \sigma_0^2$ (left column, where
	the $x-$range extends to half the box length) and its Fourier transform $\tilde{C}(k) / \sigma_0^2 \Ns $ (right column) for SLAC fermions (blue dots) and naive fermions (orange and green dots) in the symmetric, the homogeneously broken and the inhomogeneous phase ($\Nf = 8$) together with model expectations for the inhomogeneous phase (eqs.\ (\ref{EQN_C_INH}) and (\ref{EQN_Ck_INH}); gray curves).}
\end{figure}

Exemplary results for $C(x)$ and $\tilde{C}(k)$ are shown in 
Figure~\ref{FIG_C} inside the symmetric phase \\ ($(\mu / \sigma_0 , T / \sigma_0) \approx (0 , 0.993)$), inside the homogeneously broken phase ($(\mu / \sigma_0 , T / \sigma_0) \approx (0 , 0.083)$) and inside 
the inhomogeneous phase ($(\mu / \sigma_0 , T / \sigma_0) \approx (0.700 , 0.083)$ and \\ $(\mu / \sigma_0 , T / \sigma_0) \approx (0.900 , 0.083)$). 
In all cases there is reasonable agreement between our results for 
naive fermions and for SLAC fermions. Since lattice discretization 
errors are expected to be significantly larger for naive fermions, 
as discussed in section~\ref{SEC853}, we show results obtained with 
naive fermions in the inhomogeneous phase for two different lattice spacings, $a \approx 0.252 / \sigma_0$ 
and $a \approx 0.126 / \sigma_0$. Those corresponding to the finer lattice 
spacing are closer to the SLAC results, where $a \approx 0.250 / \sigma_0$. 
We interpret this as indication that both discretizations agree in the 
continuum limit. Moreover, on a qualitative level there is agreement 
with our crude expectations summarized by eqs.\ (\ref{EQN_C_SYM}) to
(\ref{EQN_C_INH}) and eqs.\ (\ref{EQN_Ck_SYM}) to (\ref{EQN_Ck_INH}), 
when the parameters in these equations are chosen appropriately. In 
particular the plots in the lower half of Figure~\ref{FIG_C} clearly 
indicate the existence of an inhomogeneous phase. $C(x)$ exhibits 
cos-like oscillations with decreasing wavelength $\lambda$ 
for increasing $\mu$, as observed in the $\Nf \rightarrow \infty$ 
limit. This is also reflected by the symmetric pair of peaks 
of $\tilde{C}(k)$ at the corresponding wave numbers $q = L / \lambda$. 
The gray curves in these plots represent the model expectations 
(eqs.\ (\ref{EQN_C_INH}) and (\ref{EQN_Ck_INH})) with parameters 
$A$, $q$, $\Delta q$ and $\epsilon$ determined by fits to the 
lattice results for $C(x)$ and $\tilde{C}(k)$.

A straightforward calculation leads to
\begin{align}
\label{EQN560} \tilde{C}(k) = \bigg\langle \frac{1}{\Nt \sqrt{\Ns}} \sum_t |\tilde{\sigma}(t,k)|^2 \bigg\rangle \, ,
\end{align}
where
\begin{equation}
\tilde{\sigma}(t,k) = \mathcal{F}_x(\sigma)(t,k)
\end{equation}
is the Fourier transform of $\sigma(\vx)$ with respect to the spatial coordinate (see eq.\ (\ref{EQN690b})). The absolute values $|\tilde{\sigma}(t,k)|$ are invariant under spatial translations $x \rightarrow x+\delta x$, because
\begin{align}
\mathcal{F}_x(\sigma(t,x+\delta x))(t,k) = e^{-\ii k \delta x} \mathcal{F}_x(\sigma(t,x))(t,k) .
\end{align}
This shows again that both $C(x)$ and $\tilde{C}(k)$ do not suffer from destructive interference, as already discussed at the beginning of this subsection. Moreover, eq.\ (\ref{EQN560}) shows in an explicit way that the Fourier transformed correlation function $\tilde{C}(k)$ also provides information about the absolute values of the Fourier coefficients of the field $\sigma(\vx)$. In particular the peaks in $\tilde{C}(k)$ at non-vanishing $k$ in the plots in the lower half of Figure~\ref{FIG_C} indicate that inside an inhomogeneous phase strong oscillations with the same wavelength are present in the majority of the generated field configurations.

\begin{figure}
	\centering
	\includegraphics[width=0.49\linewidth]{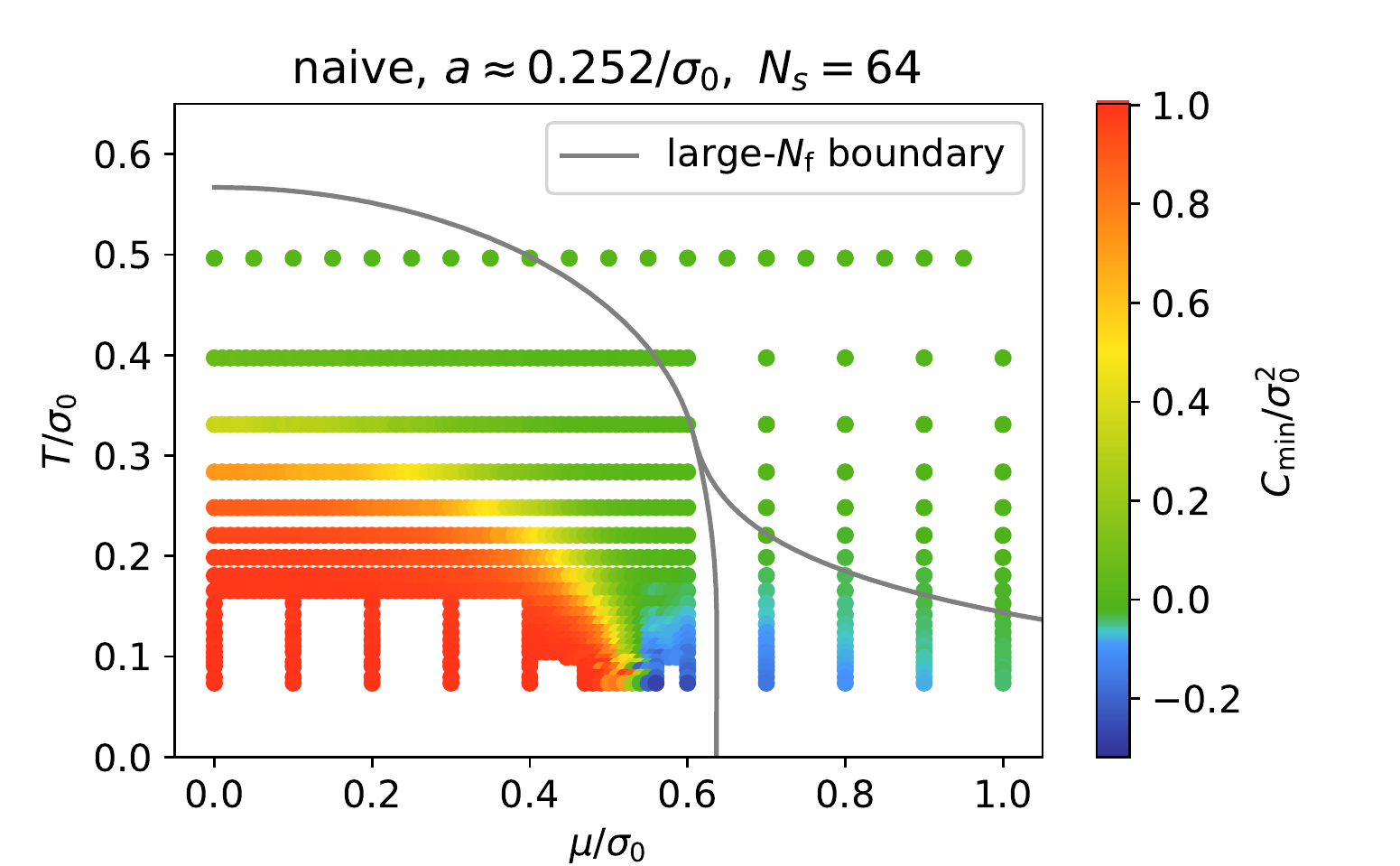}
	\includegraphics[width=0.49\linewidth]{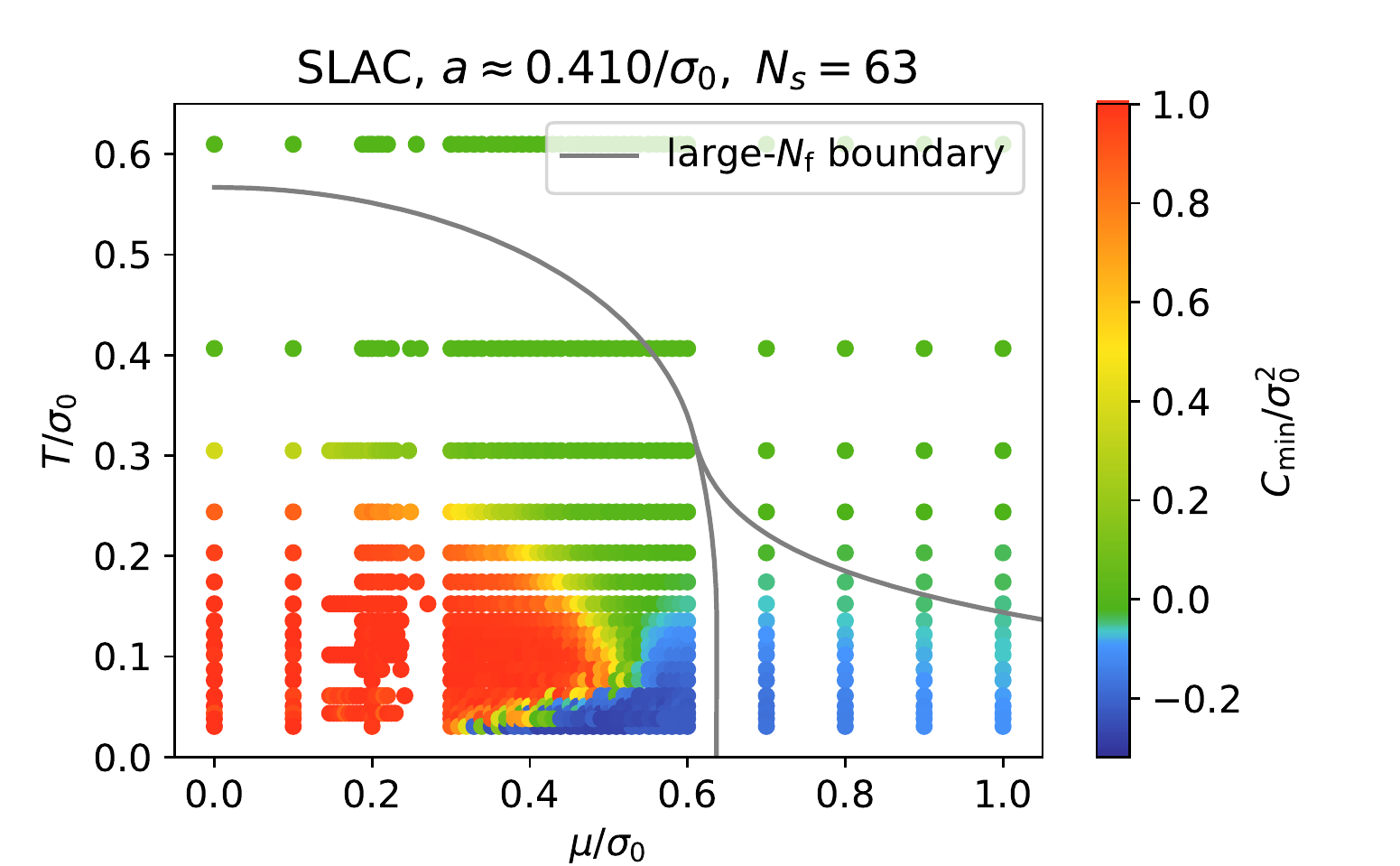}
	\caption{\label{Fig:Cmin}$C_\textrm{min} / \sigma_0^2$ in the
		$\mu$-$T$ plane for naive fermions (left plot) and SLAC 
		fermions (right plot) ($\Nf = 8$). The red regions correspond 
		to a homogeneously broken phase, the green regions to a
		symmetric phase and the blue regions to an inhomogenous 
		phase. The gray lines represent the $\Nf \rightarrow \infty$
		phase boundaries \cite{Thies:2003kk,Schnetz:2004vr}.}
\end{figure}

From Figure~\ref{FIG_C} one can see
\begin{align}
C_\textrm{min} = \min_x \, C(x) \left\{\begin{array}{rl}
\gg 0   & \text{inside a homogeneously broken phase} \\
\approx 0 & \text{inside a symmetric phase} \\
\ll 0   & \text{inside an inhomogeneous phase}
\end{array}\right. \, .
\end{align}
Thus, the minimum of the correlation function $C(x)$ is suited to plot a crude 
phase diagram as shown in Figure~\ref{Fig:Cmin} both for naive and 
for SLAC fermions. 
\begin{figure}
	\centering
	\includegraphics[width=1.05\linewidth]{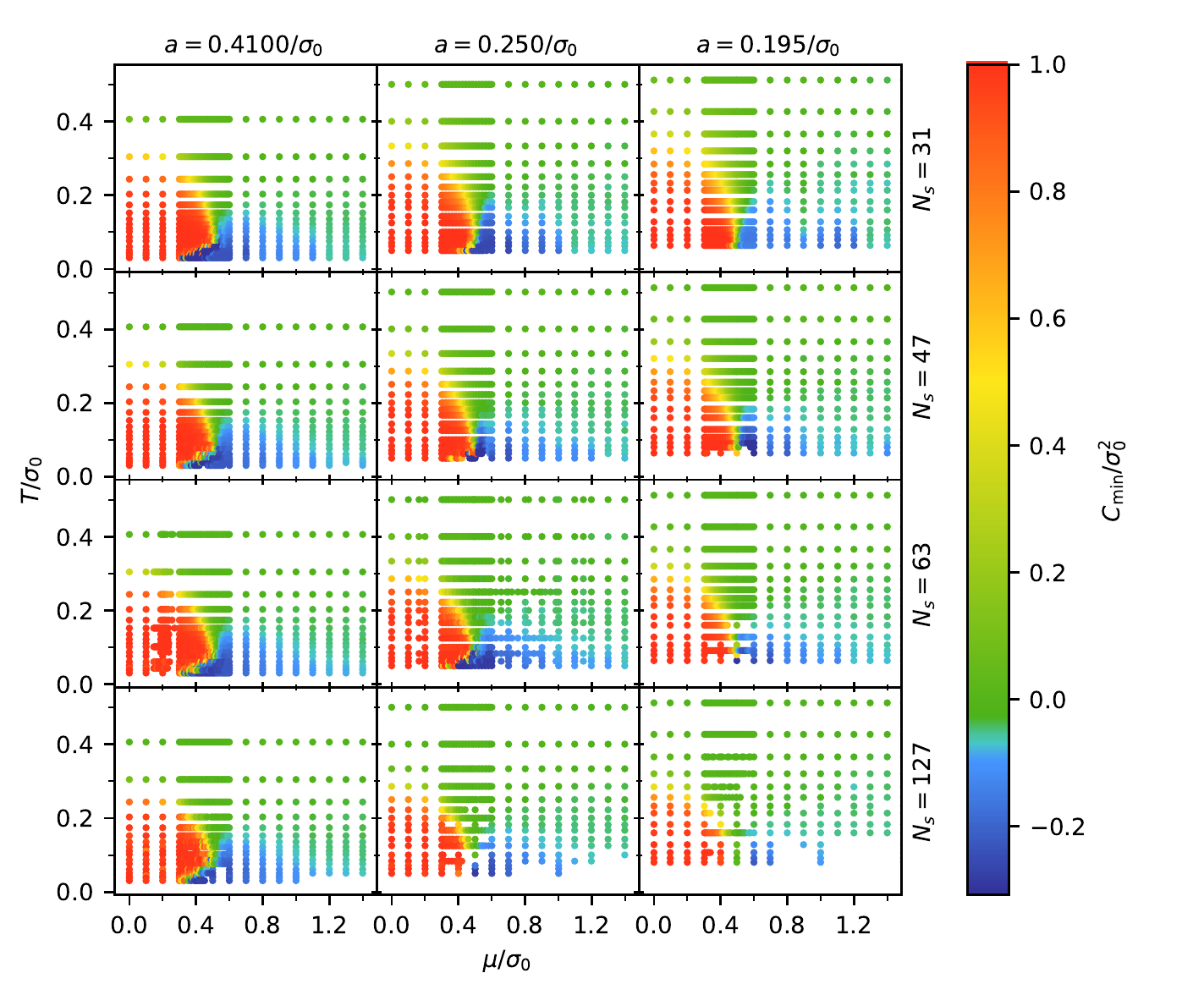}
	\caption{\label{Fig:Cmin_array}$C_\textrm{min} / \sigma_0^2$ in the $\mu$-$T$ plane for three different values of the lattice spacing $a$ (the columns) and four different numbers of lattice sites in spatial direction $\Ns$ (the rows) ($\Nf = 8$, SLAC fermions).}
\end{figure}

The red region indicates a homogeneously broken phase, 
the green region a symmetric phase and the blue region an inhomogeneous phase. 
As before, results obtained with these two different fermion discretizations 
are in fair agreement. Moreover, the phase diagram is qualitatively similar 
to the $\Nf \rightarrow \infty$ phase diagram from Refs.\
\cite{Thies:2003kk,Schnetz:2004vr}, whose phase boundaries are also shown in Figure~\ref{Fig:Cmin}.
The homogeneously broken phase and the inhomogeneous phase are, however, 
somewhat smaller for finite $\Nf$ than for $\Nf \rightarrow \infty$, 
presumably because quantum fluctuations at finite $\Nf$ increase disorder and,
thus, favor a symmetric phase. Note, however, that $C_\textrm{min}$ is not
the expectation value of a product of local operators as for example $C(x)$,
which is the two-point function of the order parameter. 
In general one must
be cautious, when using non-local quantities like $C_\textrm{min}$, since
they can fake non-existing phase transitions \cite{Kertesz:1989,Wenzel:2005nd}.
But in the present case transition lines have been localized by 
$C_\textrm{min}$ as well as the correlator $C(x)$
of the local field $\sigma$. 
%\textbf{XXXXX Transition lines = phase boundaries ...? Das haben wir %nicht gemacht. Vorschlag: ``But in the present case the three phases %have been localized by $C_\textrm{min}$ as well as by the %quasi-local correlator $C(x)$.'' XXXXX}

We also checked the stability of the phase diagram with respect to variations of the lattice spacing and the spatial volume. To this end we performed simulations using SLAC fermions at three different values of the lattice spacing, $a \approx 0.410 / \sigma_0 , 0.250 / \sigma_0 , 0.195 / \sigma_0$ (the columns in Figure~\ref{Fig:Cmin_array}), and for four different numbers of lattice sites in spatial direction, $\Ns = 31 , 47 , 63 , 127$ (the rows in Figure~\ref{Fig:Cmin_array}). Approaching the infinite volume limit at fixed lattice spacing corresponds to moving from the top to the bottom of the figure, while approaching the continuum limit at approximately fixed spatial volume corresponds to moving right and downwards at the same time. There is little difference in the crude phase diagrams shown in these twelve plots. We consider this as indication that our results, at our current level of accuracy, are consistent with results in the continuum and infinite spatial volume. 

\begin{figure}
	\centering
	\begin{minipage}{0.49\linewidth}
		{\scriptsize \begin{center} $\ \ \quad$ homogeneously broken $\leftrightarrow$ symmetric (blue points) \\ $\ \ \quad$ inhomogeneous $\leftrightarrow$ symmetric (orange points) \end{center}}
		\includegraphics[width=\linewidth]{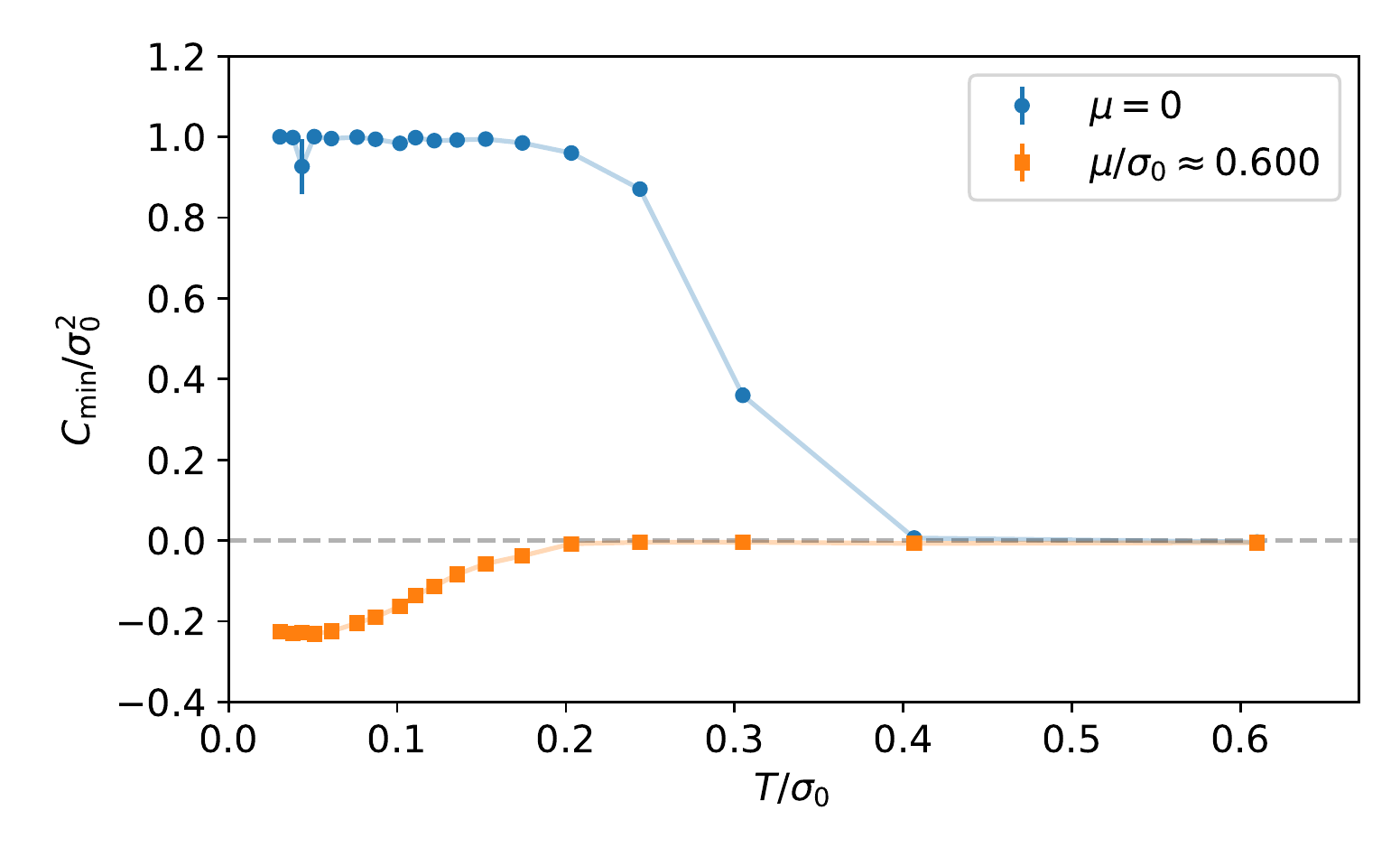}
	\end{minipage}
	\begin{minipage}{0.49\linewidth}
		{\scriptsize \begin{center} $\phantom{\textrm{homogeneously ...}}$ \\ $\ \ \quad$ homogeneously broken $\leftrightarrow$ inhomogeneous \end{center}}
		\includegraphics[width=\linewidth]{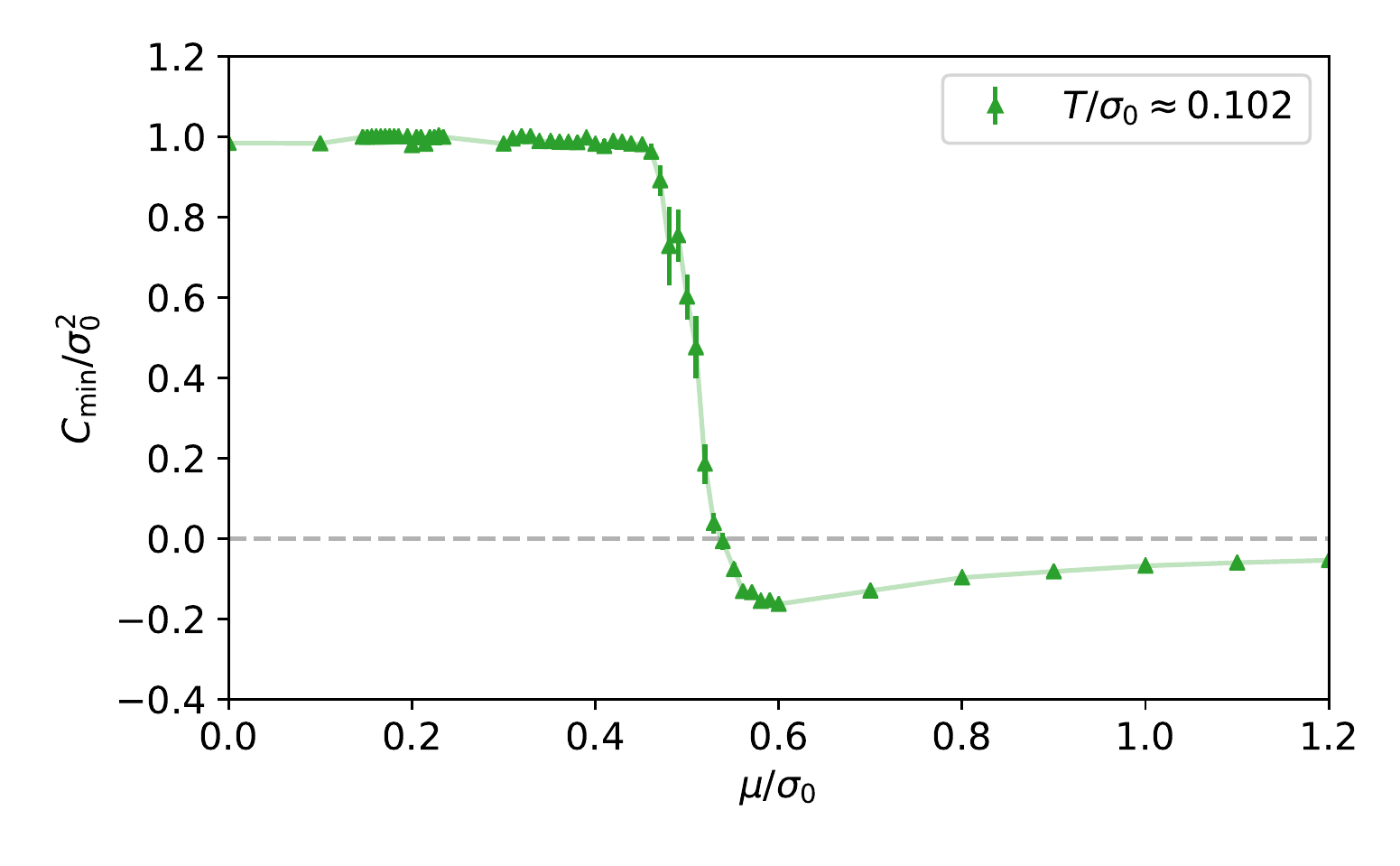}
	\end{minipage}
	\caption{\label{FIG853}$C_\textrm{min} / \sigma_0^2$ as function of $T$ for $\mu = 0$ and for $\mu / \sigma_0 \approx 0.600$ (left plot) and $C_\textrm{min} / \sigma_0^2$ as function of $\mu$ for $T / \sigma_0 \approx 0.102$ (right plot) ($\Nf = 8$, SLAC fermions, $a \approx 0.410 / \sigma_0$, $\Ns = 63$).}
\end{figure}

In the limit $\Nf \rightarrow \infty$ the phase boundaries between the three phases are of second order. In the following we present selected results for finite $\Nf$ and discuss, whether there are also phase transitions or rather 
crossovers.\footnote{As already pointed out on 
page \pageref{remark1}, we call the three regions
in $\mu$-$T$ plane ``phases'', although they may not be
phases according to the Ehrenfest classification.}
\begin{itemize}
	\item For the transition between the symmetric and the homogeneously broken phase we computed $C_\textrm{min}$ as function of the temperature $T$ for vanishing chemical potential $\mu = 0$ (see Figure~\ref{FIG853}, left plot, blue points). There is a rapid decrease of $C_\textrm{min}$ at around $T / \sigma_0 = 0.25$, which is qualitatively reminiscent to the $\Nf \rightarrow \infty$ result and, thus, might indicate that there is also a second order phase transition at finite $\Nf$.
	
	\item For the transition between the homogeneously broken and the inhomogeneous phase we computed $C_\textrm{min}$ as function of the chemical potential $\mu$ for rather low temperature $T / \sigma_0 \approx 0.102$ (see Figure~\ref{FIG853}, right plot). We observe a rapid decrease of $C_\textrm{min}$ at around $\mu / \sigma_0 = 0.5$, which indicates a phase transition similar to the $\Nf \rightarrow \infty$ case. 
	
	\item As can also be seen from the phase diagram in Figure~\ref{Fig:Cmin}, 
	the transition between the symmetric and the inhomogeneous phase is somewhat
	washed-out. This is also reflected by Figure~\ref{FIG853}, left plot, where 
	the orange points represent $C_\textrm{min}$ as function of the temperature 
	$T$ for chemical potential $\mu / \sigma_0 \approx 0.600$. 
	These results favor a weak phase transition of second or higher order or just a crossover.\footnote{A more detailed investigation of the long-range behavior of $C(x)$ presented in section~\ref{SEC598} points towards a phase transition.}
\end{itemize}
As has been pointed out earlier, these conclusions may not be
fully coherent since $C_\textrm{min}$ is a non-local quantity.

% **********

\subsection{\label{SEC598}The long-range behavior of $C(x)$}

We have argued in section~\ref{SEC499}
that SSB of a continuous (spacetime) symmetry in $1+1$ dimensions 
is a delicate issue. To decide, whether there are
NGB and if so, what type of NGB, we investigate the 
long-range correlations of the GN model with $\Nf = 2$ flavors. The question is, whether the long-range
order (which in the present context means that $C(x)$
oscillates with a constant non-zero amplitude for arbitrarily large
$\vert x\vert$),
which is necessary to form a crystal at $\Nf=\infty$,
remains at finite $\Nf$ long-range or becomes almost long-range
\`a la Berezinskii, Kosterlitz and Thouless (BKT) \cite{Berezinskii:71,Kosterlitz:1973xp} (in which case the amplitude decreases with distance like
an inverse power). 
Indeed, by studying the long-range behavior of the SU$(\Nf)$
Thirring model (this is a four-Fermi theory with current-current
interaction) in $1+1$ dimensions, Witten argued that
for finite $\Nf$ the correlations have almost long-range order,
\begin{equation}
\langle \bar{\psi}\psi(\vx)\bar{\psi}\psi(0)\rangle \sim \frac{1}{\vert \vx\vert^{1/\Nf}} \, , \qquad 
\vert \vx\vert\to\infty \, , \label{witten}
\end{equation}
such that only for $\Nf\to\infty$ the continuous chiral symmetry
is broken, i.e.\ $\langle\bar{\psi}\psi\rangle\neq 0$ \cite{Witten:1978qu}.
This way the system circumvents the no-go theorems for
SSB of continuous inner symmetries. In what follows we
try to answer the question whether a similar mechanism is at work for
translation symmetry in the GN model.

To detect SSB of translation invariance directly
we could break translation symmetry explicitly in a 
finite box with periodic BC, for example by adding a 
term $\varepsilon(x)\sigma(t,x)$
to the Lagrangian, perform the infinite volume limit
and finally remove the source $\varepsilon(x)$.
Assuming clustering in thermal equilibrium and
\begin{equation}
\lim_{\varepsilon\to 0}\lim_{L\to \infty}\langle \sigma(t,x)\sigma(t,0)
\rangle_\varepsilon=
\lim_{L\to\infty}\langle \sigma(t,x)\sigma(t,0)
\rangle_{\varepsilon=0}=C(x)
\end{equation}
we conclude that $C(x)$ is for large $|x|$ proportional to the condensate
(calculated with first $L\to\infty$ and
afterwards an adapted $\varepsilon\to 0$), i.e.\
\begin{equation}
C(x)=\big\langle \sigma(t,x)\sigma(t,0)\big\rangle \to \big\langle \sigma(t,x)\big\rangle \big\langle\sigma(t,0)\big\rangle \quad \textrm{for} \quad \vert x \vert\to\infty \, .
\end{equation}
In the inhomogeneous phase we can write 
\begin{equation}
\label{amplitude} C(x) = A(x) C_\mathrm{periodic}(x) \, ,
\end{equation}
where $A(x)$ is the non-increasing amplitude function, while $C_\mathrm{periodic}(x)$ represents the periodic oscillations. If the system forms a crystal, the amplitude function $A(x)$ should approach a non-zero constant for sufficiently large separations $\vert x \vert$. In case the system has almost long-range order \`a la BKT, the amplitude function decreases with $\vert x \vert$ as an inverse power.
To distinguish the two scenarios we study
$C(x)$ for small $\Nf = 2$, to detect a possible deviation of $A(x)$
from the asymptotically constant behavior in the large-$\Nf$ limit (for small $\Nf$ quantum fluctuations might be strong enough to
change long-range into almost long-range order as it happens in the chirally invariant SU($\Nf$) 
Thirring model for finite $\Nf$ \cite{Witten:1978qu}). Thus, we expect one of the following amplitude functions:

\begin{enumerate}
	\item In a BKT-like phase without SSB we expect the amplitude 
	function $A(x)$ to have the following behavior for large $\vert x \vert$:
	\begin{equation}
	\label{eq:decay}
	A(x)=A_{\mathrm{BKT}}(x)\sim\frac{\alpha}{|x|^{\beta}} +\ldots
	\end{equation}
	The dots indicate sub-leading terms and terms arising from the finite spatial 
	extent of the system.
	
	\item If there is SSB of translation invariance,
	$C(x)$ oscillates with constant non-zero amplitude at 
	large $|x|$, where the short-ranged contributions 
	from excited states are suppressed. The amplitude function would 
	then be
	\begin{equation}
	\label{eq:constant}
	A(x)=A_{\mathrm{SSB}}(x)\sim \gamma+\alpha e^{-m|x|}+\ldots\qquad
	\text{or}\qquad 
	A(x)=A_{\mathrm{SSB'}}(x)\sim \gamma+\frac{\alpha}{\vert x\vert^{\beta}}+\ldots
	\end{equation}
	depending on whether the excitations over the
	oscillating condensate are massive or massless.
	If the NGB decouple from the system,
	the amplitude function approaches the constant amplitude $\gamma\neq 0$
	exponentially fast. If not, $A(x)$ will
	approach $\gamma \neq 0$ with an inverse power of $\vert x\vert$.
\end{enumerate}

\begin{figure}
	\centering
	\includegraphics[width=0.65\linewidth]{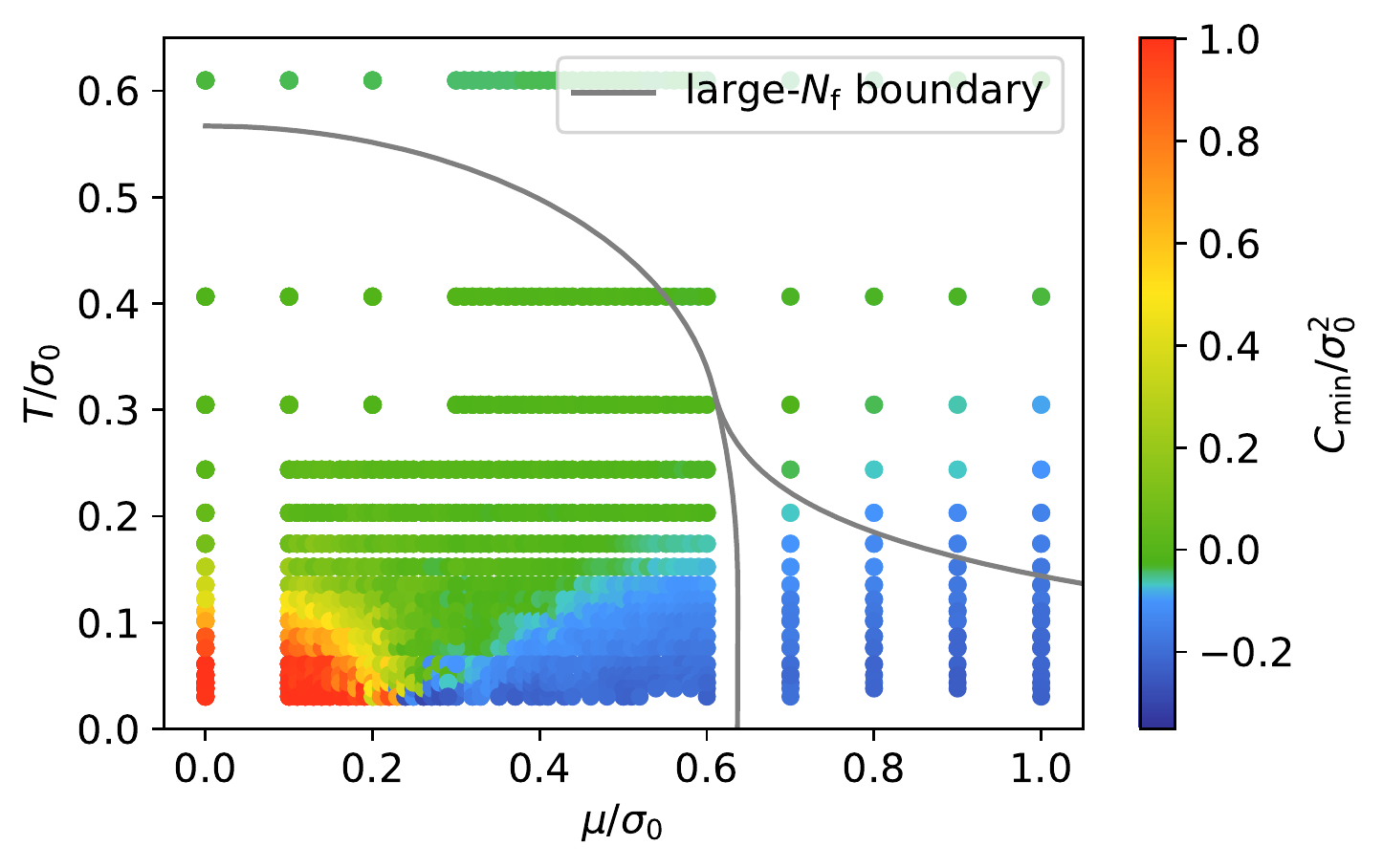}
	\caption{\label{Fig:Cmin_Nf2}$C_\textrm{min} / \sigma_0^2$ in the $\mu$-$T$ 
		plane for $\Nf = 2$. The red region corresponds to a homogeneously broken
		phase, the green region to a symmetric phase and the blue region 
		to an inhomogenous phase. The grey lines represent the $\Nf \rightarrow \infty$
		phase boundaries \cite{Thies:2003kk,Schnetz:2004vr}. }
\end{figure}

Before discussing the long-range behavior of $C(x)$ we present the
phase diagram from $C_\textrm{min}$ for $\Nf=2$ in Figure~\ref{Fig:Cmin_Nf2}.
Again we recognize the same phases as in the large-$\Nf$ limit: 
a homogeneously broken phase
which (as expected) is significantly smaller than for $\Nf=8$ and $\Nf\to\infty$,
a symmetric phase for sufficiently large 
temperature and a region, where $C_{\textrm{min}}$ is clearly negative. One unexpected and striking feature of the latter phase is 
that the temperature range 
with negative $C_\textrm{min}$ grows with increasing $\mu$, which is 
qualitatively different from the situation at large $\Nf$. This 
already happens for $\Nf=8$ in a small region of parameter space 
(see Figure~\ref{Fig:Cmin_array}) but is so pronounced at $\Nf=2$ 
that up to $\mu/\sigma_0 = 1.4$ we found no evidence that the transition 
line separating the inhomogeneous and the restored phase will bend down 
to the $\mu$ axis.

\begin{figure}
	\centering
	\includegraphics[width=0.49\linewidth]{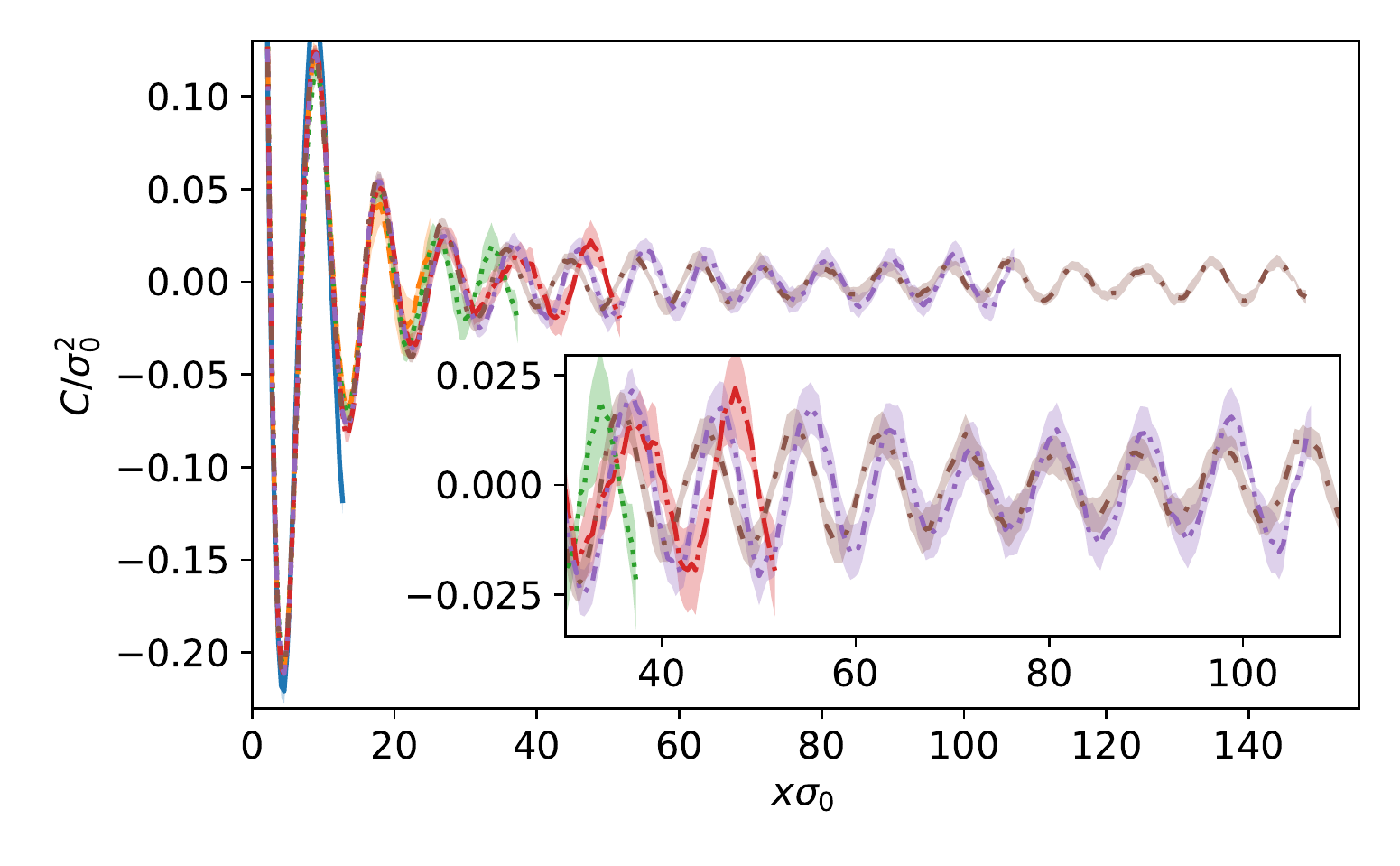}
	\includegraphics[width=0.49\linewidth]{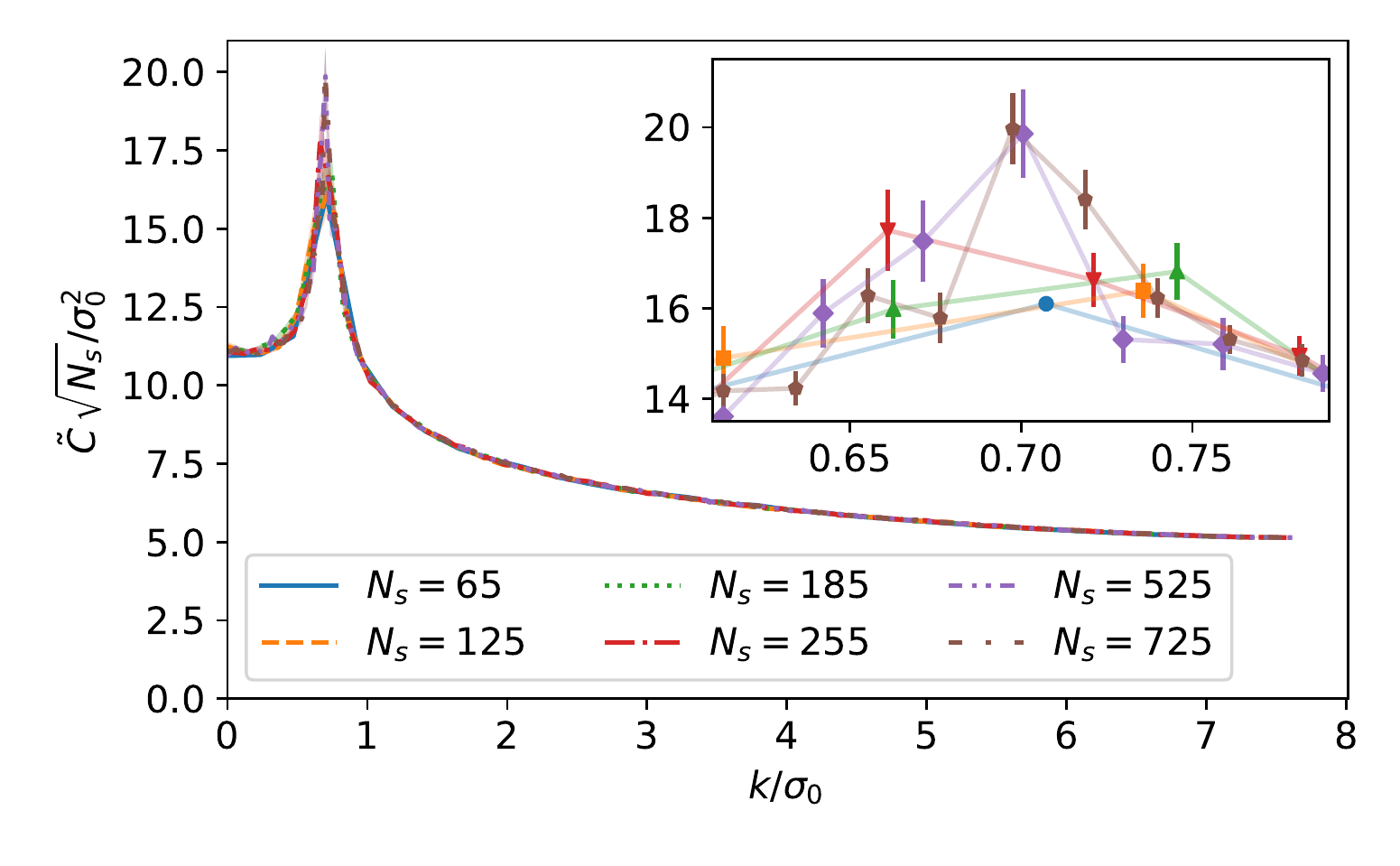}
	\caption{\label{FIG649}The spatial correlation function $C(x) / \sigma_0^2$ (left) 
		and its Fourier transform $\tilde{C}(k) \sqrt{\Ns}/ \sigma_0^2 $ (right) 
		at $(\mu/\sigma_0 , T/\sigma) = (0.500 , 0.030)$ for $\Nf=2$ and
		$\Ns = 65,125,185,255,525,$ $725$. Except for the inset 
		on the right, for clarity interpolating 
		lines are shown instead of the data points.}
\end{figure}

Now we investigate the long-range behavior of the amplitude function $A(x)$ defined in eq.~(\ref{amplitude})
for $\Nf=2$ on lattices with a rather large number of sites in spatial direction.
Figure~\ref{FIG649} shows the correlator $C(x)$ and its
Fourier transform $\tilde{C}(k)$ for $(\mu/\sigma_0 , T/\sigma) = (0.500 , 0.030)$ 
and various $\Ns$ up to $725$ corresponding to spatial extents up to $L \approx 297.3 / \sigma_0$.
We observe $32$ periods of statistically significant oscillations in $C(x)$ over the whole range of
separations and a pronounced peak of $\tilde C(k)$ at the corresponding wave number.
The position of the peak is essentially the same for all $L$ demonstrating once more that the wave length
is independent of the spatial extent.

\begin{figure}
	\centering
	\includegraphics[width=0.49\linewidth]{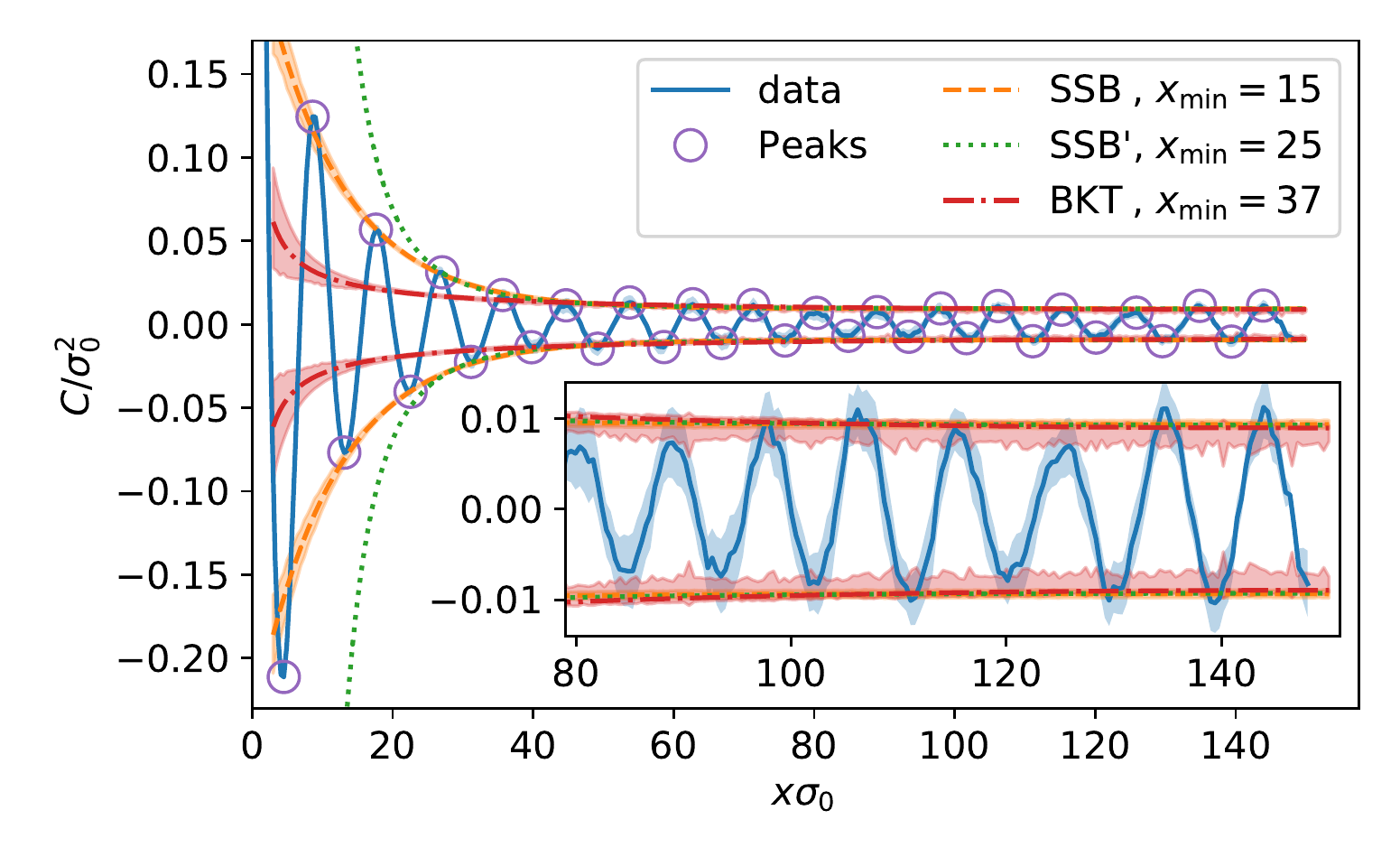}
	\includegraphics[width=0.49\linewidth]{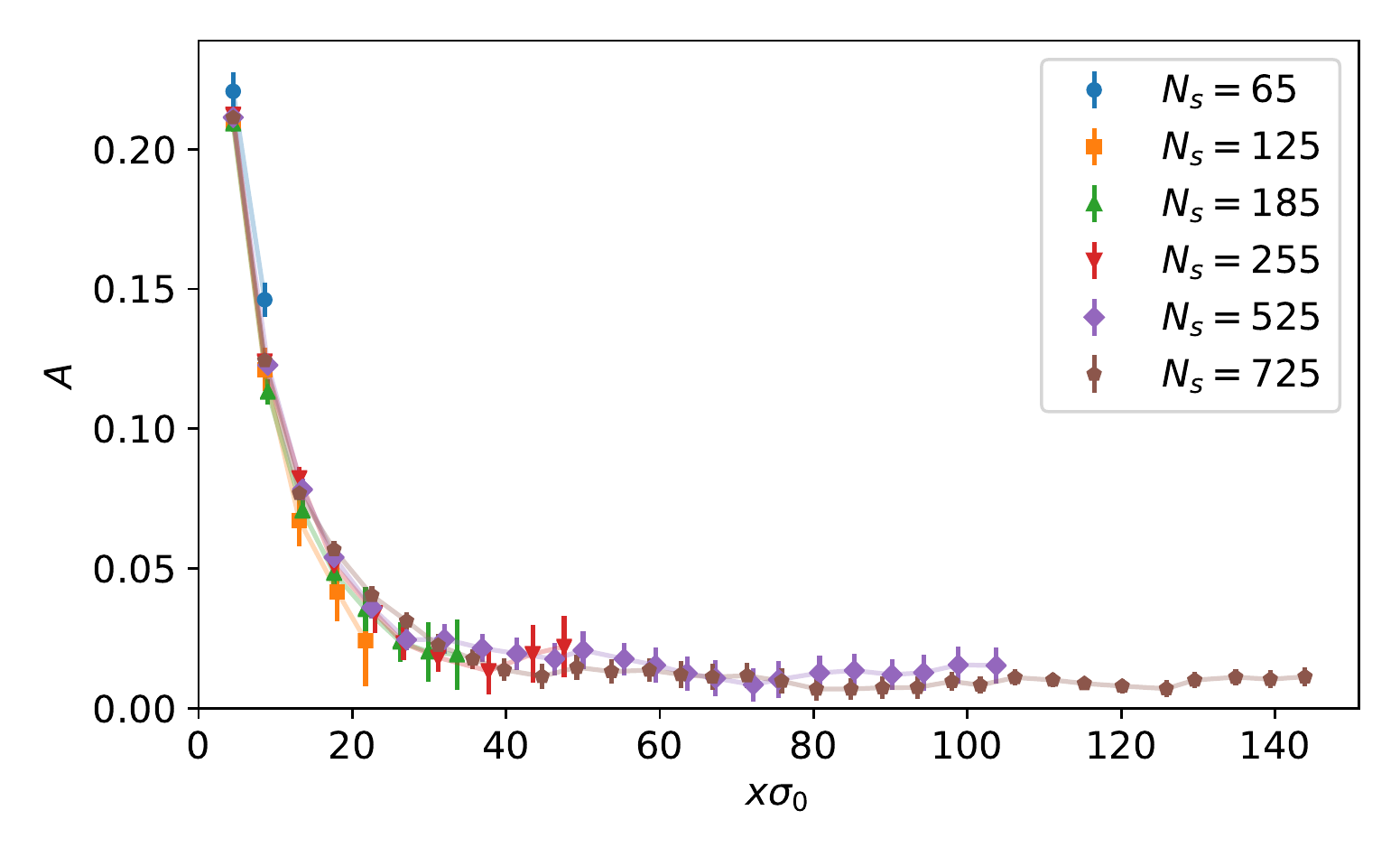}
	\caption{\label{FIG:ampDecay}The left 
		plot shows the correlation function $C(x)/\sigma_0^2$ at $(\mu/\sigma_0 , T/\sigma) = (0.500 , 0.030)$ for $\Ns=725$, the extracted peaks and fits with the functions defined in eqs.\ (\ref{eq:decay}) and (\ref{eq:constant}) (the parameters obtained by these fits are shown in Table~\ref{TAB_ampDecay}).
		For clarity interpolating lines are shown instead of the data points.
		The right plot shows the extracted peaks for $\Ns = 65,125,185,255,525,725$.}
\end{figure}

\begin{figure}
	\centering
	\includegraphics[width=0.49\linewidth]{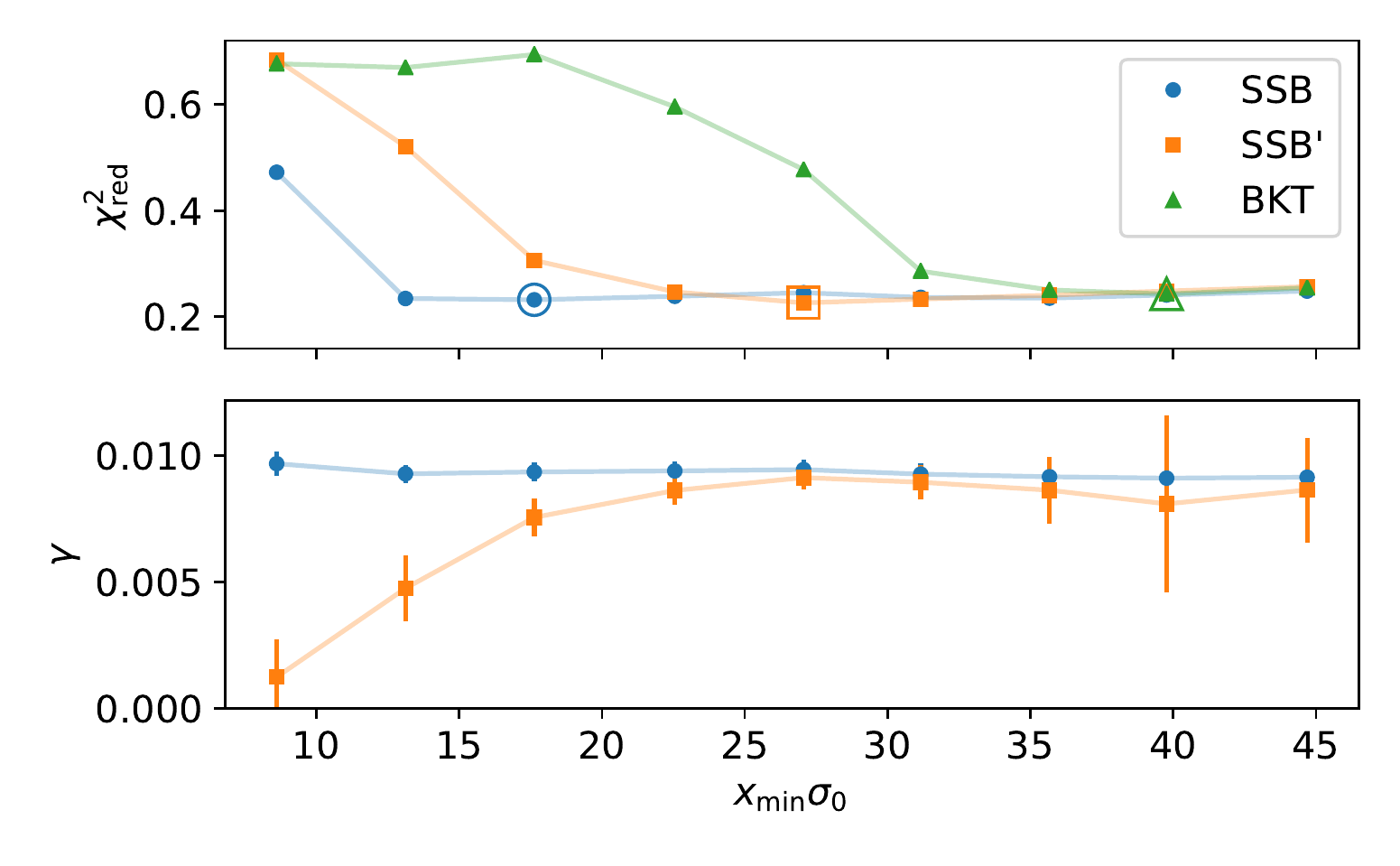}
	\caption{\label{FIG:chisq}Top: $\chi_{\textrm{red}}^2$ as a function of $x_{\min}$ for all three models ($\Ns = 725$). Bottom: Extracted parameters $\gamma$ for the SSB and SSB' model as functions of $x_{\min}$.}
\end{figure}

The amplitude function $A(x)$ is extracted from the peaks of 
the correlation function $C(x)$ (which we identified using the \texttt{scipy.signal.find\_peaks} method \cite{scipy_10_contributors_scipy_2020} with \texttt{prominence=0.01})
for all $\Ns$, as exemplified for $\Ns=725$ in the 
left plot of Figure~\ref{FIG:ampDecay}. 
The peaks of $C(x)$ for various $\Ns$ are 
depicted in the right plot of Figure~\ref{FIG:ampDecay}. There is a rapid drop for small separations $x$
that flattens out for asymptotically large $x$.
We performed $\chi^2$-minimizing fits of symmetrized versions of the expectations for the amplitude function $A(x)$ (eqs.\ (\ref{eq:decay}) and (\ref{eq:constant})) to the extracted peaks for $\Ns=725$ (see Figure~\ref{FIG:ampDecay}, left plot).
We only used data points with $x \geq x_{\min}$ in the fitting procedure, because all three fit functions are expected to model the amplitude function for large $|x|$. Since $\chi_{\textrm{red}}^2$ as a function of $x_{\min}$ is almost constant for large $|x|$ (see Figure~\ref{FIG:chisq}, upper plot), we chose (independently for each of the three fits) the minimal value $x_{\min}$, where $\chi_{\textrm{red}}^2$ is consistent with the asymptotic constant, i.e.\ that value, where the constant behavior sets in. In the lower plot of Figure~\ref{FIG:chisq} we show the extracted parameters $\gamma$
appearing in the two fit functions in eq.~(\ref{eq:decay}) as functions of $x_{\min}$. It is reassuring that both parameters $\gamma$ are essentially independent of $x_{\min}$ as long as the corresponding $\chi_{\textrm{red}}^2$ is small.
Since there are massive excitations in the SSB model
and massless excitations in the SSB' and BKT models,
one should not expect to find the same $x_{\min}$
for the three models, but a smaller $x_{\min}$ for the SSB model, where the excitations are short-range. This expectation is confirmed by our numerical results (see column ``$x_{\min}$'' in Table~\ref{TAB_ampDecay}).

\begin{table}
	\centering
	\begin{tabular}{cccccc}
		\hline\hline
		&$x_{\min}$&$\alpha$&$\beta$ resp.\ $m$&$\gamma$&$\chi_{\mathrm{red}}^{2}$\\
		\hline
		SSB&$15$&$0.231\pm0.026$&$0.0895\pm0.0054$&$0.00936\pm0.00036$&$0.23$\\
		SSB'&$25$&$\left( 1.3\pm2.2 \right)\cdot 10^{3}$&$3.33\pm0.50$&$0.00913\pm0.00046$&$0.23$\\
		BKT&$38$&$0.190\pm0.082$&$0.66\pm0.13$&-&$0.25$\\
		\hline\hline		
	\end{tabular}
	\caption{\label{TAB_ampDecay}Parameters of the amplitude functions $A(x)$ in eqs.\ (\ref{eq:decay}) and (\ref{eq:constant}) obtained by fits to the extracted peaks for $\Ns = 725$.}
\end{table}

The fit results for the parameters of the amplitude functions $A(x)$ from eqs.\ (\ref{eq:decay}) and (\ref{eq:constant}) for $\Ns = 725$ are collected in Table~\ref{TAB_ampDecay}. There are several points to note:
\begin{itemize}
	\item The SSB model admits the most stable fits with resulting parameters almost independent of $x_{\min}$ and the initial values used in the fitting algorithm. $\gamma$ is cleary different from zero.
	
	\item Fits for the SSB' model are less stable, which is reflected by the large uncertainties obtained for $\alpha$ and $\beta$. $\gamma$, however, can be determined in a reliable way and the result is again different from zero. Moreover, it is in excellent agreement with the corresponding result for the SSB model. It is also interesting to note that the SSB' model, which differs from the BKT model by the additive constant $\gamma$, leads to significantly smaller $\chi_{\mathrm{red}}^{2}$ (and $\gamma \neq 0$) than the BKT model, when the same $x_{\min}$ is used.
	
	\item The BKT model is only able to describe the amplitude function for rather large $|x|$, i.e.\ the corresponding $x_{\min}$ is significantly larger than for the SSB model and SSB' model. The fit result for the exponent is $\beta = 0.66 \pm 0.13$, which is similar to the corresponding analytically known value $\beta = 1 /  \Nf = 1/2$ of the $\textrm{SU}(\Nf)$ Thirring model.
\end{itemize}
To summarize, it seems that the excitations are probably massive, because the SSB model is able to describe the extracted peaks for significantly smaller separations $x$, than it is possible with the SSB' model or the BKT model. When allowing for a constant $\gamma$ (as in the SSB model and the SSB' model), the fits lead to stable and clearly non-zero results. Thus, our current data is best described by the SSB scenario, where the NGB completely decouples from the system. However, this does not imply that this is the physical situation in the 
thermodynamic limit, since we saw that even $\Ns = 725$ lattice sites in spatial direction are not sufficient, to rule out the $1 / |x|^\beta$ almost long-range behavior of the BKT scenario. In other words, a constant behavior $\propto \gamma$ and an inverse power $\propto 1 / |x|^\beta$ are very similar for large $|x|$, in particular in a periodic spatial volume, such that even larger lattices or higher accuracy is needed to clearly distinguish between these scenarios.

\begin{figure}
	\centering
	\includegraphics[width=0.49\linewidth]{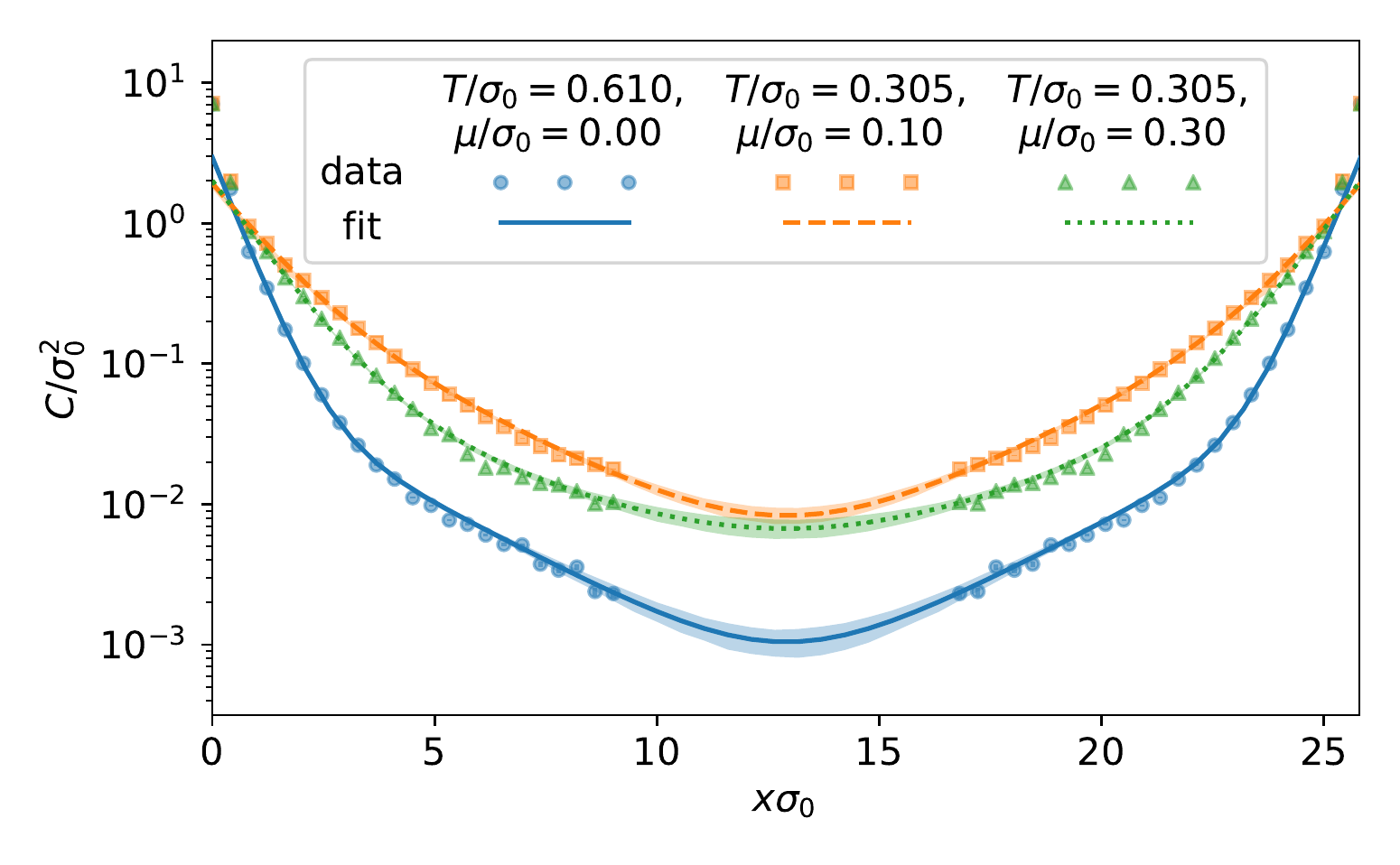}
	\caption{\label{FIG:highT_C}The spatial correlation function $C(x) / \sigma_0^2$ for three different $(\mu,T)$ in the symmetric phase and fits with $A_{1} \cosh(m_{1} (x-L/2)) + A_{2} \cosh(m_{2}(x-L/2))$. We do not show data points in the region $9 \leq x \sigma_0 \leq 16$ because of large systematic errors due to autocorrelations.}
\end{figure}

Despite the fact that we could not fully reveal the nature of the inhomogeneous phase, we can still argue that there exists a phase transition between the inhomogeneous low-temperature phase and the symmetric high temperature phase. This can be seen from Figure~\ref{FIG:highT_C}, where we show the spatial correlation function $C(x)$ together with $\cosh$-fits for three different $(\mu,T)$ in the symmetric phase. It is evident that the $\cosh$-functions perfectly fit the data points, which indicates that in the symmetric phase the excitations
are massive. In contrast to that, at low temperature in the inhomogeneous phase $C(x)$ is long-range
or almost long-range (see Figure~\ref{FIG649}). Since $C(x)$ behaves qualitatively different in the inhomogeneous phase and in the symmetric phase, i.e.\ long-range or almost long-range versus exponentially decaying, we expect a phase transition, either a symmetry-restoring transition or a BKT-like transition from the low-temperature 
inhomogeneous phase to the high-temperature symmetric phase.

% ********************

\subsection{\label{SEC599}Approaching the $\Nf \rightarrow \infty$ results with computations at finite $\Nf$}
\begin{figure}[t]
	\centering
	\includegraphics[width=0.5\linewidth]{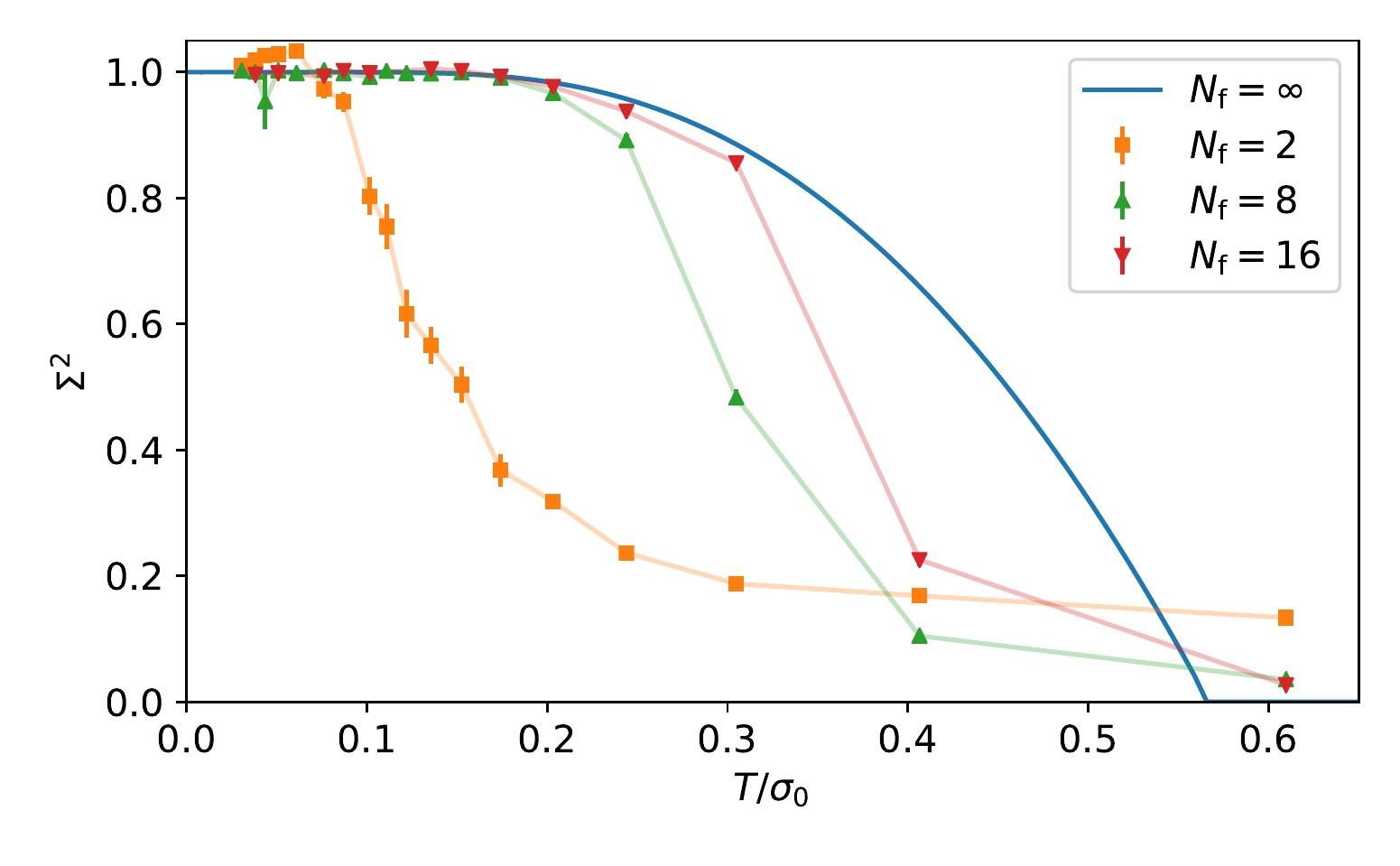}
	\caption{\label{FIG643}$\Sigma^2$ as a function of $T$ for $\mu = 0$ and $\Nf = 2 , 8 , 16 , \infty$ (SLAC fermions, $a \approx 0.410 / \sigma_0$, $\Ns = 63$).}
\end{figure}

In sections~\ref{SEC597} to \ref{SEC598} we have presented results at $\Nf = 2$ and $\Nf = 8$, which are similar to the analytically known $\Nf \rightarrow \infty$ results \cite{Thies:2003kk,Schnetz:2004vr}, e.g.\ for the phase diagram. To check and to confirm that results at finite $\Nf$ approach for increasing $\Nf$ the $\Nf \rightarrow \infty$ results, we also performed simulations at $\Nf = 16$. An exemplary plot is shown in Figure~\ref{FIG643}, where $\Sigma^{2}$ is shown as a function of the temperature $T$ for vanishing chemical potential $\mu = 0$ and $\Nf = 2 , 8 , 16 , \infty$. While results for $\Nf = 2$ agree with the $\Nf \rightarrow \infty$ result only for rather small $T / \sigma_0$, there is agreement also for larger $T / \sigma_0$, when $\Nf$ is increased, indicating that one can approach the analytically known $\Nf \rightarrow \infty$ results with computations at finite $\Nf$.

\clearpage

\section{Conclusions}

In the present work we could localize three regimes
in the space of thermodynamic control parameters
$T$ and $\mu$, in which the two-point function of the
order parameter shows qualitatively different behaviors.
We spotted a homogeneously broken phase, a symmetric
phase and a region with oscillating correlation function $C(x)$ 
defined in eq.\ (\ref{eq:Cx}).
The results of our Monte Carlo simulations with two different
types of chiral fermions 
for systems with $2$, $8$ and $16$ flavors on lattices with
sizes up to $\Nt=80$ and $\Ns=725$ were presented, analyzed
and discussed in the main body of the text.
Although we could not answer the question, whether 
in GN models with a finite number of flavors
translation invariance is spontaneously broken
at low $T$ and large $\mu$, or whether the system
is in a Berezinskii-Kosterlitz-Thouless like phase,
we clearly spotted a low-temperature and high-density
region, where the model exhibits oscillating spatial correlators. 
The wave-length of the spatial oscillation is determined 
by the chemical potential and temperature and not by the 
lattice spacing and the spatial extent of the lattice, i.e.\ spatial oscillations are neither a lattice
artifact nor a finite size effect. We argued that there
is a transition between the inhomogeneous phase and the
symmetric phase which could be an infinite order
transition (according to the Ehrenfest classification).
In an accompanying paper we shall demonstrate
that the ratio of the system size and the dominant
wave length of the condensate oscillations is equal to the number
of baryons in the systems. This further substantiates the physical
picture that the GN model in equilibrium at low
temperature and large fermion density either forms a crystal of baryons
or a viscous fluid of baryons. In this work we also showed that the amplitude of
oscillations stays constant or decays very slowly as 
suggested by a related result \cite{Witten:1978qu}. 
The first behavior is expected for a baryonic crystal
the second behavior for a viscous baryonic fluid. 
To better understand, how the long range behavior
at low temperature and high density does not
clash with the absence of Nambu-Goldstone bosons in $1+1$ dimensions, needs further high-precision results on the two-point 
function of the order parameter.
If the dispersion relation is non-relativistic or if the massless
modes fully decouple from the system then there should be no problem. 

Independent of whether the oscillating correlator $C(x)$ points to
a baryonic crystal or a baryonic liquid for finite $\Nf$
we have seen that mean-field/large-$\Nf$ approximations may keep
more information on the physics at finite $\Nf$ than one
might expect. This is reassuring 
since in particle physics and even more so in solid state physics 
we often rely on mean-field type approximations. 
%The results may be 
%of relevance in condensed matter systems, 
%e.g.\@ for large, almost one-dimensional polymers
%\cite{???}.\footnote{Of course, a mapping to condensed matter %
%	terminology is necessary beforehand.} 
An important question is, of course, whether our results have any relevance for QCD at finite baryon density.
On the one hand, we established that the interpretation 
as baryonic matter is not spoiled by taking quantum fluctuations into 
account. On the other hand, although recent 
\textit{numerical} investigations
of four-Fermi theories in $2+1$ dimensions and for $\Nf\to\infty$
spotted inhomogenous condensates \cite{Winstel:2019zfn}, the spatial modulation
is related to the cutoff scale and seems to disappear 
in the continuum limit \cite{Narayanan:2020uqt}. 
Clearly, if this happens then we cannot
expect a breaking of translation invariance for a finite 
number of flavors. Thus, extending our lattice studies to higher 
dimensions is of relevance for QCD. Simulations of 
interacting Fermi theories in $2+1$ dimensions are 
under way and we hope to report on our findings soon.

\clearpage

% **********

\appendix

\section{\label{APP567}Lattice discretization of the GN model with naive fermions}
\subsection{\label{SEC488}Free naive fermions}

The action of free naive fermions with chemical potential $\mu$ is given in eq.\ (\ref{EQN744}). The Fourier representations of the fermion fields are
\begin{equation}
\label{EQN562} \chi(\vx) = \frac{1}{\sqrt{\Nt\Ns }} \sum_{\vk} 
e^{\ii \vk\cdot\vx} \tilde{\chi}(\vk) \, , \qquad \bar{\chi}(\vx) = \frac{1}{\sqrt{\Nt\Ns }} \sum_{\vk} e^{-\ii \vk\cdot\vx} \tilde{\bar{\chi}}(\vk) \, ,
\end{equation}
where the discrete $2$-momenta $\vk = (k_0 , k_1)$ are 
in the first Brillouin zone, $-\pi \leq k_\mu a \leq \pi$, and are
chosen such that BC in $t$ direction are antiperiodic
and in $x$ direction are periodic (see section~\ref{SEC468}, in
particular eq.\ (\ref{EQN690})). Inserting these Fourier
representations into eq.\ (\ref{EQN744}) leads to
\begin{equation}
S_\textrm{free}[\tilde{\chi},\tilde{\bar{\chi}}] = -\sum_{k} \tilde{\bar{\chi}}(k) \big(\gamma^0 \ring{k}_0+ 
\gamma^1 \ring{k}_1\big) \tilde{\chi}(k)\,,\label{freelaction}
\end{equation}
where we abbreviated
\begin{equation}
\ring{k}_0= 
\cosh(\mu a) \frac{\sin(k_0 a)}{a}-\ii
\sinh(\mu a) \frac{\cos(k_0 a)}{a}
\, , \qquad 
\ring{k}_1=\frac{\sin(k_1 a)}{a}\,.
\end{equation}
In the limit $a \rightarrow 0$
the sums over $k_0$ and $k_1$ can be restricted to the ``soft modes'', 
where both $|\ring{k}_0 a| \ll 1$ and
$|\ring{k}_1 a| \ll 1$. There are four regions 
of soft modes in the first Brillouin zone, and they
are denoted by $\mathcal{R}_{uv}$ with $u,v\in\{0,1\}$.
The momenta of the soft modes in region $\mathcal{R}_{uv}$
are in the neighborhood of the four momenta
\begin{equation}
\vk_{uv}=\frac{\pi}{a}\left(u\atop v\right) \, , \qquad u,v\in\{0,1\}\,,
\end{equation}
at which (for $\mu=0$) the lattice momenta $\ring{k}_0$ and $\ring{k}_1$
vanish. For $\vk\in\mathcal{R}_{uv}$ we have
\begin{equation}
\ring{k}_0=(-1)^u k_0-\ii \mu+O(a^2) \, , \qquad 
\ring{k}_1=(-1)^v k_1+O(a^2)\,.\label{ascorr}
\end{equation}
Now we define the soft modes in the four regions 
according to
$\tilde{\chi}_{uv}(\vk)=\tilde{\chi}(\vk_{uv}+\vk)$
(and analogous for $\bar{\tilde{\chi}}$) with 
small $\vert k_\mu a \vert$.
Neglecting the $O(a^2)$ corrections in (\ref{ascorr})
we can approximate the free lattice action (\ref{freelaction}) by
\begin{equation}
S_\textrm{free}[\tilde{\chi},\tilde{\bar{\chi}}] 
\approx -\sum_{u,v}\sum_{\vk\in \mathcal{R}_{uv}}
\tilde{\bar{\chi}}_{uv}(\vk)\,\Big(
\gamma^0((-1)^uk_0-\ii \mu)+\gamma^1 (-1)^v k_1 \Big)\,\tilde{\chi}_{uv}(\vk)\,.\label{EQN639}
\end{equation}
This short calculation exhibits the well-known fermion flavor 
doubling for each spacetime dimension. It also shows that 
both $\Nt$ and $\Ns$ must be even to obey anti-periodic 
boundary conditions in $t$ direction and periodic boundary 
conditions in $x$ direction for each of the four fermion 
flavors.

It is important to note that the action (\ref{EQN639}) differs in a
couple of minus signs in front of the $\gamma$ matrices for flavors
$(u,v)\neq (0,0)$ from the corresponding continuum expression for four
free fermion flavors. These minus signs can be eliminated by changing
field coordinates via
\begin{equation}
\label{EQN167a} 
\tilde{\chi}_{uv} = (\gamma^0)^u (\gamma^1)^v \tilde{\psi}_{uv} \, , \qquad \bar{\tilde{\chi}}_{uv} = \bar{\tilde{\psi}}_{uv} (\gamma^1)^v (\gamma^0)^u \, , \qquad u,v \in \{ 0,1 \} \, .
\end{equation}
Then eq.\ (\ref{EQN639}) becomes
\begin{eqnarray}
S_\textrm{free}[\tilde{\psi},\tilde{\bar{\psi}}] \approx 
-\sum_{u,v} \ \sum_{\vk} \tilde{\bar{\psi}}_{uv}(\vk) \big(\gamma^0 (k_0-\ii\mu) + \gamma^1 k_1\big) 
\tilde{\psi}_{uv}(\vk) \, .
\end{eqnarray}
This shows that the lattice action (\ref{EQN744}) corresponds 
in the continuum limit to four massless non-interacting fermion flavors.
% **********

\subsection{Naive fermions and the GN model}

Discretizing the GN model (\ref{lsigma}) with $\Nf=4$ flavors
in a straightforward way using the fields $\chi$ and $\bar{\chi}$ via
\begin{equation}
\label{EQN389} S_{\sigma}[\chi,\bar{\chi},\sigma] = S_\textrm{free}[\chi,\bar{\chi}] + \ii\sum_{\vx} \bar{\chi}(\vx) 
\sigma(\vx) \chi(\vx) + \frac{\Nf}{2g^2} \sum_{\vx} \sigma^2(\vx) \, ,
\qquad \Nf = 4
\end{equation}
actually results in a theory different from the GN model. To show
this, we insert again the Fourier representations of the fermionic
fields (\ref{EQN562}) as well as of the real scalar field $\sigma$,
\begin{eqnarray}
\sigma(\vx) = \frac{1}{\sqrt{\Nt\Ns}} \sum_{\vk} e^{\ii \vk\cdot\vx} \,\tilde{\sigma}(\vk)\,,
\end{eqnarray}
where $\tilde{\sigma}(-\vk) = {\tilde{\sigma}}^\ast(\vk)$ and 
the discrete momenta $\vk$ are chosen such that $\sigma$
is periodic in $x^0$ and $x^1$ direction. 
The action (\ref{EQN389}) becomes
\begin{equation} S_{\sigma}[\tilde{\chi},\tilde{\bar{\chi}},\tilde{\sigma}] = S_\textrm{free}[\tilde{\chi},\tilde{\bar{\chi}}] 
+ \frac{\ii}{\sqrt{\Nt\Ns }} \sum_{\vk} \sum_{\vk'} \tilde{\bar{\chi}}(\vk) \tilde{\sigma}(\vk-\vk') \tilde{\chi}(\vk') + \frac{\Nf}{2g^2}
\sum_{\vk} \big|\tilde{\sigma}(\vk)\big|^2 \, ,
\qquad \Nf = 4 \, .
\label{EQN721} 
\end{equation}
In the limit $a \rightarrow 0$ only the soft fermion modes 
contribute, as discussed in appendix~\ref{SEC488}. Note, 
however, that there is no kinetic term for the field $\sigma$ 
and, thus, no corresponding suppression of $\sigma$ modes.
Consequently, for $a \rightarrow 0$ the interaction term 
in eq.\ (\ref{EQN721}) can be written as
\begin{equation}
\frac{\ii}{\sqrt{\Nt\Ns }}\sum_{\vk,\vk'}\sum_{u,v,u',v'}
\bar{\tilde{\chi}}_{uv}(\vk)\tilde{\sigma}_{uv,u'v'}(\vk-\vk')
\tilde \chi_{u'v'}(\vk')\,,
\end{equation}
with symmetric kernel in momentum space
\begin{equation}
\tilde{\sigma}_{uv,u'v'}(\vk)=
\tilde{\sigma}_{u'v',uv}(\vk)=
\tilde{\sigma}(\vk_{uv}-\vk_{u'v'}+\vk)\,.
\end{equation}
In terms of the usual field coordinates $\tilde{\psi}_{uv}$,
related to $\tilde{\chi}_{uv}$ via eq.\ (\ref{EQN167a}), the interaction term is
\begin{equation}
\frac{\ii}{\sqrt{\Nt\Ns }} \sum_{\vk,\vk'}
\sum_{u,v,u'v'}\bar{\tilde{\psi}}_{uv}(\vk)
(\gamma^1)^v (\gamma^0)^u \sigma_{uv,u'v'}(\vk-\vk') (\gamma^0)^{u'} (\gamma^1)^{v'} \tilde{\psi}_{u'v'}(\vk') \, . \label{EQN400}
\end{equation}
Now it is obvious that the action (\ref{EQN389}) 
is not a discretization of the GN model with $\Nf = 4$ fermion 
flavors. While the four terms with $uv=u'v'$ in eq.\
(\ref{EQN400}) represent the correct GN interaction for the four fermion flavors,
\begin{eqnarray}
\frac{\ii}{\sqrt{\Nt\Ns }} \sum_{u,v} \ \sum_{\vk,\vk'} \tilde{\bar{\psi}}_{uv}(\vk) \tilde{\sigma}(\vk-\vk') \tilde{\psi}_{u'v'}(\vk') ,
\end{eqnarray}
there are twelve additional terms not present in the GN model, where the field $\sigma$ couples two different fermion flavors,
\begin{equation}
\frac{\ii}{\sqrt{\Nt\Ns }} \sum_{\vk,\vk'}\tilde{\bar{\psi}}_{10}(\vk) \gamma^0 \tilde{\sigma}(\pi/a+k_0-k_0',k_1-k_1')
\tilde{\psi}_{00}(\vk')+\text{eleven more terms}\,,
\end{equation}
as was already pointed out in Ref.\ \cite{Cohen:1983nr}. Including
these twelve terms in a numerical simulation, i.e.\ using the action
(\ref{EQN389}), corresponds to studying a different theory and leads
to results significantly different from those obtained with a
correct discretization of the GN model (examples are shown at the
end of this section).

\begin{figure}[t]
	\centering
	\includegraphics[width=0.6\linewidth]{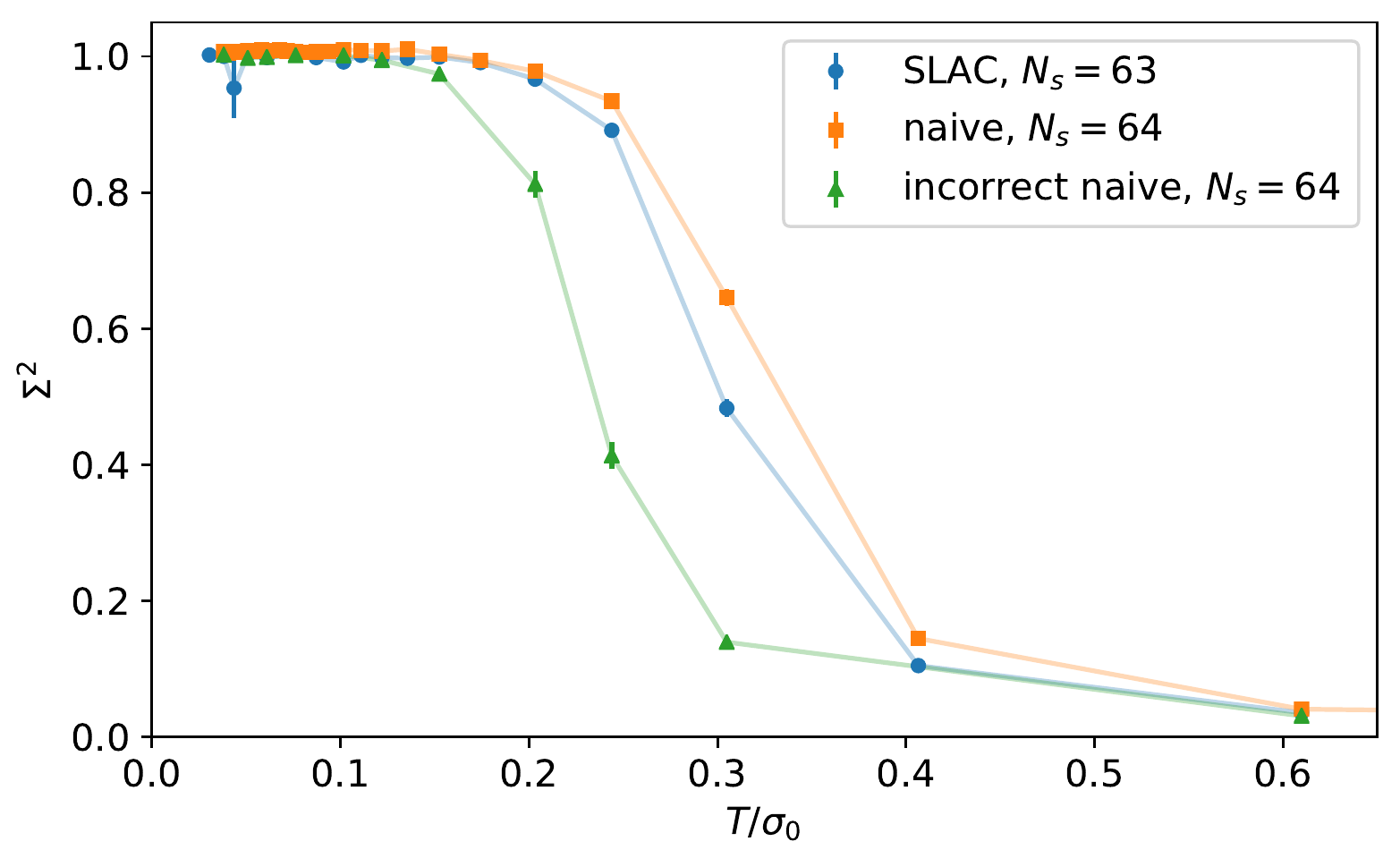}
	\caption{\label{FIG954}$\Sigma^2$ as a function of $T$ at $\mu = 0$ for SLAC fermions, naive fermions and the straightforward, but incorrect naive discretization (\ref{EQN389}) ($\Nf = 8$, $a \approx 0.410/\sigma_0$).}
\end{figure}
Now we derive a proper lattice discretization of the GN model. To this
end, we note that only the soft fermion modes contribute in the limit
$a \rightarrow 0$ and that the four correct interaction terms are
proportional to $\tilde{\sigma}_{uv,uv}$, while the twelve spurious
interaction terms are proportional to $\tilde{\sigma}_{uv,u'v'}$ 
with $uv\neq u'v'$. Thus, one can eliminate the spurious terms 
by replacing $\tilde{\sigma}(\vk)$ in the
interaction term in eq.\ (\ref{EQN721})
by $\tilde{W}(\vk) \tilde{\sigma}(\vk)$. Here
$\tilde{W}$ is a weight-function with
\begin{itemize}
	\item $\tilde{W}(\vk) \rightarrow 1$ for $\vk \approx \vk_{00} = (0,0)$ 
	(i.e.\ in region $\mathcal{R}_{00}$),
	\item $\tilde{W}(\vk) \rightarrow 0$ for $\vk \approx \vk_{uv}$ 
	with $(u,v)\neq (0,0)$ (i.e. in the other regions $\mathcal{R}_{uv}$).
\end{itemize}
A simple choice, which we use for our numerical simulations, 
is
\begin{equation}
\tilde{W}(\vk) = \frac{1}{4} 
\prod_{\mu=0}^{1}\Big(1+\cos(ak_\mu)\Big) \, .
\end{equation}
Expressing this modified action in terms of 
$\chi(x)$, $\bar{\chi}(x)$ and $\sigma(x)$ is straightforward,
\begin{equation}
\label{EQN892a} S_\textrm{GN}[\chi,\bar{\chi},\sigma] = S_\textrm{free}[\chi,\bar{\chi}] + 
\frac{\ii}{\sqrt{\Nt\Ns}}\sum_{\vx,\vx'} \bar{\chi}(\vx) W(\vx-\vx') \sigma(\vx') \chi(\vx) 
+ \frac{\Nf}{2 g^2} \sum_{\vx} \sigma^2(\vx) \, ,
\qquad \Nf = 4 \, ,
\end{equation}
where $W$ is the Fourier transform of $\tilde{W}$ and given in eq.\ (\ref{EQN892b}).
All calculations and arguments presented in this section
apply to $\Nf = 8, 12, 16, \ldots$ flavors in a trivial way. 
The generalization of eq.\ (\ref{EQN892a}) is 
eq.\ (\ref{EQN892c}).

In Figure~\ref{FIG954} we show numerical evidence that using the straightforward naive discretization of the GN model (\ref{EQN389}) leads to incorrect results, i.e.\ results not corresponding to the GN model. We plot $\Sigma^2$ as a function of the temperature $T$ for chemical potential $\mu = 0$. The blue and orange curves correspond to the SLAC discretization (see section~\ref{SEC432}) and the correct naive discretization (\ref{EQN892c}) (or equivalently (\ref{EQN892a})). These curves are rather similar and get closer, when decreasing the lattice spacing, indicating that they have the same continuum limit. The green curve, on the other hand, corresponding to the straightforward naive discretization (\ref{EQN389}) is quite different and does not approach the blue and orange curves, when decreasing the lattice spacing. We obtained similar results also for non-vanishing chemical potenial.

% **********

\clearpage

\section*{Acknowledgements}
We acknowledge useful discussions with Michael Buballa, Holger Gies, Felix Karbstein, Adrian K\"onigstein, Maria Paola Lombardo, Dirk Rischke, Alessandro Sciarra,
Lorenz von Smekal, Stefen Theisen, Michael Thies, Marc Winstel and Ulli Wolff
on various aspects of fermion theories and spacetime symmetries.
Special thanks go to Philippe de Forcrand for his constructive
remarks and to Martin Ammon
who shared his knowledge on SSB and Goldstone bosons and 
for his steady encouragement in the
past 15 months.

J.J.L. and A.W. have been supported by the Deutsche 
Forschungsgemeinschaft (DFG) under Grant No. 406116891 within the 
Research Training Group RTG 2522/1. L.P.\ and M.W.\ acknowledge support by the Deutsche Forschungsgemeinschaft (DFG, German Research Foundation) through the CRC-TR 211 ``Strong-interaction matter under extreme conditions'' -- project number 315477589 -- TRR 211. M.W.\ acknowledges support by the Heisenberg Programme of the Deutsche Forschungsgemeinschaft (DFG, German Research Foundation) -- project number 399217702.

This work was supported in part by the Helmholtz International Center for FAIR within the framework of the LOEWE program launched by the State of Hesse. Calculations on the GOETHE-HLR and on the on the FUCHS-CSC high-performance computers of the Frankfurt University were conducted for this research. We would like to thank HPC-Hessen, funded by the State Ministry of Higher Education, Research and the Arts, for programming advice.

We used the python programming language, most notably pandas \cite{mckinney_data_2010} and numpy \cite{van_der_walt_numpy_2011} for data analysis and matplotlib \cite{hunter_matplotlib_2007} for plotting but also further algorithms from the SciPy ecosystem \cite{scipy_10_contributors_scipy_2020}.

% ********************
% ********************
% ********************
% ********************
% ********************

\clearpage

% **********

% ********************
% ********************
% ********************
% ********************
% ********************

\end{document}